%% file: ppnp_EP_DI_IS.tex
\documentclass[twoside,12pt]{elsarticle}

\input{header}

\journal{Progress in Particle and Nuclear Physics}
\begin{document}

\title{ Photoproduction Reactions and Non-Strange Baryon Spectroscopy}

\author[uglasgow]{David~G.~Ireland}
\address[uglasgow]{SUPA, School of Physics and Astronomy, University of Glasgow G12 8QQ, UK}
\author[jlab]{Eugene~Pasyuk\corref{correspondingauthor}}
\cortext[correspondingauthor]{Corresponding author.}\ead{pasyuk@jlab.org}
\address[jlab]{Thomas Jefferson National Accelerator Facility, Newport News, VA 23606, USA}
\author[gwu]{Igor~Strakovsky}
\address[gwu]{The George Washington University, Washington, DC 20052, USA}

\begin{abstract} 
We review the last two decades of using photon beams to measure the production of mesons, and in particular the information that can be obtained on the spectrum of light, non-strange baryons. 
This is a compendium of experimental results, which should be used as a complement to theoretical reviews of the subject. Lists of data sets are given, together with a comprehensive set of 
references. An indication of the impact of the data is presented with a summary of the results. 
\end{abstract}
\maketitle
\tableofcontents

\section{Introduction}

Measurements of pion photoproduction on both proton and quasi-free neutron targets have a very long history, starting about 70 years ago with the discovery of the pion by the University of Bristol group~\cite{pion}. Two years later, at the 1949 Spring Meeting of the National Academy of Sciences, a preliminary account was given of some observations of mesons produced by the 335-MeV photon beam from the Berkeley synchrotron~\cite{pi+1}. Starting with the use of bremsstrahlung facilities, pioneering results for $\gamma p\to\pi^0
p$~\cite{pi0a,pi0b,pi0c,pi0d,pi0e}, $\gamma p\to\pi^+ n$~\cite{pi+a,pi+b,pi+c,pi+d}, and $\gamma n\to\pi^-p$~\cite{pi-} were obtained. Despite all the shortcomings of the first measurements (such as large normalization uncertainties, wide energy and angular binning, limited angular coverage and so on), these data were crucial for the discovery of the first excited nucleon state, the $\Delta(1232)3/2^+$,~\cite{Delta}. 

Whilst the ability of photoproduction measurements to deliver information on baryon resonances had been shown from an early stage, most of the light baryon spectrum states and their properties were subsequently obtained by pion-nucleon scattering. Until the end of the 1970s, meson photoproduction was essentially only able to confirm pion scattering data, without adding a substantial amount of additional information. Indeed, the evolution of particle physics towards energies beyond the regime in which hadronic states are the relevant degree of freedom suggested to some that the study of the light baryon spectrum had come to an end, if not a conclusion. This was summarized in a 1983 review article ``Baryon Spectroscopy" by Hey and Kelly~\cite{Hey83} who stated in their introduction:
\begin{quote}
``Baryon spectroscopy is now thirty years old and perhaps approaching a mid-life crisis. For it is inevitable in such a fast-moving field as high energy particle physics, that experiments have moved on beyond the resonance region to higher energies and different priorities. Thus it is probably no exaggeration to say that we now have essentially \emph{all} the experimental data relevant to the low-energy baryon spectrum, that we are \emph{ever} likely to obtain." 
\end{quote}
\begin{figure}[hbt] 
\begin{center}
\includegraphics[width=0.8\columnwidth,trim=4cm 2.5cm 0 0]{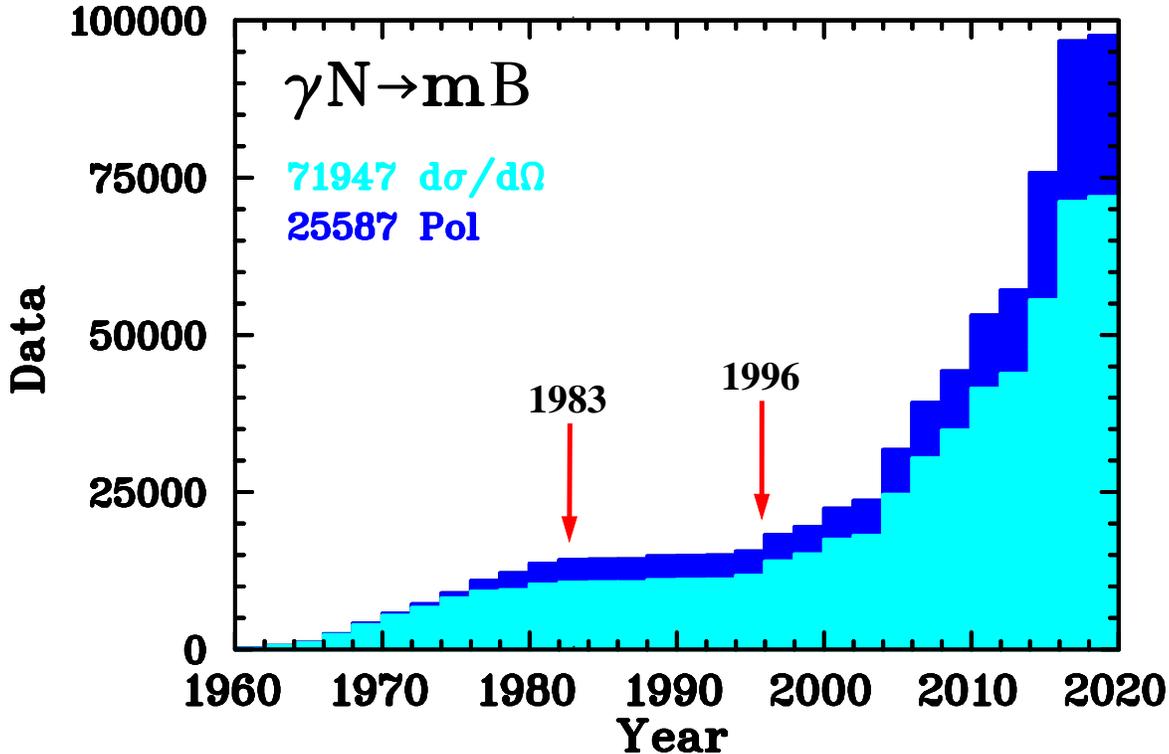}
\caption{\label{fig:evolution}Stacked histogram of full database for single meson photoproduction $\gamma N \to m B$. m~=~($\pi$,~$\eta$,~$\eta'$,~$K$,~$\omega)$, B~=~($n$,~$p$,~$\Lambda$,~$\Sigma$). Light shaded -- cross sections, dark shaded -- polarization data. Experimental  data from the SAID database~\protect\cite{SAID}. }
\end{center}
\end{figure}

Armed with the benefit of hindsight, we beg to differ! The 1980s saw several advances in accelerator technology that enabled the production of photon beams of the order of a GeV in energy, whose energy could be accurately enough determined through the tagging of degraded electrons in bremsstrahlung, or via laser backscattering from electron beams. These facilities initially concentrated on photonuclear research, but as soon as the threshold for pion production was reached, it became clear that photon beams for hadron physics research was a reality.

Nevertheless, it took a while for this potential to be realized, which is clearly demonstrated in Figure~\ref{fig:evolution}. This plot shows the increase in the worldwide dataset for photoproduction reactions as a function of year. One can readily see that by the time of the Hey and Kelly 
review~\cite{Hey83} (1983), the amount of new data being obtained was indeed tailing off, so their pessimism about more data was at the time well-founded. It took until the turn of the 21st century before a substantial increase was seen. The beginning of the exponential rise in the number of data points around 1996 therefore serves as a starting point for this current review. 

The plot also does not indicate the relative improvement in the accuracy of the data, which can only be appreciated by delving into the relevant literature. Where initial measurements showed rough energy and angular dependencies, more recent results have been obtained that allow energy scans and fits to angular distributions that allow sophisticated partial wave analyses (PWA) that were previously only possible with pion scattering data. 

The scope of this review may seem to be somewhat narrow (a particular set of reactions and only the lightest sector of the baryon spectrum). However, we have limited ourselves to this scope not only to avoid an enormous task of covering all of baryon spectroscopy, but to point out that our knowledge of the light baryon spectrum is not yet complete and that there is a vigorous amount of activity devoted to extracting as much information as possible from the most recent, precise and statistically accurate measurements. In addition, measurements of photoproduction reactions, and in particular those on pseudoscalar meson photoproduction including polarization have now been carried out. It is therefore timely to review this work.

In this review, we concentrate on the measurements of physical quantities, and the information that can be extracted from them. We are less concerned with theoretical interpretations other than the identification of new resonances, and leave a discussion of different models to other excellent reviews (e.g., Ref.~\cite{CrRo13}). In this sense, we are taking a \emph{phenomenological} point of view, but our aim is to tie together the many different experimental results over the last couple of decades, and present this unified overview as a starting point for further serious assaults on the understanding of the light baryon spectrum from first principles.

We start with an overview of formalism for dealing with measured data in Section~\ref{sec:formalism}, followed in Section~\ref{sec:howto} by a description of how information can be extracted from the data. In Section~\ref{subsec:exper}, we review various experimental facilities that have been used to obtain the data sets, which are described and sorted by final state in Section~\ref{sec:available}. Some concluding remarks are given in Section~\ref{sec:conclusion}.

\section{\label{sec:formalism}Formalism for Photoproduction Reactions}

Experiments only ever measure counts. For a specific beam intensity, hitting a target with a specific density of scattering centers in a specific state of polarization reacting to give a specific final state, whose particles have specific spin orientations, all that an experiment will do is to register counts. 
The registered counts are subject to the efficiency of the detection apparatus, both in sensitivity and in correctly identifying the desired combination of particles. Advances in experimental technologies are aimed at improving this efficiency so that more complicated measurements can be performed. In the last couple of decades there have been many such advances that have been relevant to photoproduction reactions, including: control and polarization of photon beams, development of polarized gas and solid targets, construction of large solid angle detectors, development of higher rate data acquisition systems and of data analysis and statistical techniques. 

What is recorded by an experiment is most likely a distribution of counts in the space of independent kinematic variables, which includes the effect of potentially complicated resolution effects due to the detection apparatus. The data analysis process tries to minimize the resolution effects and to quantify the associated uncertainties (systematic uncertainties). The processed data are then used to estimate physically meaningful quantities, either by binning the counts in one or more dimensions, or by treating the data event-by-event. In any case, there is always uncertainty associated with a finite number of counts (statistical uncertainties).

What are commonly referred to as \emph{observables} are usually theoretical constructs of physically meaningful quantities, and are derived from a consideration of the contributing quantum mechanical amplitudes. Being able to extract information at the amplitude level is therefore seen as a goal of these campaigns, since no more information is available to us, even in principle. Since amplitudes are complex functions, there is always an unknown phase.

A number of amplitude schemes are commonly employed, and the concept of combining
observables to realize a \emph{complete} experiment has arisen over the years, which would allow the extraction of all relevant amplitudes up to an unknowable phase. However, given that the observables themselves are related to distributions of measured counts, it is worth stressing that the concept of a complete experiment is only mathematically meaningful. 

In practical terms, one does not ``observe'' observables. One measures counts, either as total intensities or as asymmetries for experimental configurations that can be constructed with combinations of polarized beam, target and recoils. The extensive work done to study the theoretically complete experiment~\cite{Barker1975,fasano_spin_1992,Chiang1997}  can perhaps best be utilized by combining it with an approach to quantify the information content of polarization measurements~\cite{ireland_information_2010}, as a guide to developing the most informative measurements.

In this section we describe both the formalism and how to extract estimates of observables from measurements. We then indicate how this information can be utilized to gain insight into the light baryon resonance spectrum. We concentrate on single pseudoscalar meson photoproduction, since it is the most straightforward reaction in terms of measurement and formalism, to give a flavour of the relevant issues. Double pseudoscalar meson and vector meson photoproduction require more complicated formalisms, and we will refer the reader to the relevant literature in the interests of saving space.

\subsection{Amplitudes and Observables}

\subsubsection{Single Pseudoscalar Meson Photoproduction}

Single pseudoscalar meson photoproduction involves the interaction of a photon with a free proton, a bound neutron or a whole nucleus. For studies of the baryon spectrum, we are normally interested in the first two of these. So a spin-1 particle (the photon, two helicity states) and a spin-$\frac{1}{2}$ particle (the nucleon) react to give a spin-0 particle (the pseudoscalar meson) and a spin-$\frac{1}{2}$ particle (the recoiling baryon). This gives eight spin combinations, of which four are possible within the parity-conserving strong interaction that has taken place. The four combinations are represented as amplitudes, the exact form of which is a matter of choice. Common options are CGLN~\cite{chew_relativistic_1957}, helicity amplitudes~\cite{jacob_general_1959} and 
transversity amplitudes~\cite{Barker1975}. Within any of these bases, there are 16 possible bilinear combinations that are referred to as the ``observables''.

To illustrate this in detail, a completely general expression for the cross section of these reactions following Ref.~\cite{Sandorfi11}, with the explicit dependence on the observables, is given below: 


\stepcounter{equation}
\setcounter{exno}{\theequation}
\newcounter{subexno}
\renewcommand\thesubexno{\alph{subexno}}
\renewcommand\theexno{\arabic{exno}\thesubexno}

\input{ps-cross-section}
\begin{figure}[hbt]
\begin{center}
\includegraphics[width=0.8\columnwidth]{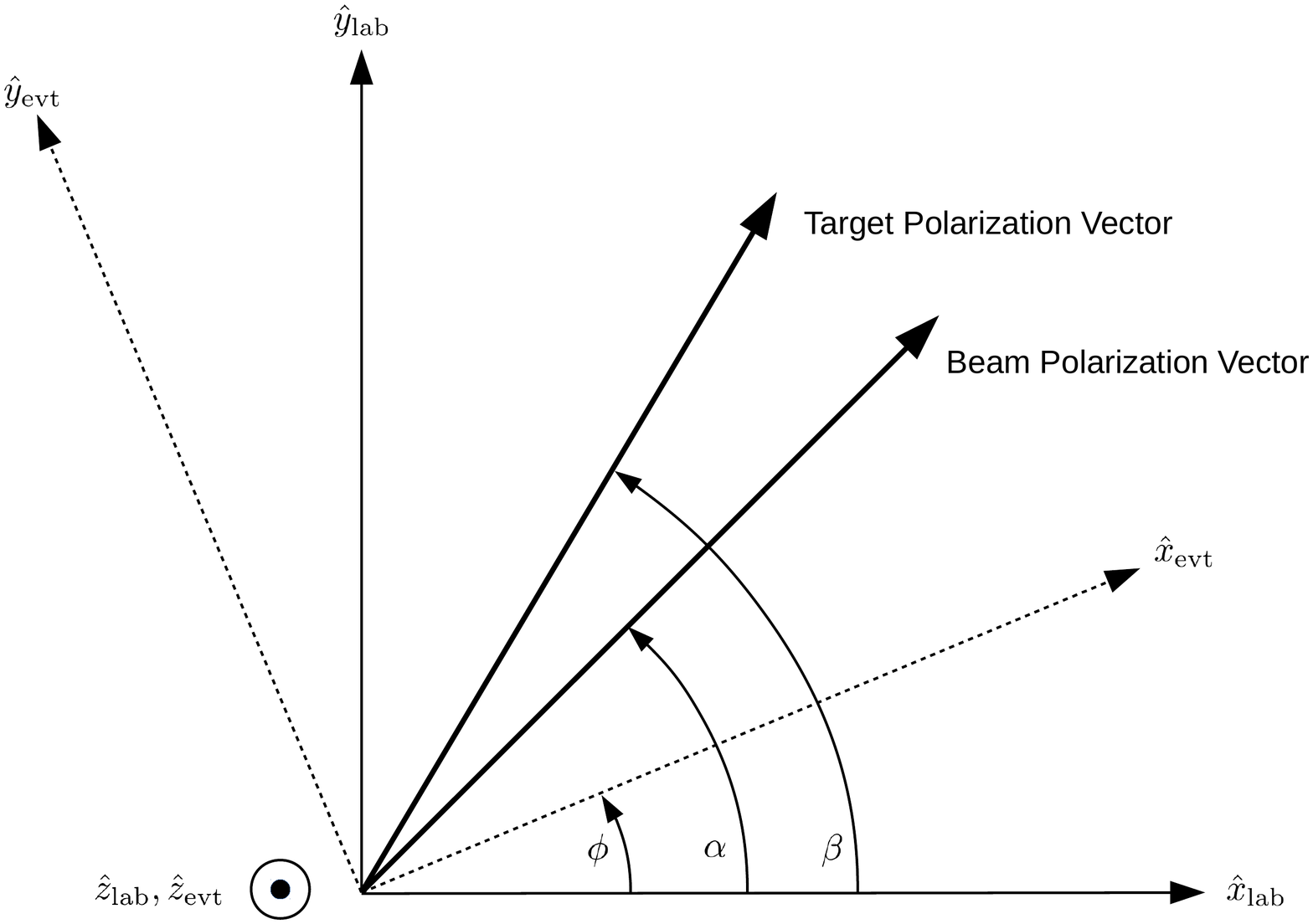}
\caption{\label{fig:axes} The definitions of laboratory and event axes, as well 
    as azimuthal angles. The common laboratory, center-of-mass and 
    event $z$-axis is directed out of the page. The lab $x$- and $y$-axes 
    are in the horizontal and vertical directions, and the event $y$-axis 
    is normal to the reaction plane.}
\end{center}
\end{figure}

In these equations, $\sigma_0$ denotes unpolarized cross section, $P^\gamma_L$ denotes degree of linear photon polarization, $P^\gamma_{\odot}$ denotes degree of circular photon polarization, $P^T_{x,y,z}$ and $P^R_{x^\prime,y^\prime,z^\prime}$ describe target and recoil baryon polarization components.  The angle $\phi$ is the azimuthal angle of the reaction plane, which is defined in the diagram in Figure~\ref{fig:axes}. 

$\obs{\Sigma}$, $\obs{T}$, and $\obs{P}$ are the single beam, target and recoil spin asymmetries. $\obs{E}, \obs{G}, \obs{H}$ and $\obs{F}$ are the beam-target double spin asymmetries; 
$\obs{C_{x^\prime}}, \obs{C_{z^\prime}}, \obs{O_{x^\prime}}$ and $\obs{O_{z^\prime}}$ are the beam-recoil double spin asymmetries; $\obs{T_{x^\prime}}, \obs{T_{z^\prime}}, \obs{L_{x^\prime}}$, and $\obs{L_{z^\prime}}$ are the target-recoil double spin asymmetries. The primes refer to a coordinate system in which $\hat{z^{\prime}}$ is parallel to the pseudoscalar meson momentum, $\hat{y^{\prime}}$ is normal to the scattering plane and $\hat{x^{\prime}} = \hat{y^{\prime}}\times \hat{z^{\prime}}$. 
The unprimed coordinate system has $\hat{z}$ parallel the photon momentum, $\hat{y}$ is normal 
to the scattering plane and $\hat{x} = \hat{y}\times \hat{z}$. 

In Equations~(\ref{eq:xsec}\ref{eqn:cs1}) to~(\ref{eq:xsec}\ref{eqn:cs16}), it can be seen that each observable enters twice. This means that there are always experimental configurations that can be used to extract the values, some of which require triple polarization measurements. Whilst not strictly required, the extraction of observables from two experimental configurations is desirable in order to reduce systematic uncertainties.

\subsubsection{Two Pseudoscalar Meson Photoproduction}

\begin{figure}[hbt]
\begin{center}
\includegraphics[width=0.8\columnwidth]{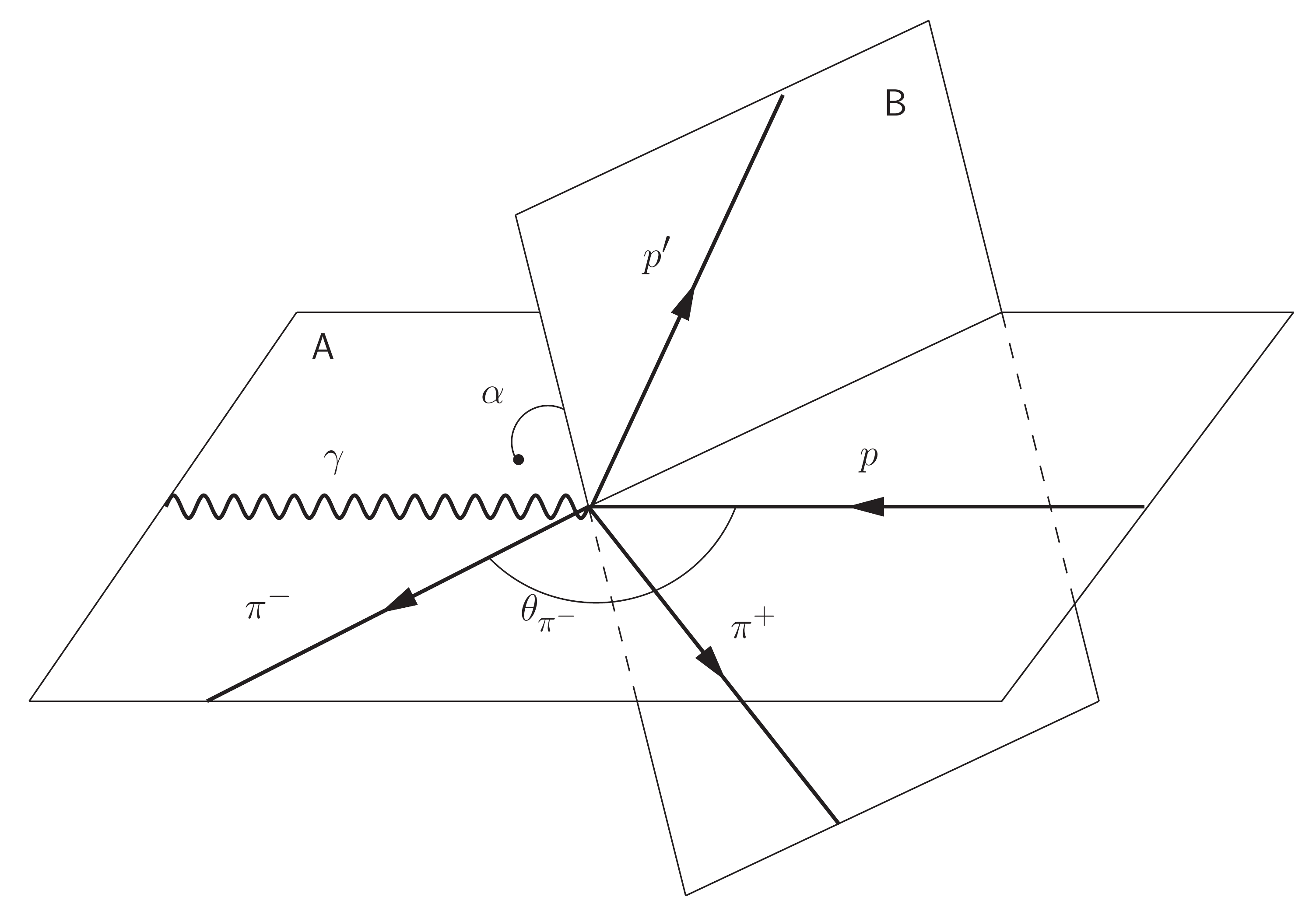}
	\protect\caption{\label{fig:2pikin} Angular kinematic variables for the 
    reaction $\gamma p\to\pi^+\pi^-p'$ in the CM frame. The set with 	
    $i$=$\pi^-$, $j$=$\pi^+$, and $k$=$p'$ includes the angular 		
    variables for $\theta_{\pi^-}$, the polar angle of the $\pi^-$, and 	
    $\alpha{[\pi^-p][\pi^+p']}$, which 	is the angle between the planes $A$ 
    and $B$, where plane $A$ ($[\pi^-p]$) is defined by the 3-momenta of the 
    $\pi^-$ and the initial state proton and plane $B$ ($[\pi^+p']$) 
    is defined by the 3-momenta of the $\pi^+$ and the final state proton 
    $p'$. The polar angle $\theta_{p'}$  is relevant for the set with 
    $i$=$p'$, $j$=$\pi^+$, and $k$=$\pi^-$, while the polar angle 
    $\theta_{\pi^-}$ belongs to the set with $i$=$\pi^+$, $j$=$p'$, and 
    $k$=$\pi^-$.} 
\end{center}
\end{figure}
Reactions such as $\gamma p\to p\pi^+\pi^-$, as well as any other two pseudoscalar mesons reactions, have three body final states and therefore have more independent kinematic variables and observables. 
Figure~\ref{fig:2pikin} illustrates the momenta involved. Formalism relating amplitudes and observables for these reactions is discussed detail in Ref.~\cite{Roberts05}. Writing a full cross section formula in component form is not practical, so we illustrate a more special case with a vector notation. In case of polarized photons and polarized target, using notation consistent with the previous section, the cross section can be written as :
\begin{align}
    \obs{d\sigma^{B,T}}(\vec{P^\gamma},\vec{P^T},x_i)
	= \obs{d\sigma_0} \{(1 &+ 		
    \vec{P^T}\cdot\obs{\vec{P}}) \nonumber \\
	&+  \circpol (\obs{I^{\odot}} + \vec{P^T}\cdot\obs{\vec{P}^{\odot}}) 	   
    \nonumber\\
	&+ \linpol [\sin 2(\alpha - \phi) (\obs{I^s} + \vec{P^T}\cdot\obs{\vec{P}^s}) \nonumber\\
	&+ \cos 2(\alpha - \phi) (\obs{I^c} + \vec{P^T}\cdot\obs{\vec{P}^c})], 	  \}
\end{align}
where:\\
$\obs{d\sigma_0}$ is the unpolarized cross section; \\
$\alpha - \phi$ is the angle between photon polarization and reaction plane; \\
$x_i$ represents all the kinematic variables; \\
$\circpol, \linpol$ are the degrees of circular or linear photon polarization; \\
$\vec{P^T}$ is the target nucleon polarization $(P^T_x, P^T_y, P^T_z)$;\\
Track changes is on

The observables in this case are:\\
$\obs{I^{\odot , s, c}}$ single spin beam asymmetries associated with polarized photons;\\
$\obs{\vec{P}}$ target asymmetry $\obs{(P_x, P_y, P_z)}$;\\
$\obs{\vec{P}^{\odot ,s,c}}$ double spin asymmetries 
$\obs{(P^{\odot}_x, P^{\odot}_y, P^{\odot}_z)}$,
$\obs{(P^s_x, P^s_y, P^s_z)}$,$\obs{(P^c_x, P^c_y, P^c_z)}$.\\
In these reactions there are a total of 64 possible observables. In practice, however, it would be extremely challenging to extract all of these with reasonable accuracy, so published experiments 
tend to concentrate on a few of them.

\subsubsection{Vector Meson Photoproduction}

With a spin-1 vector meson in the final state, the number  of underlying helicity amplitudes is 12, which would require a total of 23 independent observables at each energy and angle to extract. As with the full suite of two-pion spin observables, it may never be practical to extract all of them.

The decay angular distributions of the vector mesons can be examined to extract some of the spin density matrix elements (SDMEs). 
 A comprehensive guide to this formalism is given in Ref.~\cite{schilling_analysis_1970}. The SDMEs are defined in the rest frame of the vector meson, however the effects of resonances and other mechanisms relevant to the low energy baryon spectrum require that spin observables be measured in the $\gamma$N center of mass frame~\cite{kloet_spin_1998}.

In this review, we restrict ourselves to $\omega$ photoproduction; $\rho$ photoproduction is predominantly analysed in the context of two pion photoproduction, and other light vector mesons such as the $\phi(1020)$ and $K^{\ast}(892)$ have hitherto had limited impact on studies of the light baryon spectrum. 

\section{\label{sec:howto} How to Extract Observables and Amplitudes from the Data}

The number of counts $N$ registered in a detector of efficiency $\eff$, subtending solid angle $d\Omega$ and in a measurement of luminosity $\lum$ for a total time $T$, is given 
by 
\begin{equation}
	N = \eff^{-1} \int_0^T \lum dt \int 
	\dfrac{d\sigma \left( \theta, \phi \right)}{d\Omega} 	 d\Omega,
\end{equation}
where the \emph{efficiency} is the ratio of the number of particles of interest identified by the detector to the number of the particles passing through the solid angle, the \emph{luminosity} is a (possibly time-dependent) product of beam flux and density of scattering centers. The process also depends on beam energy. To simplify notation we write that for a specific experimental configuration $i$,
\begin{equation}
	N_i = \eff_i^{-1} \lum_i \sigma_i,
\end{equation}
where it is implicit that $\lum_i$ is an integrated luminosity for the configuration, and that $\sigma_i$ is the differential cross section, which could depend on energy and scattering angles\footnote{We will simply refer to these quantities as ``luminosity'' and ``cross section'' hereafter.}.

The efficiency and the luminosity are experiment-dependent, whereas the cross section contains all the physics information and is a link to theoretical models of the reaction.

The main observable for any reaction of interest is the cross section, and its determination as a function of energy and angle requires careful setup and handling of the beam, target and detector systems, in order to obtain an accurate value for the luminosity and efficiency of the experiment. If the experiment is setup so that the spin configuration of beam, target or recoils is not fixed then the cross section represents a sum over initial spins and an average over final spins. If the experiment does contain an element of polarization, then the distribution of cross section will contain additional dependence on the kinematics of the reaction and the degrees of polarization. Since theoretical models of cross sections are calculated from coherent sums of amplitudes that are dependent on the individual spin combinations of beam, target and recoiling products, it is desirable to evaluate these as well. 

Table~\ref{tab:cross-sections} summarizes the distributions for the various experimental configurations, where we again limit the discussion to single pseudoscalar meson photoproduction. The main point of this table is to illustrate that as more elements of the experimental configuration are polarized, the more complicated is the dependence of the intensity distribution on the number of observables. \pagebreak[10]

\stepcounter{equation}
\setcounter{exno}{\theequation}
\setcounter{subexno}{0}
\vspace{-\parskip}
\input{exp-config-table}

Rather than measuring cross-sections for specific polarization configurations, a common technique is to access them by measuring \emph{asymmetries}. Defining in general the notation for asymmetry in the number of counts between two experimental configurations $i$ 
and $j$
\begin{equation}
	A_N = \dfrac{N_i - N_j}{N_i + N_j}
	= \dfrac{\eff_i^{-1} \lum_i \sigma_i - \eff_j^{-1} \lum_j \sigma_j}	
    {\eff_i^{-1} \lum_i \sigma_i + \eff_j^{-1} \lum_j \sigma_j},
\end{equation}
and introducing the further notation
\begin{equation}
	A_\lum=\frac{\lum_i-\lum_j}{\lum_i+\lum_j};
	\qquad
	A_\eff=\frac{\eff_i-\eff_j}{\eff_i+\eff_j};
	\qquad
	A_\sigma=\frac{\sigma_i-\sigma_j}{\sigma_i+\sigma_j};
\end{equation}
we find
\begin{equation}
	A_N = \dfrac{A_\sigma + A_\lum - A_\eff
	- A_\sigma A_\lum A_\eff}{1 - A_\lum A_\eff - A_\sigma A_\eff + 	
    A_\sigma  A_\lum}.
\end{equation}

In most cases, the difference in efficiency between two settings will be 
close to, if not identically, zero, and the expression simplifies to
\begin{equation}
	A_N = \dfrac{A_\sigma + A_\lum}{1 + A_\sigma  A_\lum},
\end{equation}
which shows that if $A_\lum$ can be made small (i.e., the luminosity in 
the two settings is roughly equal), the main driver in the asymmetry of counts will be in $A_\sigma$, which contains the physics quantities of interest.

For a given setting $\sett$ of a configuration of beam and target polarization, the cross section formula can be written in a simple form
\begin{equation}
	\sigma = u + \sett v,
\end{equation} 
where $u$ is a function of everything that does not depend on the setting $\sett$ and $v$ is a function of everything that does depend on it. If we have two settings, $\sett_i$ and $\sett_j$ then
\begin{equation}
	A_\sigma = \frac{\left( \sett_i - \sett_j \right) v}	
	{2u + \left( \sett_i + \sett_j \right) v},
\end{equation}
so that if we can arrange $\sett_j = - \sett_i$  this would maximally 
isolate the function $v$ in the asymmetry. This may not be possible to 
achieve in practice, so if the best we can do is $\sett_j = 2\delta - \sett_i$, 
where $\delta$ represents half the difference in \emph{degree} 
of polarization between the two settings, then
\begin{equation}
\label{eq:asymm-sig}
	A_\sigma = \frac{\left( s_i  + \delta \right) v}{u + 	\delta v},
\end{equation}
where $s_i \in [0,1]$ is the degree of polarization in setting $\sett_i$.

To make this less abstract, we give in Table~\ref{tab:asymmetries} some examples 
of $A_\sigma$s for a range of beam and target polarization 
settings. For clarity we take $\delta = 0$, so that $A_\sigma = s_i v / u$ 
but note the straightforward extension to Eq.~(\ref{eq:asymm-sig}) if the 
degree of polarization is different between settings. We include the 
terms related to recoil polarization measurement, which can be removed 
if recoil polarization is not determined (i.e.,~set $\recpol_{x^\prime}=
\recpol_{y^\prime}=\recpol_{z^\prime}=0$). Note that in some cases, such as 
the identification of $\Lambda$s from the decay to $\pi p$ by detecting 
the pion or proton, there will be sensitivity to recoil polarization, so 
those terms cannot be removed.

\stepcounter{equation}
\setcounter{exno}{\theequation}
\setcounter{subexno}{0}

\input{asymm-config-table}

Tables~\ref{tab:cross-sections} and \ref{tab:asymmetries} show that in practice 
observables are always measured in combinations. The final, but most technically 
challenging measurement, given by Eq.\,(5\ref{row18}) in 
Table~\ref{tab:cross-sections} is perhaps the nearest one could claim 
to being a ``complete experiment" as it is sensitive to a ``complete 
set" of observables, but note that it is additionally sensitive to 
several more observables. The more important challenge is to perform measurements 
with sufficient accuracy. A rule of thumb is that 
pseudoscalar photoproduction observables need to be measured to 
better than $\pm \sim 0.5$ to provide any information.

\subsection{How to Extract Parameters of Nucleon Resonances from the Photoproduction Data} \label{subsec:extract}

Very simply put, one constructs a data model whose parameters are explicitly or implicitly related to physical parameters such as masses, branching ratios and coupling constants. The data model can be constructed from a physics model of the reaction. Physics models can vary from simply describing a single reaction channel at the tree level, to complicated coupled-channel models that require the analysis of \emph{any} reaction that can kinematically contribute to a final state. The advantage of a single-channel reaction model is that it is relatively straightforward to calculate and to obtain a rough idea of the main contributions from resonances. The disadvantage of this is that the extracted parameters are more difficult to interpret when comparing results for different channels. A coupled-channels approach on the other hand allows one to extract coupling constants and other parameters in such a way as to be consistent between channels, at the expense of having to estimate sometimes hundreds of parameters, which requires heavy computational resource.

In doing this there are a number of complications. For instance, how does one choose which resonant states to include? This is a model comparison problem, since adding more resonances will mean the addition of more parameters, thereby making a fit to the data easier. On the other hand, an Occam's razor approach to keep the model as simple as possible should act to reduce the number of resonances that require to be invoked.

Alternatively one may want to extract information in a ``model independent'' way. By analyzing distributions in energy and angle, a partial-wave analysis (PWA) can be carried out in which the intensity and phase of each partial wave can be examined to determine the contributions of different resonances. Again, there is a model comparison issue with the question of how many partial waves to include in fits.

Originally, PWA arose as the technology to determine the amplitudes of a reaction through fitting scattering data. This is a non-trivial mathematical problem -- looking for a solution of an 
ill-posed problem, as described in Hadamard~\cite{Ha02} and Tikhonov~\cite{Ti77}. Resonances appeared as a by-product (bound states objects with definite quantum numbers, mass, lifetime and so on). Standard PWA reveals resonances that are not too wide ($\Gamma<$ 500~MeV) and possess  a large enough elastic branching ratio (BR $>$ 4\%). It is possible, however,  to miss narrow resonances with $\Gamma<$ 30~MeV~\cite{Ar09}.

Whether one wants to extract physics from the data by fitting model parameters or projecting out partial waves, there is a choice as to how to use the data. If the phenomenology group is well enough connected with the experiments, it can be possible to construct likelihood functions on an event-by-event basis. This approach does require high numbers of events for the results to be robust, but means that quantities are not averaged over regions of phase space. A more common interface between experiment and theory is for the experimenters to report the values of observables, which have been binned in energy and angles. At the current levels of accuracy, both approaches are yielding similar results.

\subsubsection{Resonance Parameters}

The main objectives of PWA schemes, apart from establishing the existence of resonances, are to derive estimates of resonance properties such as mass, width, branching ratios, couplings, etc. Calling an object a resonance implies that there is a resonant frequency and an associated width that characterizes the state. By analogy with mechanical resonances  Breit-Wigner (BW) parameters, mass and width, can be used to describe each resonance, but their exact values depend on the model-dependent method of extraction.  The preferred approach, as described in the Review of Particle Physics~\cite{PDG_2018}, is for an analysis to estimate the position of poles in the complex energy plane.

\subsection{Reactions on Neutron Targets}

Only with good data on both proton and neutron targets, can one hope to disentangle the isoscalar and isovector electromagnetic couplings of various $N^\ast$ and $\Delta^\ast$ resonances, as well as the isospin properties of non-resonant background amplitudes~\cite{Wa52,Wa69}.

Unfortunately, there is no free neutron target. The radiative decay width of neutral baryons may be extracted from $\pi^-$ and $\pi^0$ photoproduction from neutrons, but in practice one can only use a target containing a bound neutron. To extract relevant information one requires the use of model-dependent final-state interaction (FSI) corrections~\cite{Mi55,Wa52}. There is no way to isolate FSI experimentally~\cite{FSIpi-,FSIpi0}.

At lower energies (E < 700~MeV), there are data for the inverse $\pi^-$ photoproduction reaction,  $\pi^-p\to\gamma n$. This process is free from complications associated with a deuteron target. However, there is a major disadvantage of using $\pi^-p\to\gamma n$: there is a large background from $\pi^-p\to\pi^0n\to\gamma\gamma n$ reactions, whose cross section is 5 to 500 times larger than $\pi^-p\to\gamma n$.   

\begin{figure}[H]
\begin{center}
\includegraphics[width=0.9\columnwidth,trim={0 5cm 0 4cm},clip]{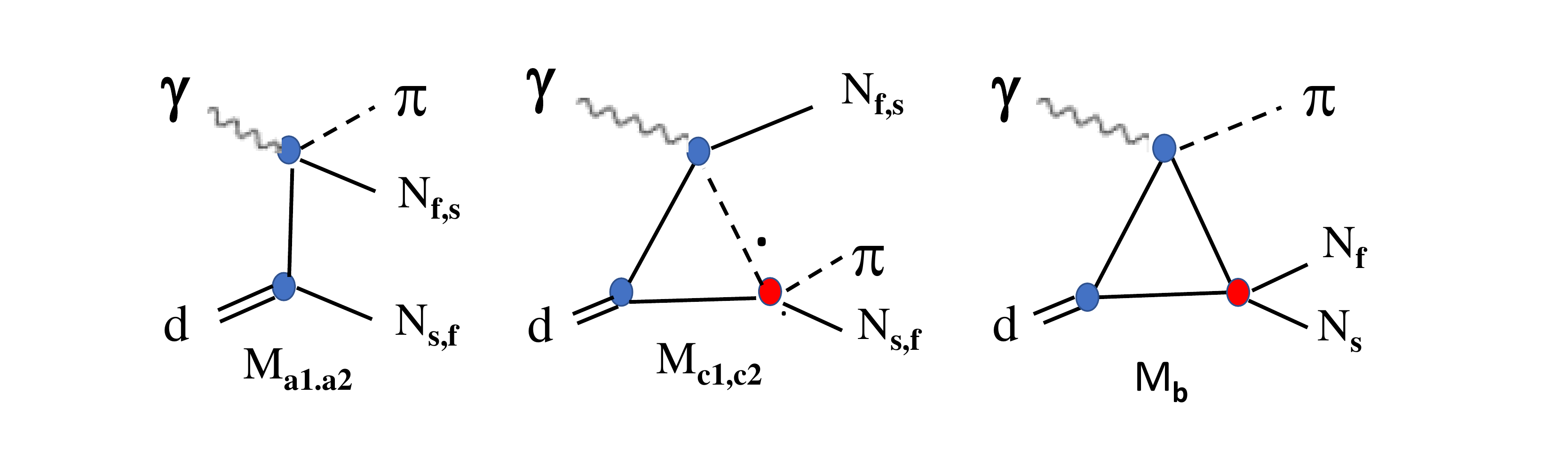}
\caption{\label{fig:fs1} The IA ($M_{a1}$, $M_{a2}$), NN-FSI ($M_b$), 	
    and $\pi N$ ($M_{c1}$, $M_{c2}$) diagrams for the reaction $\gamma 	
	d\to\pi N$. Wavy, solid, dashed and double lines correspond to the 
    photon, nucleons, pion, and deuteron, respectively.}
\end{center}
\end{figure}

Studies of the $\gamma n\to\pi^-p$ and $\gamma n\to\pi^0n$ reactions can be carried out in quasi-free kinematics with deuteron targets. The reactions $\gamma d\to\pi^-p(p)$ and $\gamma d\to\pi^0n(p)$ in these kinematics have a fast, knocked-out nucleon and a slow proton spectator, and the slow proton is assumed not to be involved in the pion production process. In this quasi-free region, the reaction mechanism corresponds to the ``dominant" impulse approximation (IA) diagram in Figure~\ref{fig:fs1}(a) with the slow proton emerging from the deuteron vertex.  Here, the differential cross section on the deuteron can be related to that on the neutron target in a well understood way~\cite{FSIpi-,FSIpi0}. Figure~\ref{fig:fs1} illustrates this dominant IA diagram, as well as the leading terms of FSI corrections. 

An energy and angle dependent FSI correction factor, $R(E,\theta)$, can be defined as the ratio 
between the sum of three dominant diagrams in Figure~\ref{fig:fs1} and IA (the first of the diagrams). This can then be applied to the experimental $\gamma d$ data to get a two-body cross section for $\gamma n\to\pi^-p$ and $\gamma n\to\pi^0n$.

The GWU SAID database contains phenomenological amplitudes for the reactions $\pi N\to\pi N$~\cite{Ar06}, $NN\to NN$~\cite{Ar07}, and $\gamma N\to\pi N$~\cite{DU07}.  The GW-ITEP group, for example, used these amplitudes as inputs to calculate the dominant diagrams of the GWU-ITEP FSI approach. The full Bonn potential~\cite{Bonn} was then used for the deuteron description, which includes the Fermi motion of nucleons. 

The GWU-ITEP FSI calculations~\cite{FSIpi-} are available over a broad energy range (threshold to E = 2.7~GeV), and
for the full CM angular range ($\theta = 0^\circ$ to $180^\circ$). Overall, the FSI correction factor $R < 1.00$, while its value varies from 0.70 to 0.90 depending on the kinematics.  The behavior of $R$ is very smooth vs pion production angle. There is a sizable FSI effect from the S-wave part of pp-FSI at small angles. 

$R(E,\theta)$ is used as the FSI correction factor for the CLAS quasi-free $\gamma d\to\pi^-pp$ cross section averaged over the laboratory photon energy bin width~cite{CH12,MA17}. Note  that the FSI correction grows rapidly to the forward direction ($\theta < 30^\circ$). There are currently few measurements in this regime, so the uncertainty due to FSI for this reaction at forward angles does not cause too much concern. The contribution of uncertainty in FSI calculations to the overall systematic normalization uncertainty is estimated to be about 2-3\% (the sensitivity to the deuteron wave-function is 1\% and to the number of steps in the integration of the five-fold integrals is 2\%). For the CLAS measurements, no sensitivity was found to the value of proton momentum used to determine whether or not it is a spectator. 

The $\gamma n\to\pi^0n$ measurement is much more complicated than the case of $\gamma n\to\pi^-p$ because the $\pi^0$ can come from both neutron and proton initial states. The GW-ITEP studies have shown that photoproduction cross sections from protons and neutrons are generally not equal~\cite{FSIpi0}. For  $\pi^0$ photoproduction on  proton and neutron targets we have
\begin{equation}
	A(\gamma p\to\pi^0p) = A_v + A_s~~~~{\mathrm and}~~~~ 
    A(\gamma n\to\pi^0n) = A_v - A_s,
\end{equation}
where $A_v$ and $A_s$ are the isovector and isoscalar amplitudes, respectively.  Therefore, if  $A_s\neq 0$ the $\gamma p$ and $\gamma n$ amplitudes are not equal.

\begin{figure}[H]
\begin{center}
\includegraphics[width=0.3\columnwidth,trim=5cm 1.5cm 0 0]{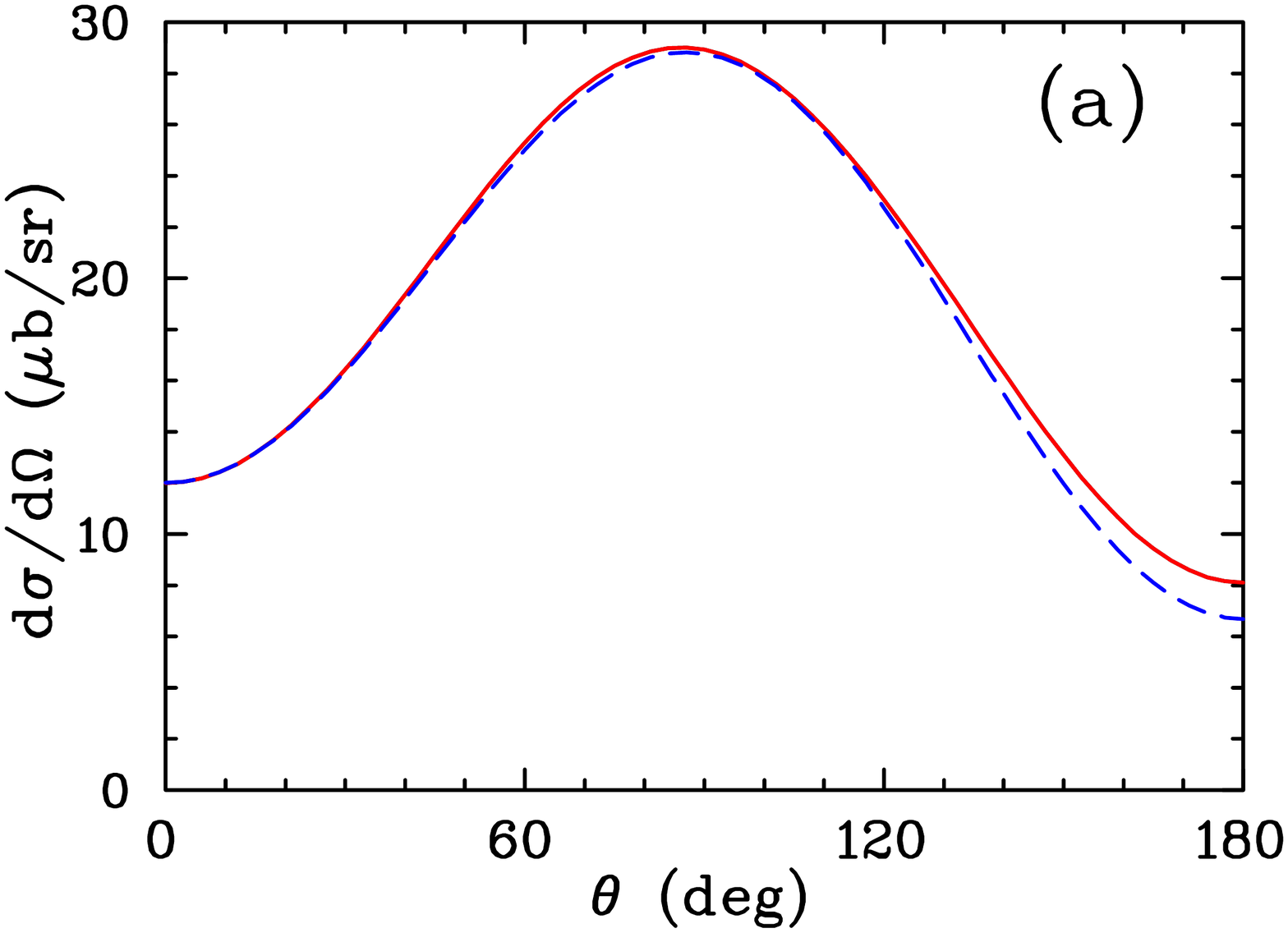}
\includegraphics[width=0.3\columnwidth,trim=5cm 1.5cm 0 0]{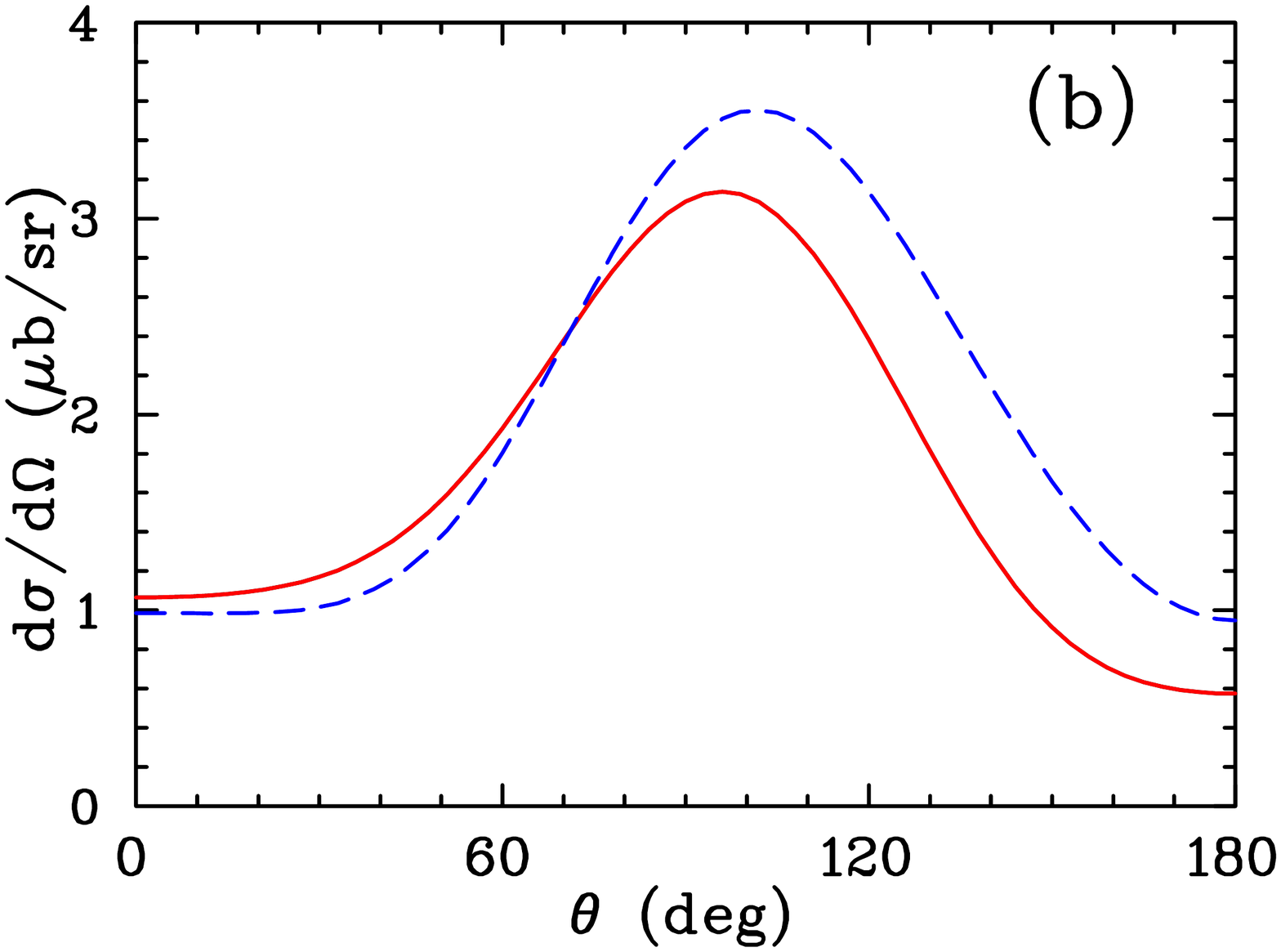}
\includegraphics[width=0.3\columnwidth,trim=5cm 1.5cm 0 0]{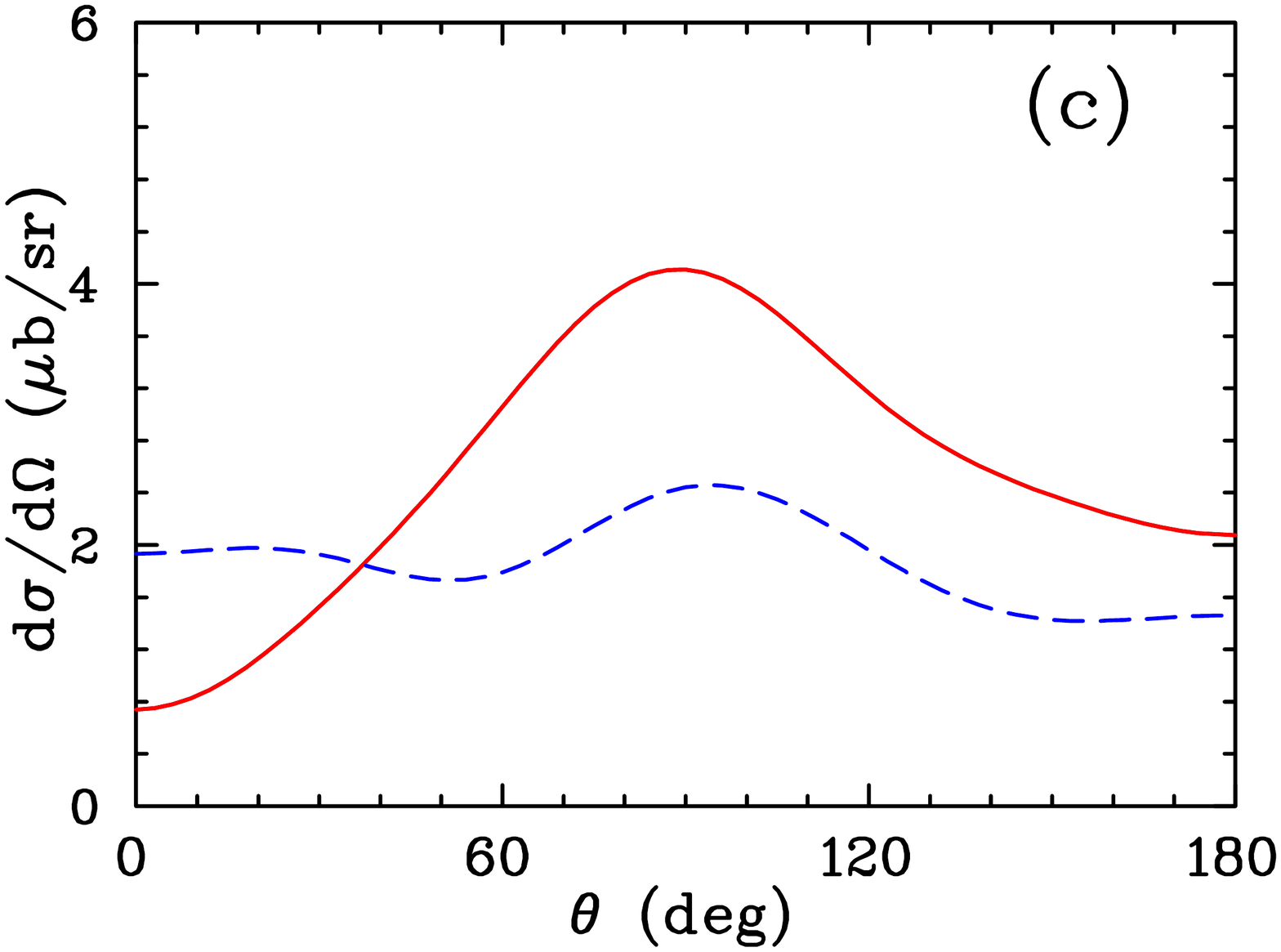}

\caption{\label{fig:fs2} The differential cross sections of the $\gamma 
    p\to\pi^0p$ (red solid curves) and $\gamma n\to\pi^0n$ (blue dashed 
    curves) reaction reactions at several photon energies (a) E = 340~MeV, 
    (b) E = 630~MeV, and (c) E = 787~MeV, which correspond to 
    $\Delta(1232)3/2^+$, $N(1440)1/2^+$, and $N(1535)1/2^-$ regions, 
    respectively (Ref.~\protect\cite{FSIpi0}).} 
\end{center}
\end{figure}

Figure~\ref{fig:fs2} shows that proton and neutron cross sections are very close to each other in the $\Delta(1232)3/2^+$ region ($A_s = 0$). At higher energies, however, the contributions from $N(1440)1/2^+$ and $N(1535)1/2^-$ become important, the isoscalar amplitude does not equal zero, and the difference between proton and neutron differential cross sections becomes more clearly visible.  That means in general that one cannot simply use the ratio between free and bound proton data to be indicative of the ratio between free and bound neutron data.   Measurements using bound neutrons 
will thus always carry significant model-dependent uncertainty.


Unfortunately, there are currently no FSI calculations for polarized measurements on neutron targets. In the absence of these calculations, for PWA one can only assume that the effects of FSI on polarization observables are small. There is some indirect proof that this assumption is reasonable, since several PWAs can successfully fit the polarized measurements in the world database (see, for instance,~\cite{HO17}).

\section{Experimental Facilities \label{subsec:exper}}

In this section, we provide a brief description and references to experimental facilities that were the main contributors of the photoproduction data over last two decades. Some of them used bremsstrahlung to generate real photons, others used laser Compton backscattering. Some detectors were optimized for charged particles, others for neutrals. In that respect they are complimentary to each other.

\subsection{CEBAF}

The Thomas Jefferson National Accelerator Facility (TJNAF) commonly known as Jefferson Lab or JLab is the home of the Continuous Electron Beam Accelerator Facility, CEBAF. This is a race track shaped machine that consists of two linear accelerators joined together with a pair of arc sections. For the results reported here, the electron beam made up to five passes through the machine and gained energy up to 6~GeV. The extracted beam was delivered to end stations known as Hall~A, Hall~B and Hall~C. The electron beam can be highly polarized. The majority of photoproduction data at CEBAF was obtained in Hall~B with CLAS detector.

Recently CEBAF was upgraded and its energy doubled. Now it can accelerate electrons up to 12~GeV. One more experimental hall, Hall~D, was added.

\subsubsection{CLAS}

The CEBAF Large Acceptance Spectrometer (CLAS) was a magnetic spectrometer with a toroidal magnetic 
field~\cite{CLAS}. It has since been upgraded to CLAS12 to cope with the increased electron beam energy in Hall~B. The new detector has a slightly different configuration to the older CLAS, although some of the original detector subsystems 
have been refurbished and retained. 
The toroidal field bends particles of different charge either towards or away from the beam direction, which results in some asymmetry of the acceptance for opposite charges. The magnetic field is produced by six superconducting coils positioned around the beam. Essentially it may be considered as six independent spectrometers. The gaps between each pair of the coils are filled with detector packages. Each package has six multilayer drift chambers  for charged particle tracking. The momentum resolution for charge particles from tracking depends on the angle and magnetic field setting and on average was $\Delta p/p \sim 0.5-1\%$. Polar angle resolution is about 1~mrad or better. Azimuthal angle resolution is about 4~mrad. They drift chambers followed by gas \v Cerenkov counters for electron pion separation covering forward angles up to $45^\circ$. Further out there is an array of TOF scintillation counters that were used for charged particles identification. TOF counters cover polar angle rage from $8^\circ$ to $142^\circ$ and full range of azimuthal angles. The solid angle for charged particles was about 60\% of $4\pi$. The last detector in a package is an electromagnetic calorimeter. It is a sampling calorimeter made of alternating layers of lead and plastic scintillators. The total thickness is 16 radiation lengths. The sampling fraction is approximately 0.3 for electrons of 3 GeV and greater, and for smaller energies, there is a monotonic decrease to about 0.25 for electrons of 0.5~GeV. The energy resolution was $\sigma/E = 10.3\% / \sqrt{E(GeV)}$. In order to get coordinates of the shower the scinitillator strips were arranged to provide three views crossing each other at $60^\circ$. The calorimeters covered angles from $8^\circ$ to $45^\circ$.
The design of CLAS was optimized for charged particles.

The unpolarized or circularly photons were produced via bremsstrahlung on a thin gold foil. Coherent bremsstrahlung on a diamond radiator was used to produce linearly polarized photons. Tagging of bremsstrahlung photons was done by the Hall B tagging spectrometer~\cite{HallB_tagger} with a tagging range from 20\% to 95\% of the electron beam energy. The focal plane was instrumented with a two-layer scintillation hodoscope. The first layer consisted of 384 overlapping counters providing the energy of the post-bremsstrahlung electron with an accuracy of $\sim 0.001$ of the electron beam energy. The second layer of 61 counters provided timing information.

The target was placed in the center of the detector and was surrounded by a scintillation start counter. CLAS could operate with various types of targets: unpolarized gas, liquid and solid targets. Two different frozen spin polarized targets were used in photoproduction experiments. One, FROST~\cite{FROST}, with butanol as a target material was used for experiments with polarized protons. It allowed for longitudinal and transverse polarization of protons. The second target, HDIce~\cite{HDICE}, was used for experiments with longitudinally polarized protons and deuterons.

\subsection{ESRF}

The European Synchrotron Radiation Facility (ESRF) is the most intense source of synchrotron-generated light. After the ESRF pre-injector LINAC a 200~MeV electrons injected into the booster synchrotron which accelerate them to 6~GeV. They then injected in a 
6~GeV storage ring where they can be used for physics. 

\subsubsection{GRAAL}

One of the Collaborative research beam lines at ESRF hosted GRenoble Anneau Acc\'el\'erateur Laser (GRAAL) facility~\cite{GRAAL}. Photons were produced by Compton backscattering of laser light from the electron beam. The tagged photon energy spectrum at GRAAL extended from 600~MeV to 1500~MeV. The core of the facility was a large solid angle detector (La$\gamma$range). The central part of La$\gamma$range was a BGO calorimeter which covered polar angles $25^\circ - 155^\circ$ and full range of azimuthal angle. In the center of the calorimeter there was a plastic scintillator barrel and internal tracker made of two cylindrical multiwire proportional chambers (MWPC). The forward polar angles below $25^\circ$ were covered by two pairs of planar MWPC and double wall of plastic scintillators followed by shower wall consisting of four layers of lead and plastic scintillators. The calorimeter had excellent energy resolution for photons and electrons, 3\% at 1~GeV. It also had good response for protons below 300~MeV. Charged particles could be tracked by MWPCs. Neutrons could be detected either in BGO calorimeter or forward wall. The entire apparatus was optimized for the detection of mesons decaying to photons but could also detect charged particles. GRAAL is no longer in operation. The BGO calorimeter has been moved to Bonn and became a part of the new BGO-OD setup~\cite{BGOOD}.

\subsection{MAMI}

The Mainz Microtron, MAMI, is an accelerator for electron beams run by the Institute for Nuclear Physics of the University of Mainz, and is used extensively for hadron physics experiments. It is a continuous wave accelerator system. Over the years it went through a chain of upgrades. The latest incarnation is MAMI-C, which can accelerate electrons up to 1508~MeV. Experimental area A2 is dedicated to experiments with tagged bremsstrahlung photons. Linearly polarized photons are produced via coherent bremsstrahlung on a diamond radiator. The tagging is done by the Glasgow tagger~\cite{Glasgow_tagger}. It was originally built for MAMI-B with maximum energy of 833~MeV. To improve energy resolution it was later complimented by a microscope~\cite{Glasgow_microscope} with increase energy resolution over a smaller range of electron energies. After MAMI-C went into operation the tagger was upgraded for use with beam energy of 1500~MeV~\cite{Glasgow_C}. The tagging range is 5 -- 93\%  of the electron beam energy. The energy resolution without the microscope is 4~MeV for a 1500~MeV incident beam. The microscope improves energy the resolution by a factor of 6 in the 60~MeV energy range.

\subsubsection{DAPHNE}

DAPHNE (Detecteur \`a grande Acceptance pour la PHysique photoNucleaire Experimentale) is a large acceptance tracking detector for intermediate-energy hadrons comprising a vertex detector surrounded by a segmented calorimeter~\cite{DAPHNE}. The detector consists of three principal parts, arranged as a set of coaxially . In the center there is a vertex detector which is surrounded by a charged-particle detector consisting of several layers of scintillator which is itself surrounded by a lead-aluminium-scintillator sandwich designed to detect neutral particles. It covers polar angles from $21^\circ$ to $159^\circ$ and has full azimuthal angle coverage. Now DAPHNE is no longer in operation.

\subsubsection{TAPS}

TAPS (Two Arm Photon Spectrometer)~\cite{TAPS} is a detector array of 384 individual modules of hexagonal shaped detectors. Each detector module is a telescope consisting of a BaF$_2$ crystal and a separate plastic scintillator in front of it. It can be used for charged/neutral separation and charged particle identification. The energy resolution of TAPS is $\sigma/E = 0 .59\% /\sqrt{E_\gamma} +1.9\%$ where $E_\gamma$ is given in GeV. The position resolution is about 2~cm. TAPS was originally designed to detect two photon decays of $\pi^0$ and $\eta$ mesons. Recently TAPS was split in two pieces which were used separately with other detectors, the Crystal Ball and the Crystal Barrel.

\subsubsection{Crystal Ball/TAPS}

The latest experimental setup in A2 is a combination of Crystal Ball and half of TAPS. The details of the most recent configuration of the setup can be found in Ref.~\cite{CBTAPS}. The Crystal Ball (CB) was originally built by Stanford Linear Accelerator Center (SLAC)~\cite{CB}. It consists of 672 optically isolated NaI(Tl) crystals with a thickness of 15.7 radiation lengths . The crystals are arranged to form a sphere covering 93\% of the full solid angle. The energy resolution for electromagnetic showers is described as $\Delta E/E = 0.02/(E/{\rm GeV})^{0.36}$. The accuracy of the shower direction reconstruction is about $\sigma_\theta \sim 2-3^\circ$ in polar angle and $\sigma_\varphi \sim 2^\circ / \sin \theta$. In the center of the CB there is a barrel of 24 scintillation counters surrounding the target. It measures energy losses of the charged particle and can be used in $\Delta E/E$ analysis for charged particles identification and also to separate charged particles from neutrals. The forward angles $\theta = 1 - 20^\circ$ are covered by a half of the TAPS which is placed 1.5~m downstream of CB center. The combined solid angle of CB and TAPS is 97\% of 4$\pi$. This setup can be used with both polarized and unpolarized targets. This facility is operational and continues data taking.

\subsection{ELSA}

The electron accelerator ELektronen-Stretcher-Anlage (ELSA)~\cite{ELSA} is operated by the university of Bonn. It has three stages: injector LINACs, booster synchrotron and stretcher ring. It can deliver beams of polarized or unpolarized electrons with energies up to 3.5~GeV. Real photon beam is produced via bremsstrahlung. The linearly polarized beam is produced via coherent bremsstrahlung. The bremsstrahlung photons are tagged with tagging hodoscope. The accuracy of the photon energy is 0.4\% of electron beam energy.  

\subsubsection{SAPHIR}

SAPHIR (Spectrometer Arrangement for PHoton Induced Reactions)~\cite{SAPHIR} was a large solid angle detector at the Bonn accelerator ELSA. SAPHIR was a magnetic spectrometer with a dipole magnet. The photon beam entered through a hole in the magnet yoke. The space between the magnet poles was occupied by the Central Drift Chamber (CDC) for charged particle tracking. The target was placed in the center of the CDC. For better tracking and momentum resolution there were also three planar drift chambers, two on the sides and one in the forward direction. The momentum resolution of about 6.5\% was achieved at 1.0~GeV/$c$ particle momentum. The use of the forward drift chamber improved the momentum resolution considerably up to 2\% at 1.8~GeV/$c$. There were three planes of scintillation counter hodoscopes, two on the sides and one in the forward direction. The hodoscopes in coincidence with tagging system produced the trigger and were used for particle identification by measuring time-of-flight (TOF). Downstream of the forward TOF there was an array of electromagnetic shower counters (EMC). The energy resolution of the EMC was found to be $13\%/\sqrt{E}$, where E in GeV. Now SAPHIR is no longer in operation.

\subsubsection{CBELSA}

The central part of the setup is the Crystal Barrel~\cite{Cbarrel}, the calorimeter that was used at the Low Energy Antiproton Ring (LEAR) at CERN. In its original configuration it consisted of 1380 CsI(Tl) crystals. The length of each crystal is 16.1 radiation lengths. The crystals are grouped in 26 rings ($\Delta\theta = 6^\circ$), where the larger rings consist of 60 crystals ($\Delta\varphi = 6^\circ$), the six smallest rings contain 30 crystals ($\Delta\varphi = 12^\circ$). It covers angles from $12^\circ$ to $168^\circ$ with respect to the beam direction resulting in 97.8\% coverage of the solid angle. During the first configuration change the three forward rings were taken out and part of TAPS, MiniTAPS, was installed to extend coverage to smaller angles down to $1^\circ$. During the second configuration change the forward crystals ($\theta < 27^\circ$) were covered by plastic scintillators in front of each crystal for charged particle identification.  Inside the calorimeter, a three-layer inner detector with 513 scintillating fibers was installed. More details about the most recent version of the setup can be found in Ref.~\cite{CBELSA}. This setup is optimized for detection of multiphoton events. CBELSA is active and continues data taking.

\subsubsection{BGO-OD}

The BGO-OD~\cite{BGOOD} is a new experiment at ELSA. It consists of a central detector enclosing the target in the angular range $10 - 155^\circ$. This is complemented by a large aperture forward magnetic spectrometer covering the angular range from approximately $2^\circ$ to $12^\circ$. The main component of the central detector is BGO calorimeter formerly used at GRAAL. A segmented plastic scintillator barrel and a double layer cylindrical MWPC placed inside the calorimeter enable tracking and identification of charged particles. The forward spectrometer consists of a large aperture dipole magnet sandwiched between tracking detectors. Front tracking upstream of the magnet is performed with two sets of scintillating fibre detectors. Eight double layers of drift chambers serve for rear tracking downstream of the magnet. Several new components are to be added. The BGO-OD was commissioned in 2016.

\subsection{Spring-8}

SPring-8 is a large synchrotron radiation facility located in Harima Science Park City, Japan. The name ``SPring-8" is derived from ``{\bf S}uper {\bf P}hoton {\bf ring-8}~GeV". As the name implies it is an 8~GeV electron storage ring. Among many other applications it is used for hadronic physics and photoproduction in particular. 

\subsubsection{LEPS}

Backward Compton scattering of laser light from a high energy electron beam is used to produce high energy photons. This type of beam line was constructed at Spring-8 and is called ``Laser-electron-photon" (LEP). If the laser light is polarized then the produced high energy photons also polarized. The photons were tagged by detecting scattered electron. The initial version of this facility could provide photons with energies up to 2.4~GeV. The first detector, LEPS~\cite{LEPS}, was designed to study $\phi$-meson photoproduction in forward angles. It is a magnetic spectrometer with a dipole magnet. The vertex detector is located upstream of the magnet and consists silicon strip detectors and drift chambers. Downstream of the magnet there were two sets of drift chambers, one on each side of the beam. Particle identification is done using TOF. The LEP beam line has been upgraded to increase the intensity of the photon beam and extend the energy range up to 2.9~GeV~\cite{LEPS1}. 

\subsubsection{LEPS2}

This approach was used to construct the second LEP beam line, LEPS2~\cite{LEPS2}. 
LEPS1 had acceptance limitation to forward angles only. To overcome this limitation a new detectors needed to be constructed for LEPS2. One of the detectors aimed to study $\eta^\prime$ mesic nuclei is BGOegg~\cite{BGOEGG}. The detector is optimized for detection of photons. It is an egg-shaped electromagnetic calorimeter. It consists of 1320 BGO crystals of 20 radiation lengths. It has a polar angle coverage from $24^\circ$ to $144^\circ$ and complete azimuthal coverage. The energy resolution is 1.3\% at 1~GeV and position resolution is 3.1~mm. To detect charged particles the scintillation hodoscopes and cylindrical drift chambers are installed in the center of the calorimeter. 

The second detector for LEPS2 is a solenoid spectrometer~\cite{LEPS2}. It is designed to detect both charged particles and photons. It is a solenoid magnet with a 0.9~T field. Tracking of charged particles is done by the Time projection chamber and forward drift chambers. The tracking detectors are surrounded by a barrel of resistive plate chambers (RPC). RPCs have very good timing resolution and are used for particle identification by measuring TOF. For particle momenta above 1~GeV in addition to TOF the aerogel \v Cerenkov counters are used. The outer most detector is barrel electromagnetic calorimeter, Barrel $\gamma$. It is a sampling lead/plastic scintillator calorimeter with a thickness of 14.3 radiation lengths and covers polar angles $30 - 110^\circ$.

\section{\label{sec:available} Available Experimental Data on Meson Photoproduction}

In this section, we  give an overview of available experimental data of 
meson photoproduction. The source of the data is SAID database~\cite{SAID} which is to date the most comprehensive. The data are organized by the final state. The number of data points accumulated thus far makes it pointless to try to plot each of them in this review. Instead we plot for each channel, in the style of Figure~\ref{fig:evolution}, the number of data points as a function of hadronic mass $W$, and as a function of year. The data are split into unpolarized and polarized stacked histograms and are mean to convey the relative amount available from each channel, as well as an indication of the progress in measurements over time.

\begin{figure} 
\begin{center}
\includegraphics[width=0.4\columnwidth,trim=4cm 2.5cm 0 0]{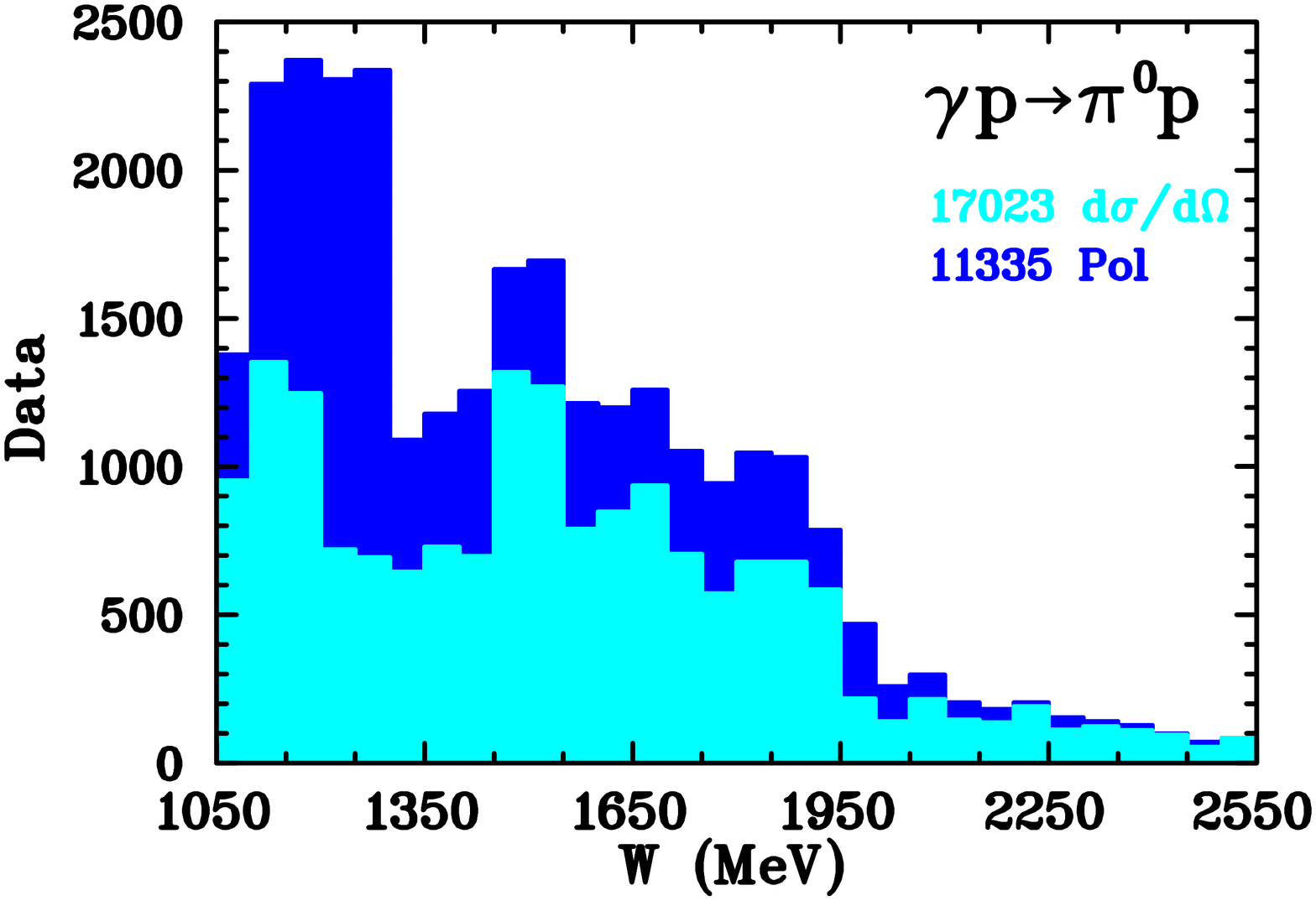}
\includegraphics[width=0.4\columnwidth,trim=4cm 2.5cm 0 0]{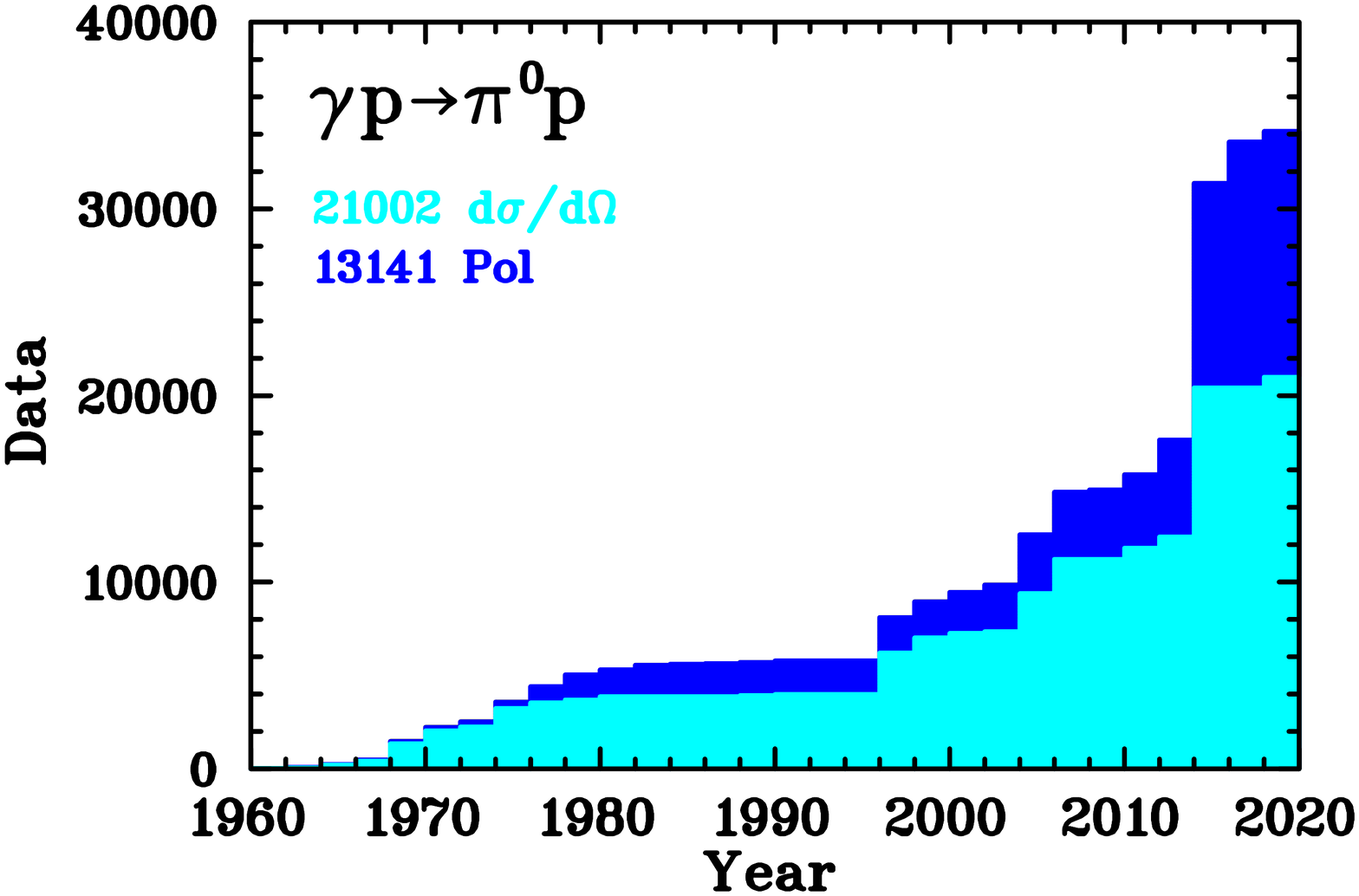} 
\protect\caption{\label{fig:pi0p}Database for $\gamma p\to\pi^0p$. \underline{Left}:  Experimental data from the SAID database~\protect\cite{SAID} selected for 1996 through 2018. \underline{Right}: Amount of data as a function of time. Full SAID database.  The data shown as stacked histogram. Light shaded~--~cross sections, dark shaded~--~polarization data.}
\end{center}
\end{figure}

\begin{figure} 
\begin{center}
\includegraphics[width=0.4\columnwidth,trim=4cm 2.5cm 0 0]{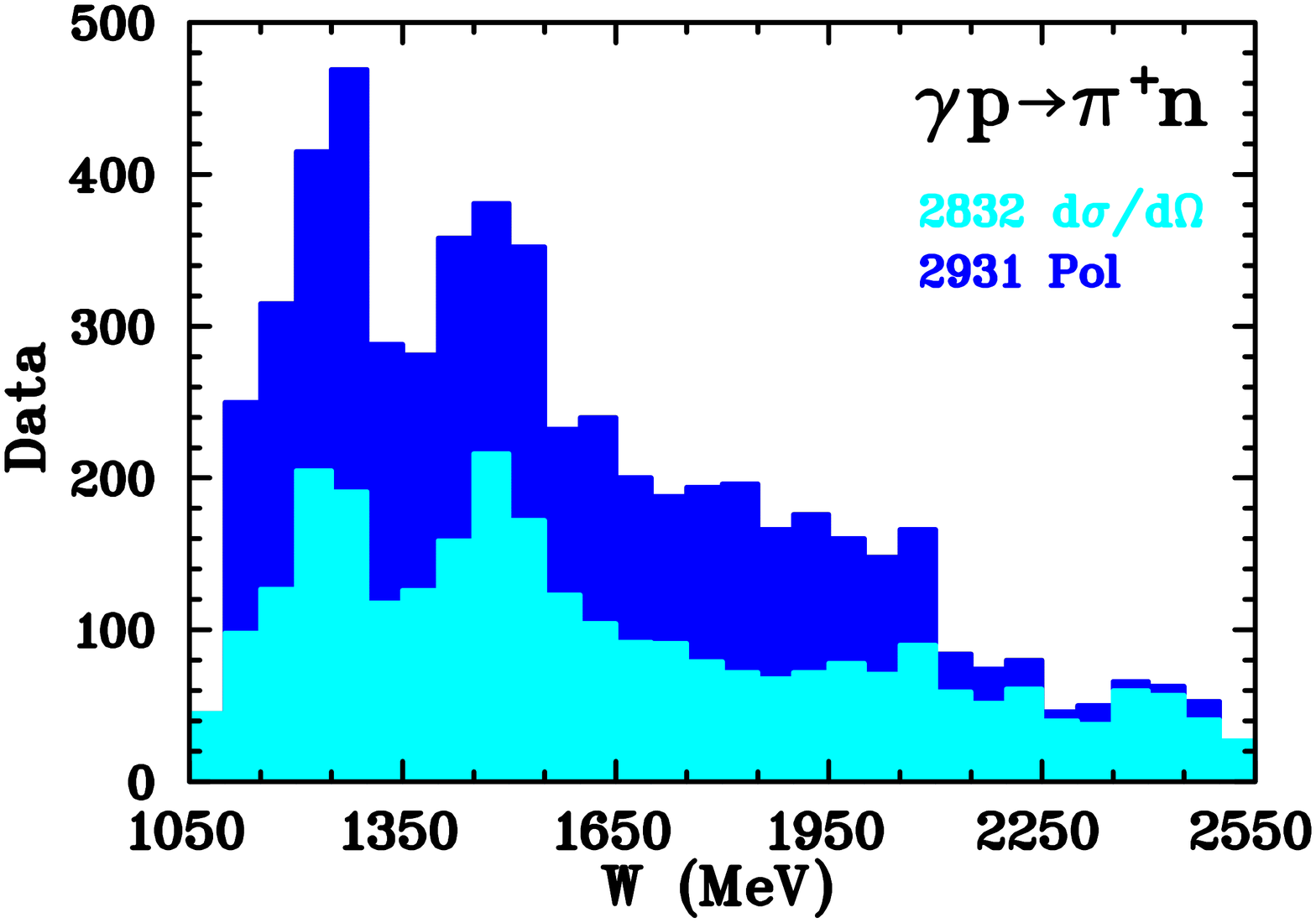}
\includegraphics[width=0.4\columnwidth,trim=4cm 2.5cm 0 0]{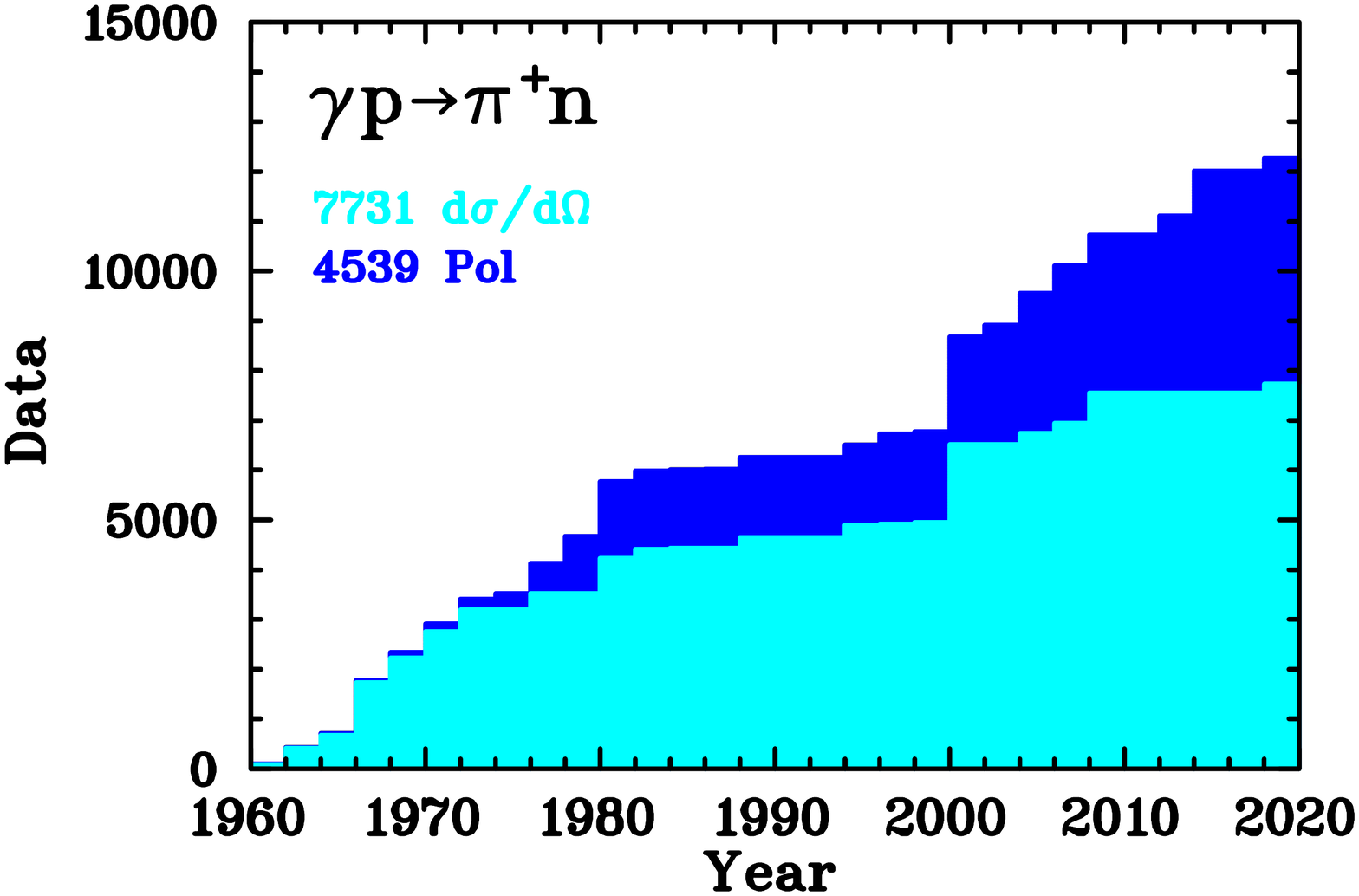}
	\protect\caption{\label{fig:pi+n}Database for $\gamma p\to\pi^+n$. 
    The notation is the same as in Figure~\protect\ref{fig:pi0p}.}
\end{center}
\end{figure}

\begin{figure} 
\begin{center}
\includegraphics[width=0.4\columnwidth,trim=4cm 2.5cm 0 0]{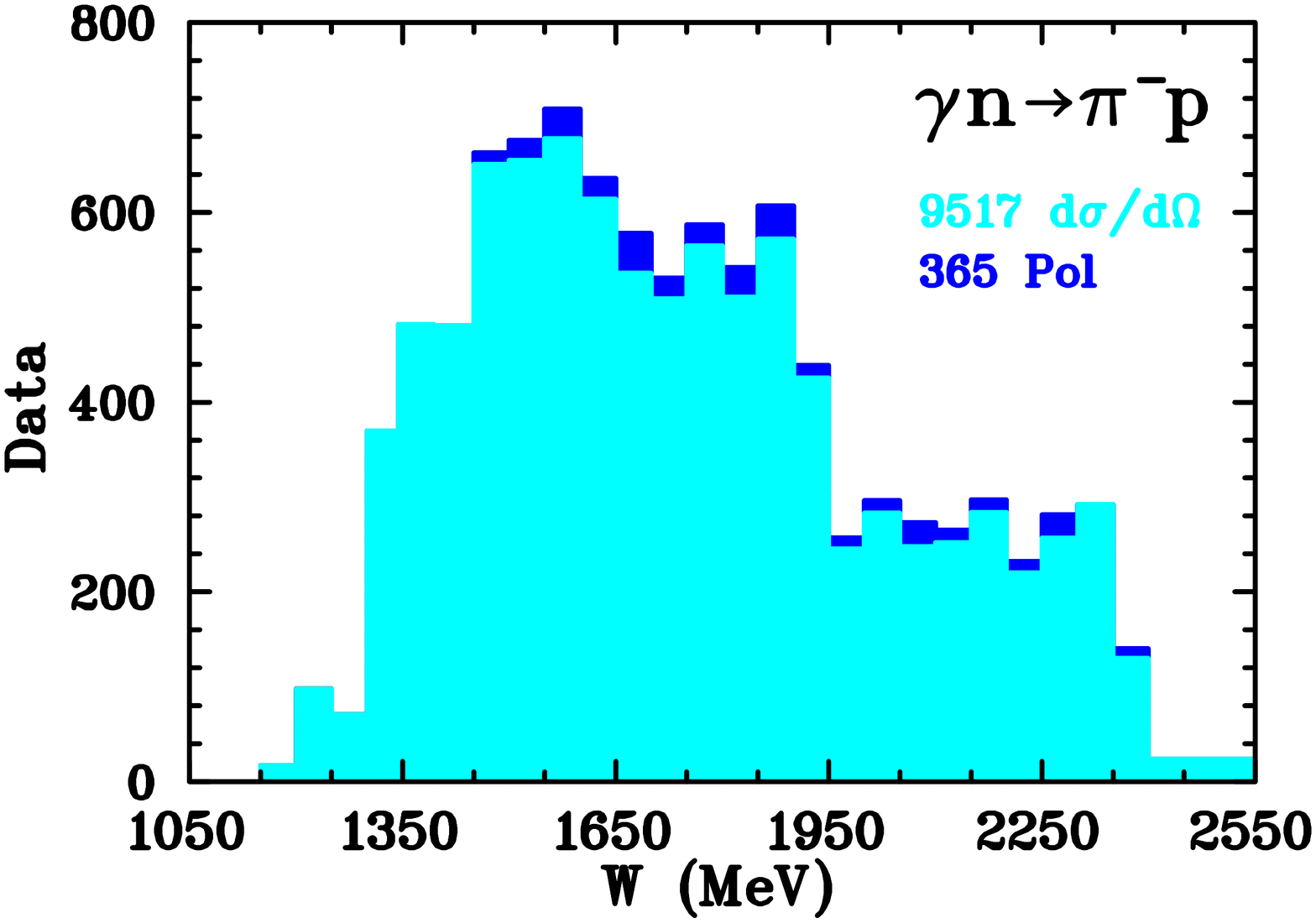}
\includegraphics[width=0.4\columnwidth,trim=4cm 2.5cm 0 0]{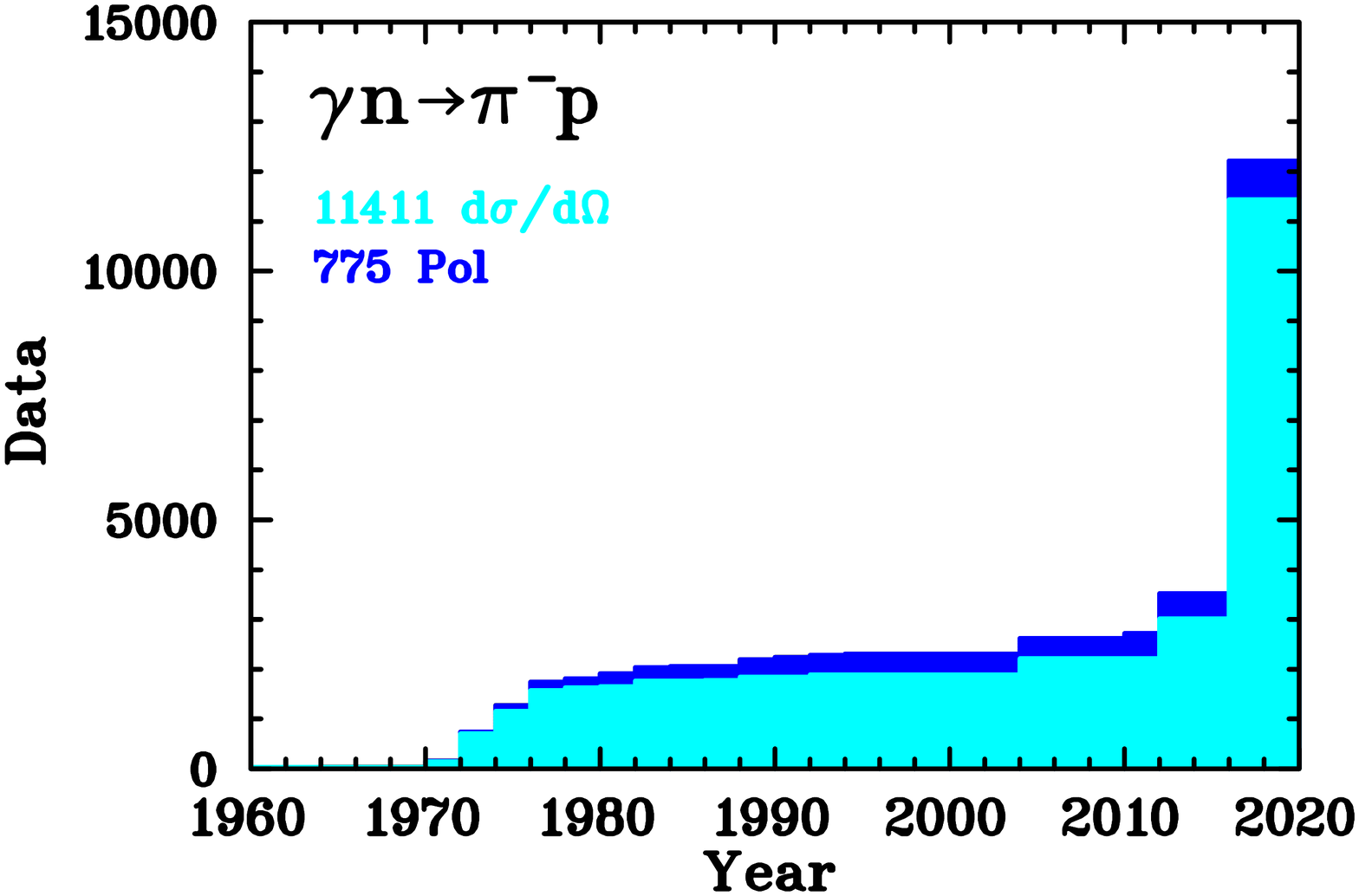}
	\protect\caption{\label{fig:pi-p}Database for $\gamma n\to\pi^-p$. 
    The notation is the same as in Figure~\protect\ref{fig:pi0p}.}
\end{center}
\end{figure}

\begin{figure}[H] 
\begin{center}
\includegraphics[width=0.4\columnwidth,trim=4cm 2.5cm 0 0]{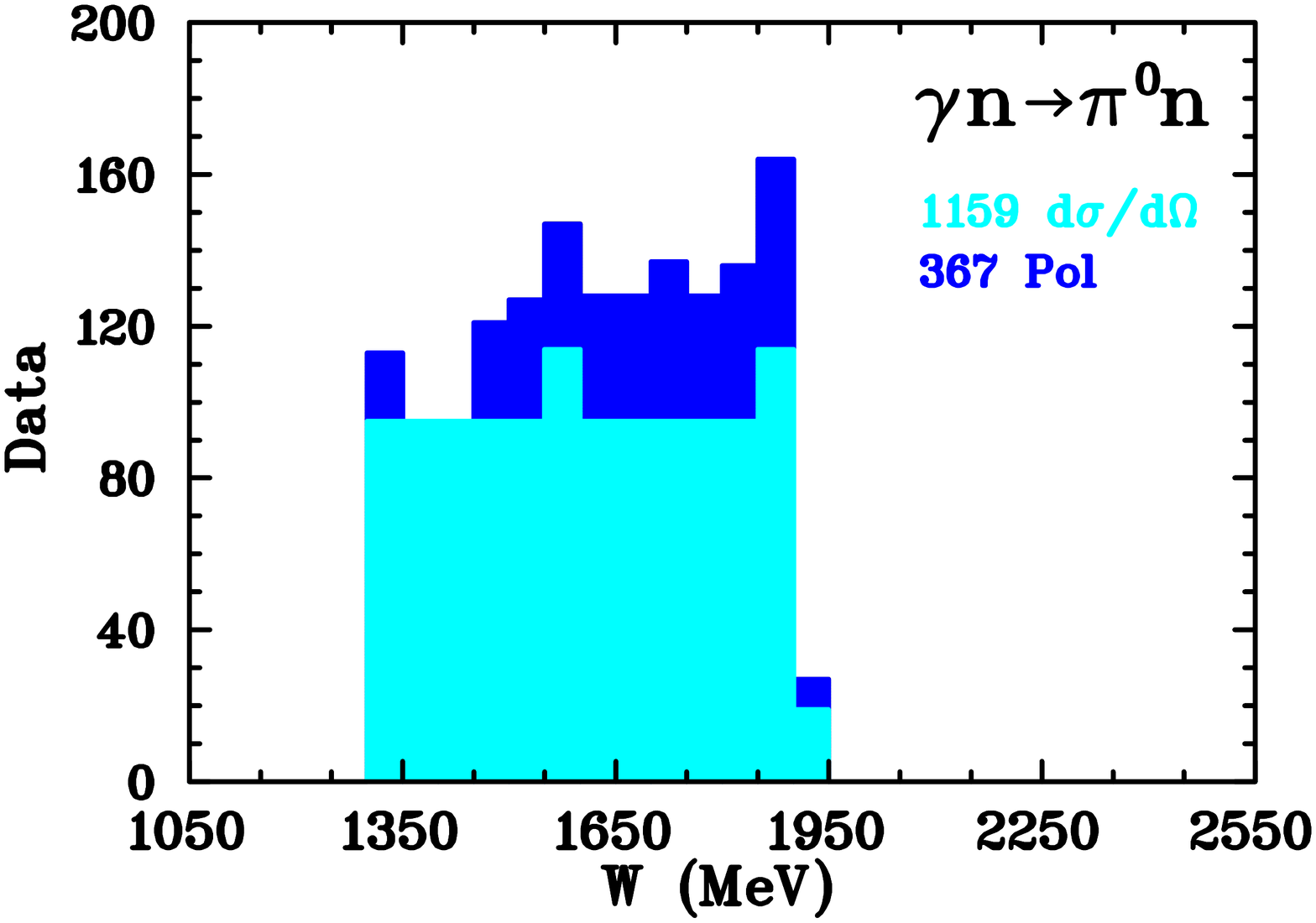}
\includegraphics[width=0.4\columnwidth,trim=4cm 2.5cm 0 0]{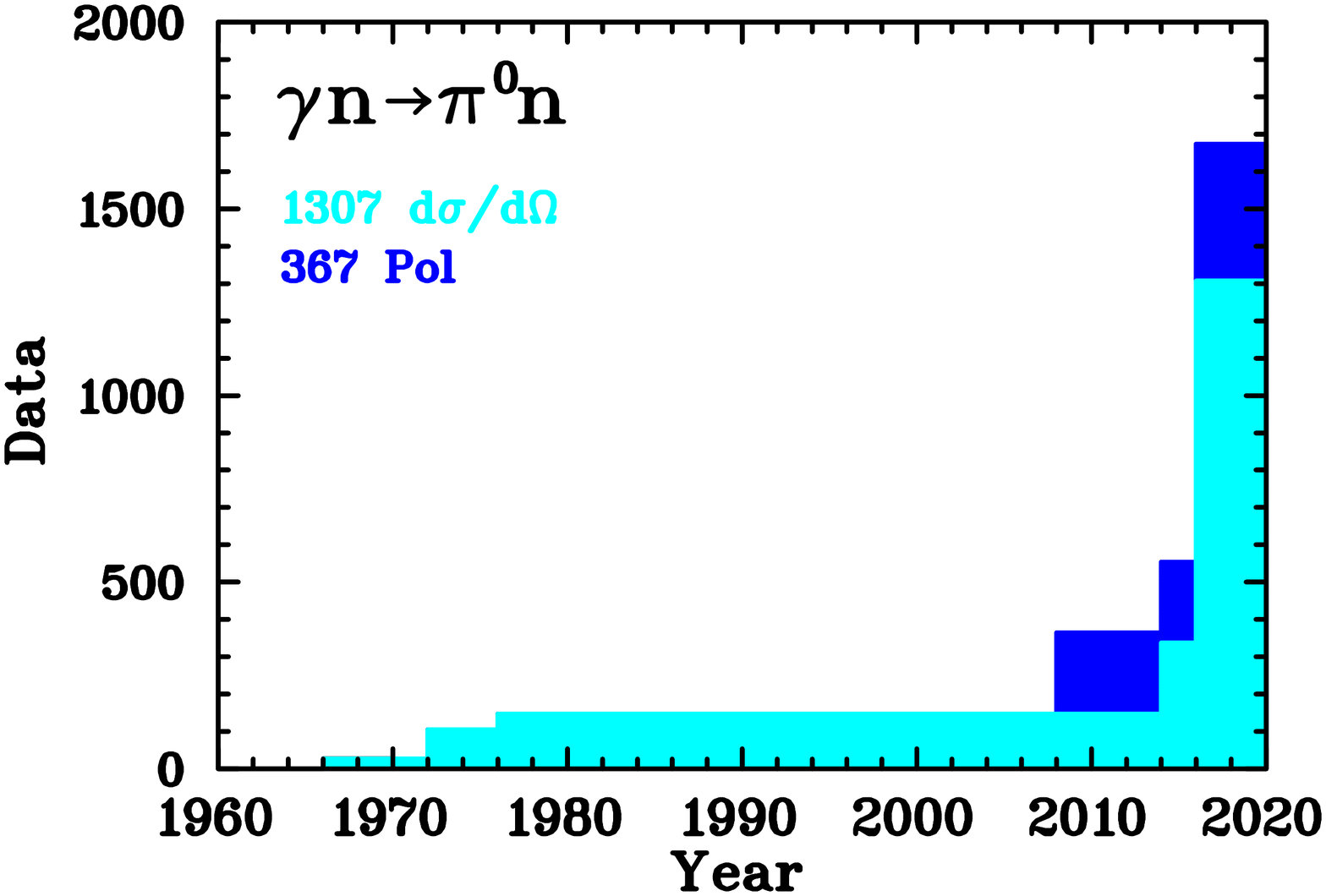}
	\protect\caption{\label{fig:pi0n}Database for $\gamma n\to\pi^0n$. 
    The notation is the same as in Figure~\protect\ref{fig:pi0p}.}
\end{center}
\end{figure}

\begin{figure}[H] 
\begin{center}
\includegraphics[width=0.4\columnwidth,trim=4cm 2.5cm 0 0]{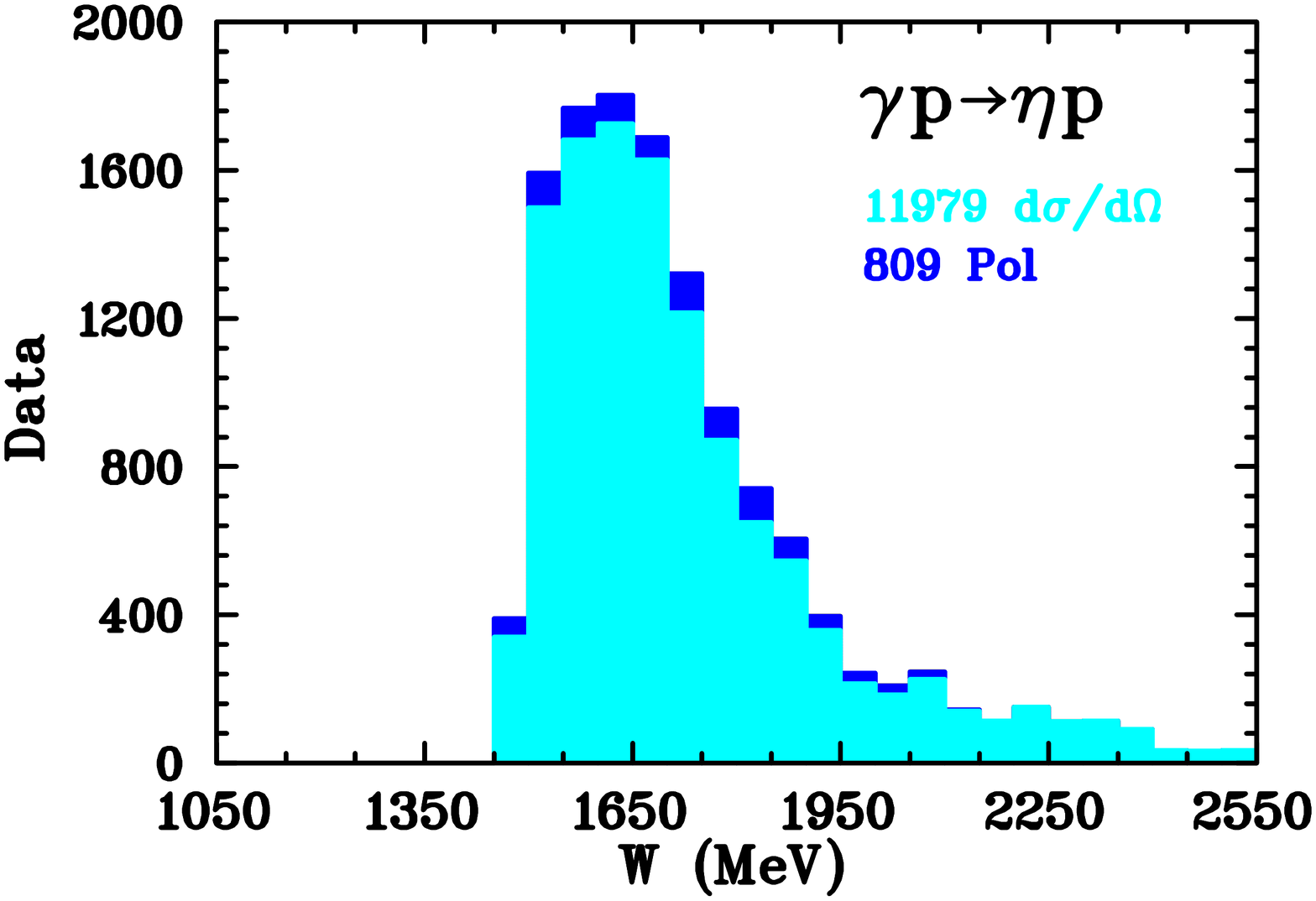}
\includegraphics[width=0.4\columnwidth,trim=4cm 2.5cm 0 0]{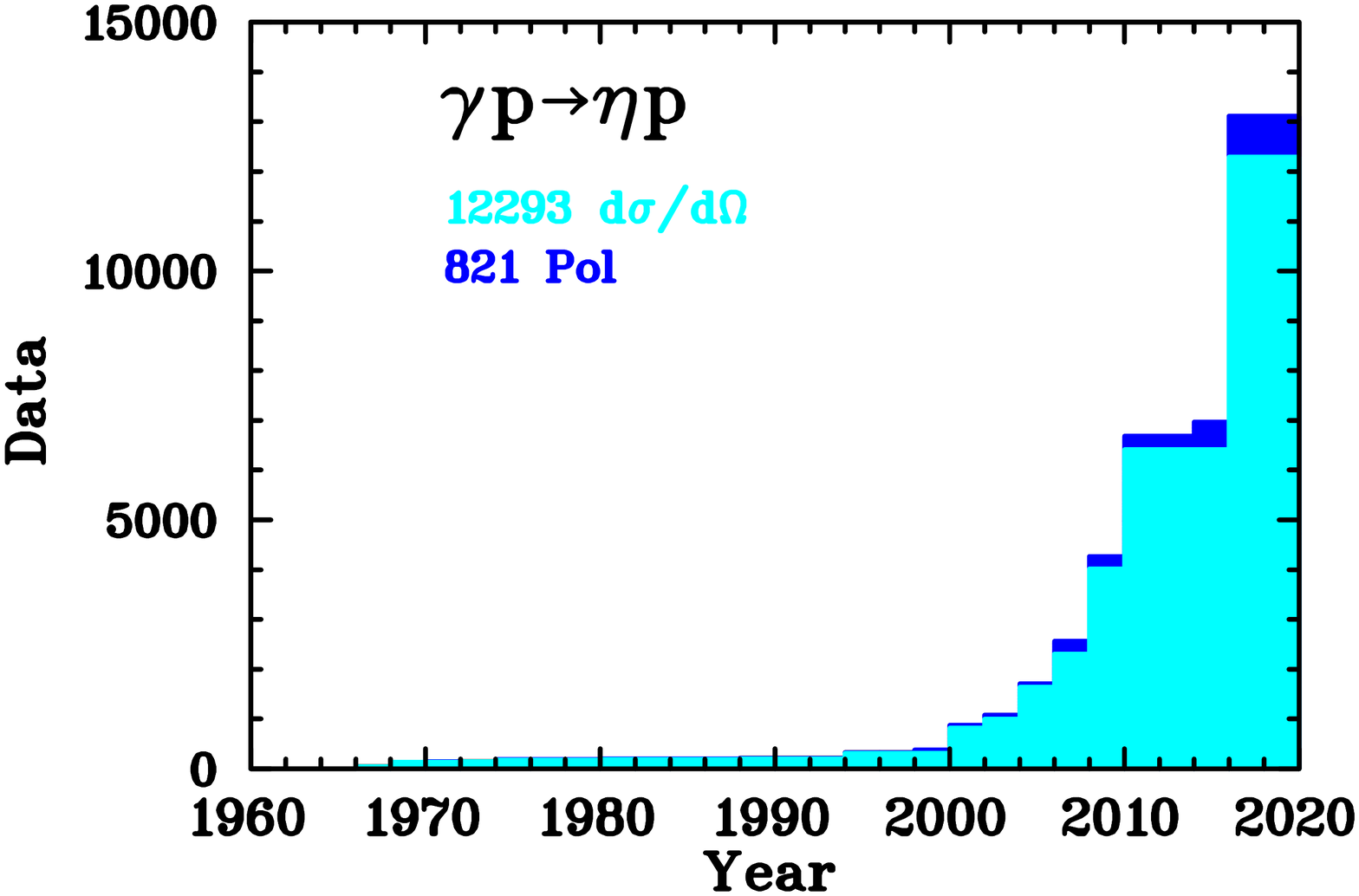}
	\protect\caption{\label{fig:etap}Database for $\gamma p\to\eta p$. 
	The notation is the same as in Figure~\protect\ref{fig:pi0p}.}
\end{center}
\end{figure}

\begin{figure} 
\begin{center}
\includegraphics[width=0.4\columnwidth,trim=4cm 2.5cm 0 0]{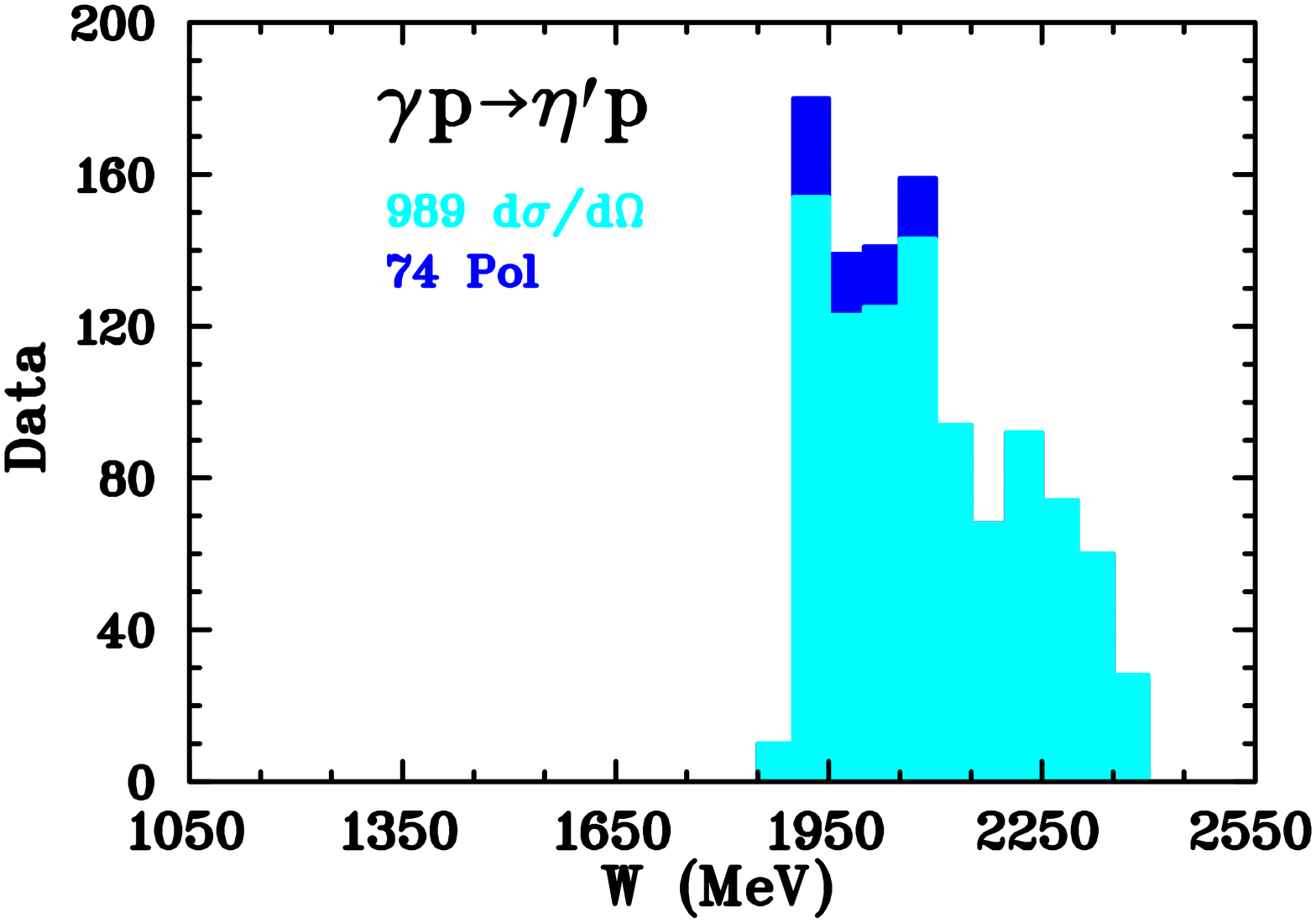}
\includegraphics[width=0.4\columnwidth,trim=4cm 2.5cm 0 0]{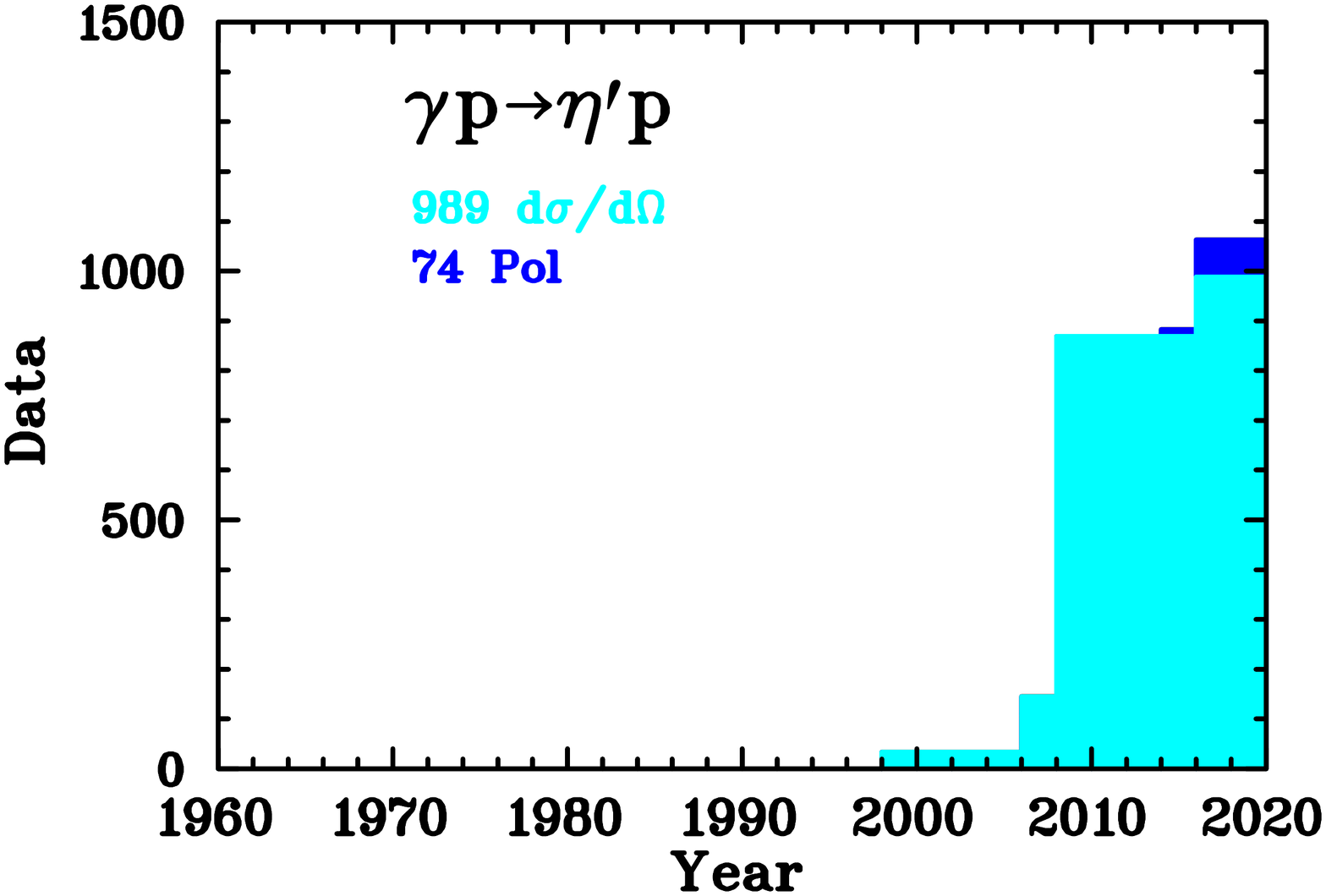}
	\protect\caption{\label{fig:etapp}Database for $\gamma 		
    p\to\eta^\prime p$. The notation is the same as in 		
    Figure~\protect\ref{fig:pi0p}.}
\end{center}
\end{figure}

\begin{figure} 
\begin{center}
\includegraphics[width=0.4\columnwidth,trim=4cm 2.5cm 0 0]{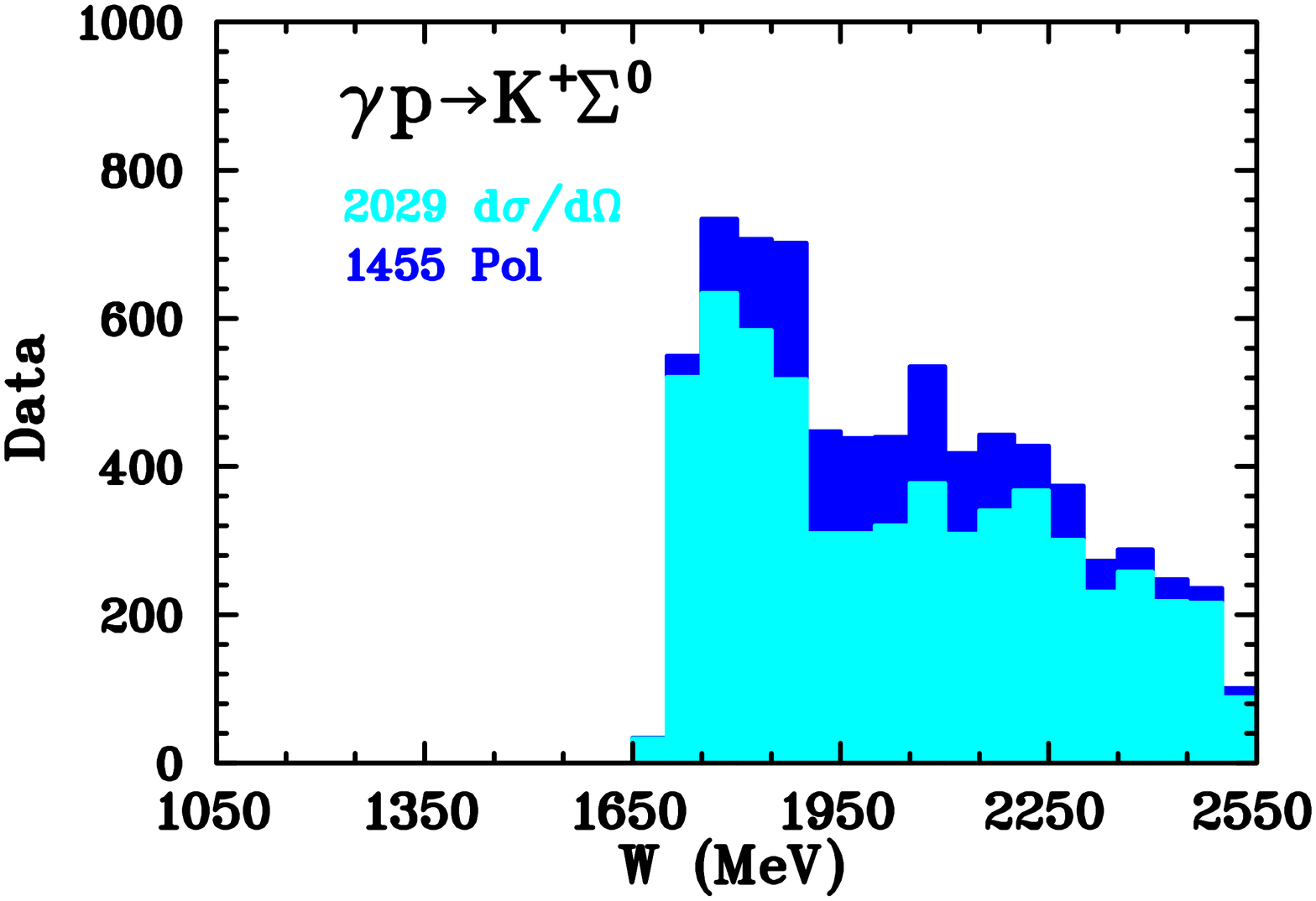}
\includegraphics[width=0.4\columnwidth,trim=4cm 2.5cm 0 0]{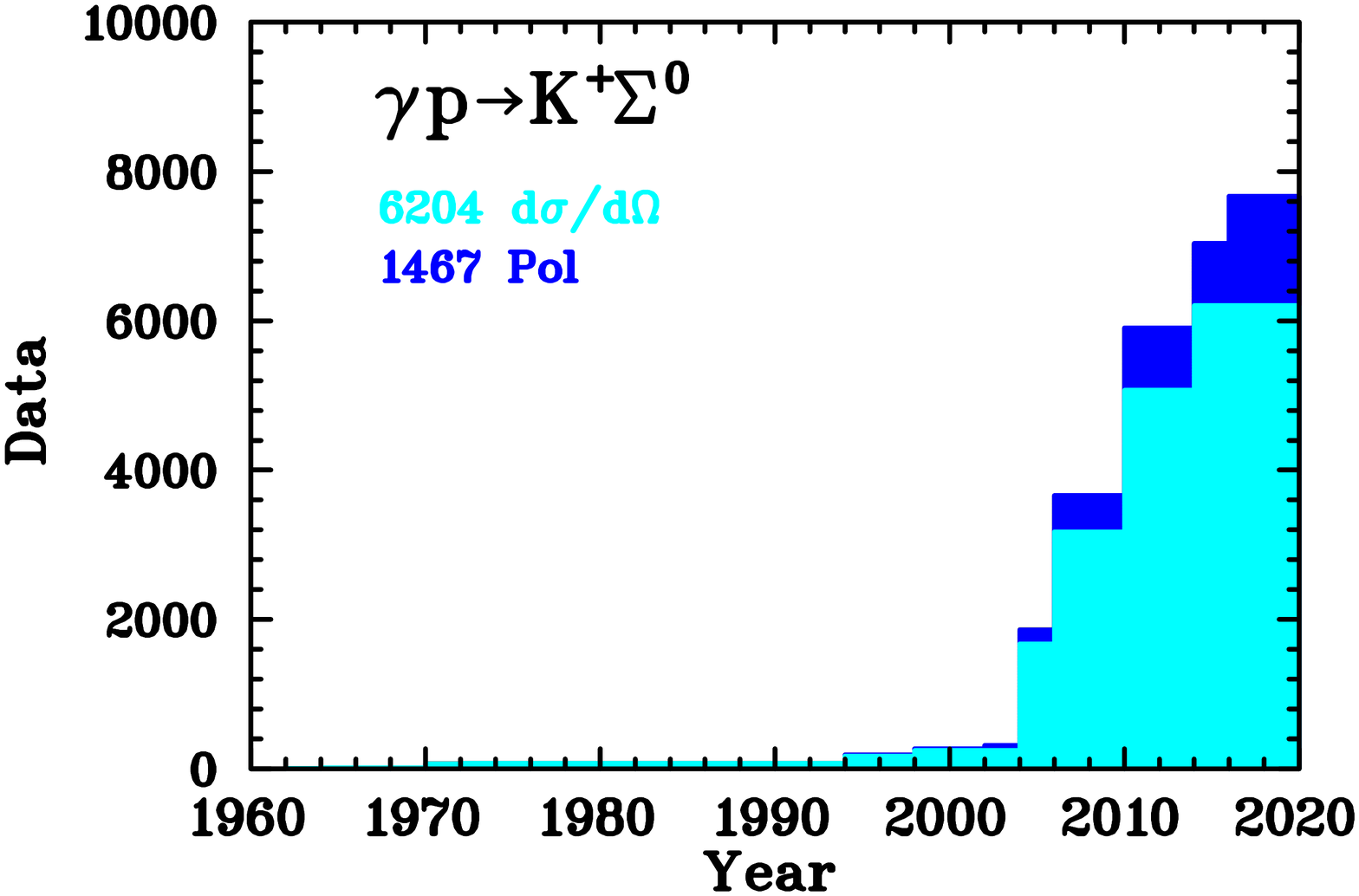}
	\protect\caption{\label{fig:K+S0}Database for $\gamma p\to 
    K^+\Sigma^0$. The notation is the same as in Figure~\protect\ref{fig:pi0p}.}
\end{center}
\end{figure}

\begin{figure}[H] 
\begin{center}
\includegraphics[width=0.4\columnwidth,trim=4cm 2.5cm 0 0]{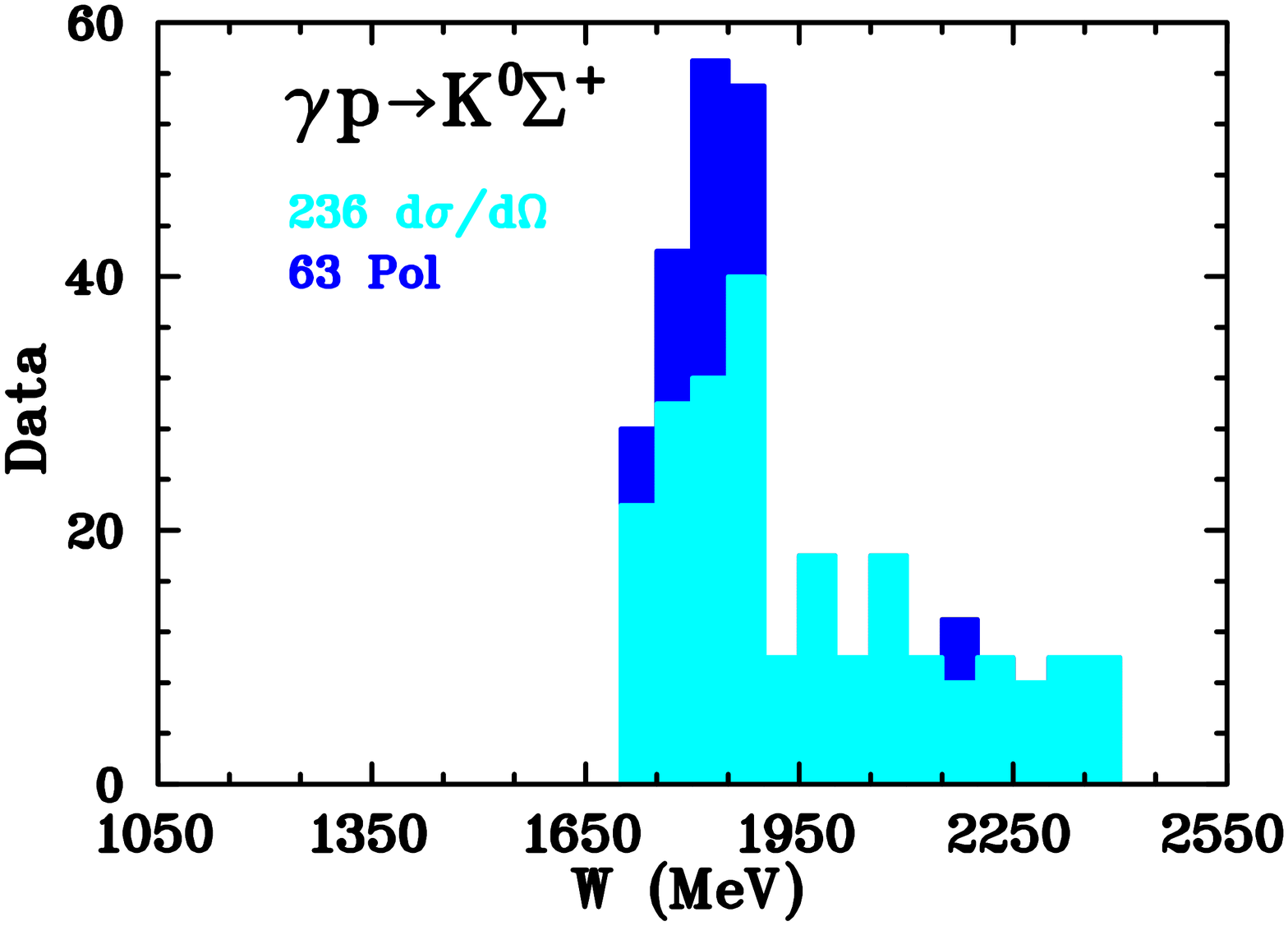}
\includegraphics[width=0.4\columnwidth,trim=4cm 2.5cm 0 0]{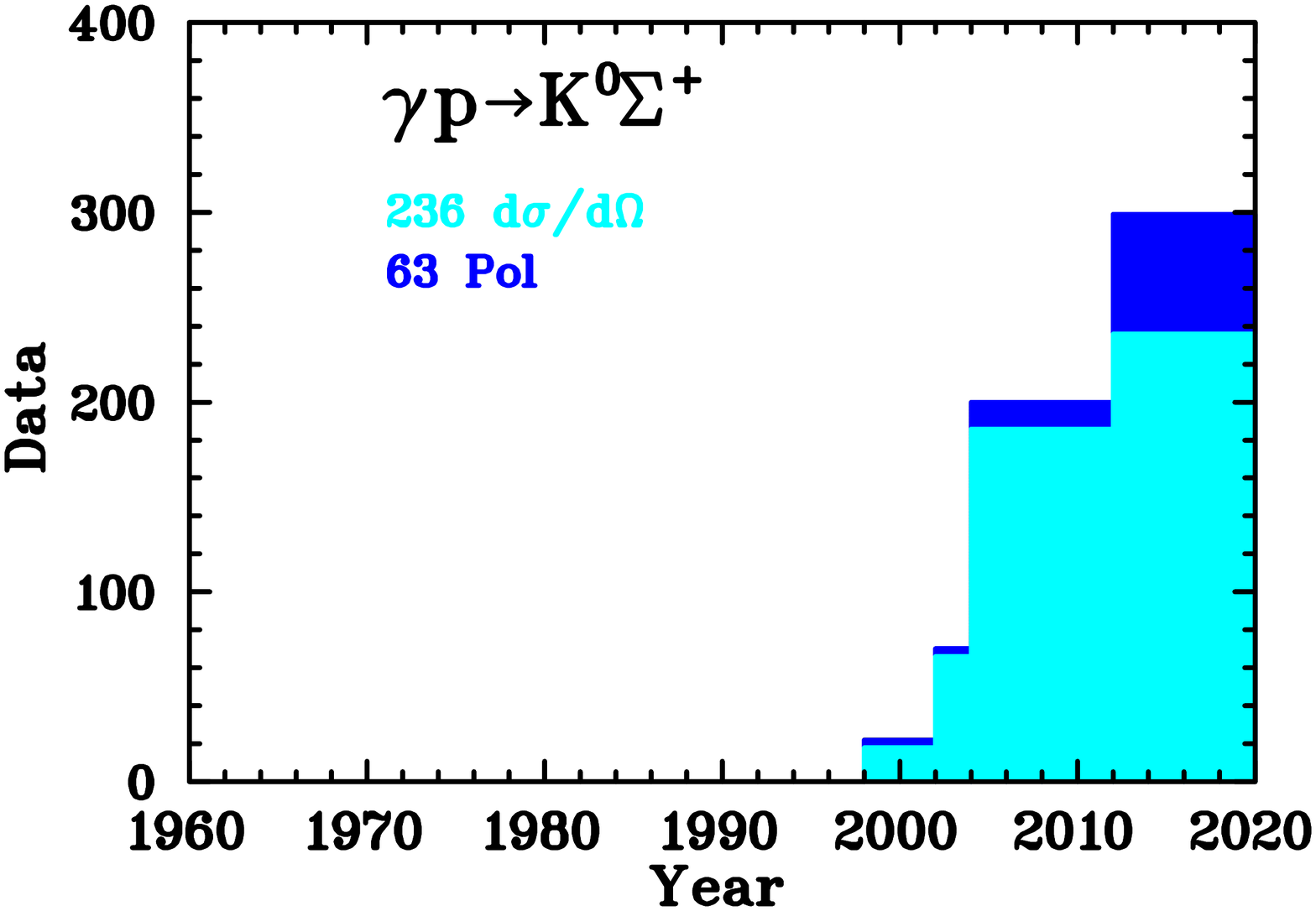}
	\protect\caption{\label{fig:K0S+}Database for $\gamma p\to 
    K^0\Sigma^+$. The notation is the same as in Figure~\protect\ref{fig:pi0p}.}
\end{center}
\end{figure}

\begin{figure} 
\begin{center}
\includegraphics[width=0.4\columnwidth,trim=4cm 2.5cm 0 0]{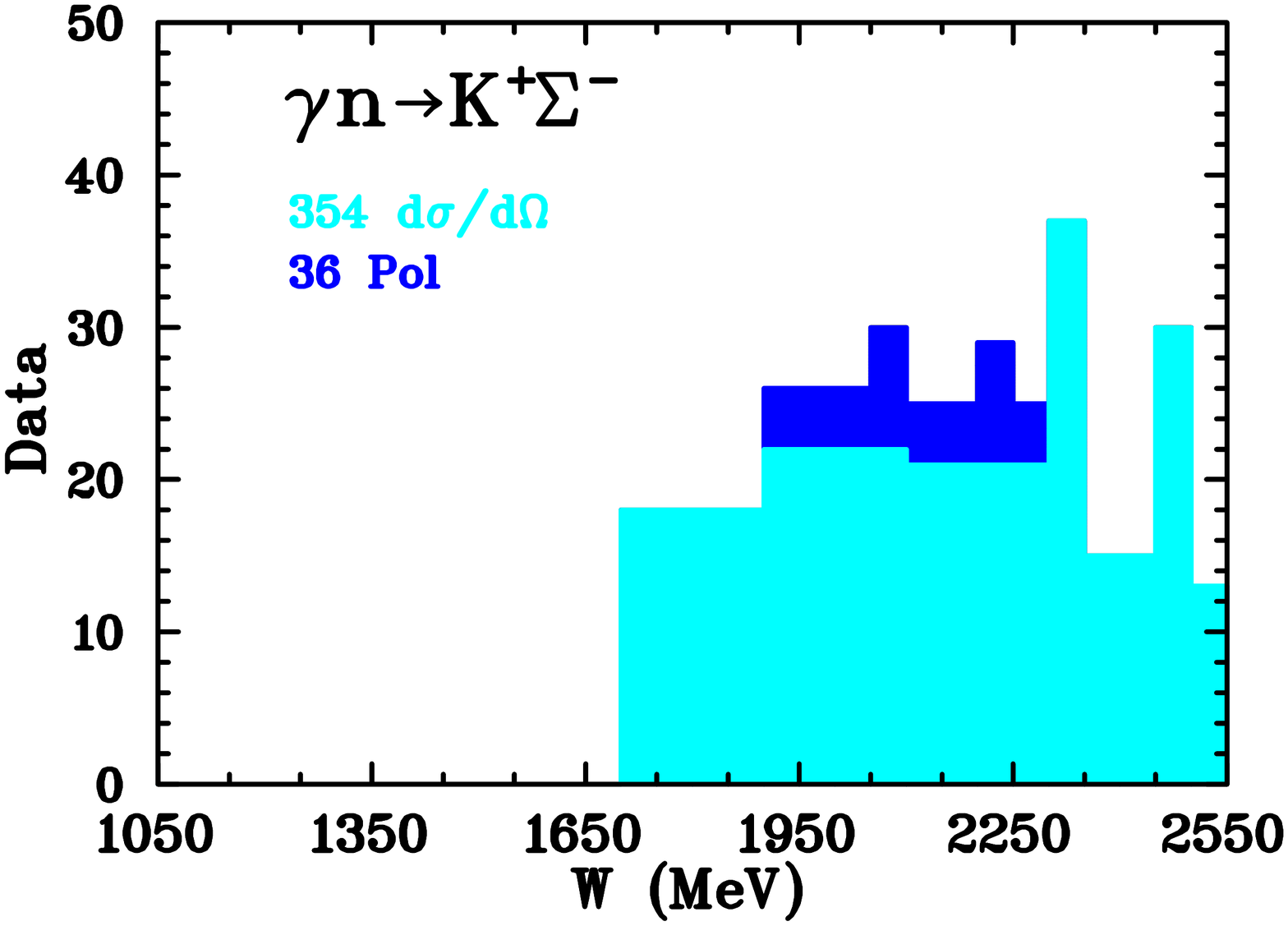}
\includegraphics[width=0.4\columnwidth,trim=4cm 2.5cm 0 0]{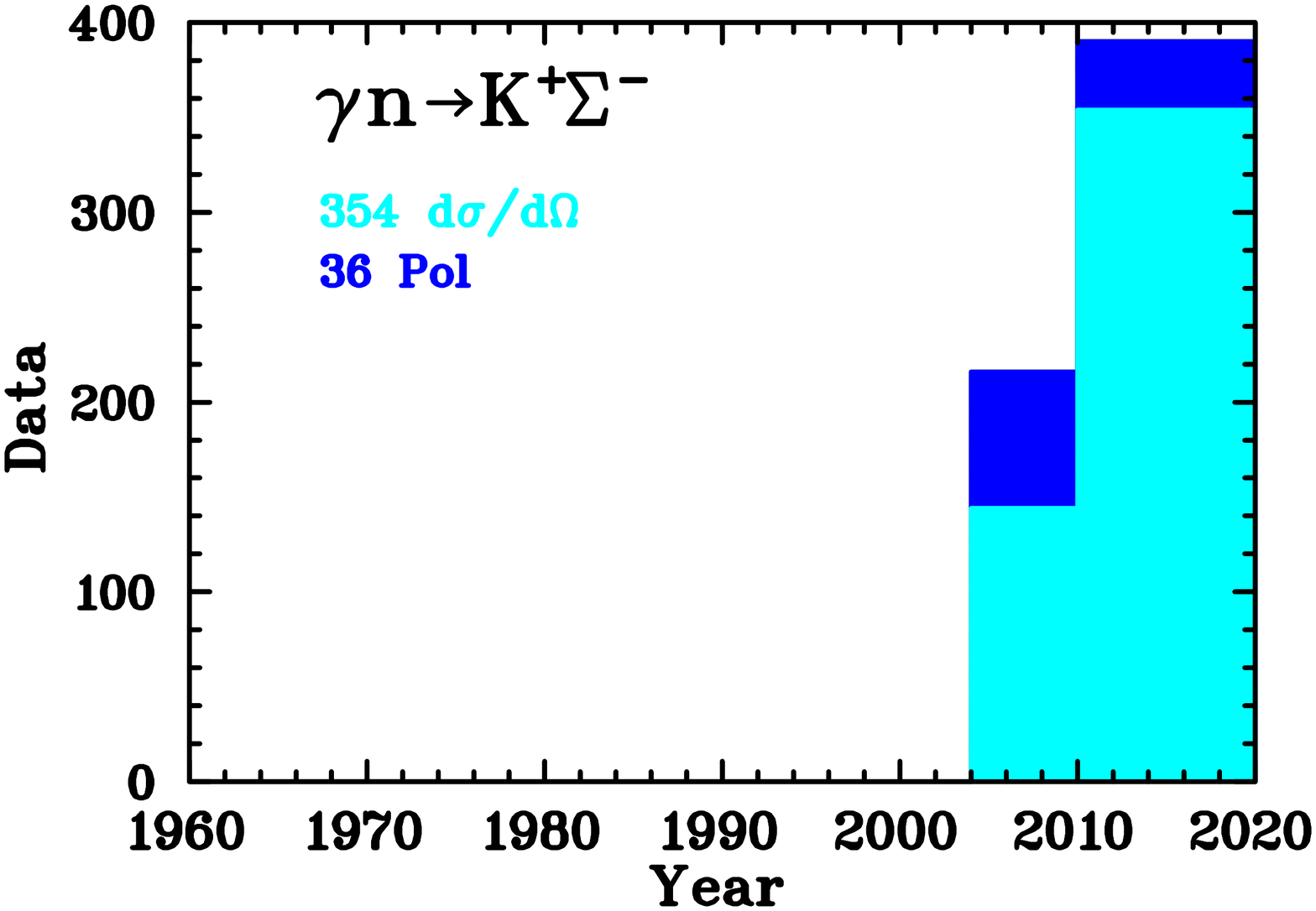}
	\protect\caption{\label{fig:K+S-}Database for $\gamma n\to 
    K^+\Sigma^-$. The notation is the same as in Figure~\protect\ref{fig:pi0p}.}
\end{center}
\end{figure}

\begin{figure}[hbt] 
\begin{center}
\includegraphics[width=0.4\columnwidth,trim=4cm 2.5cm 0 0]{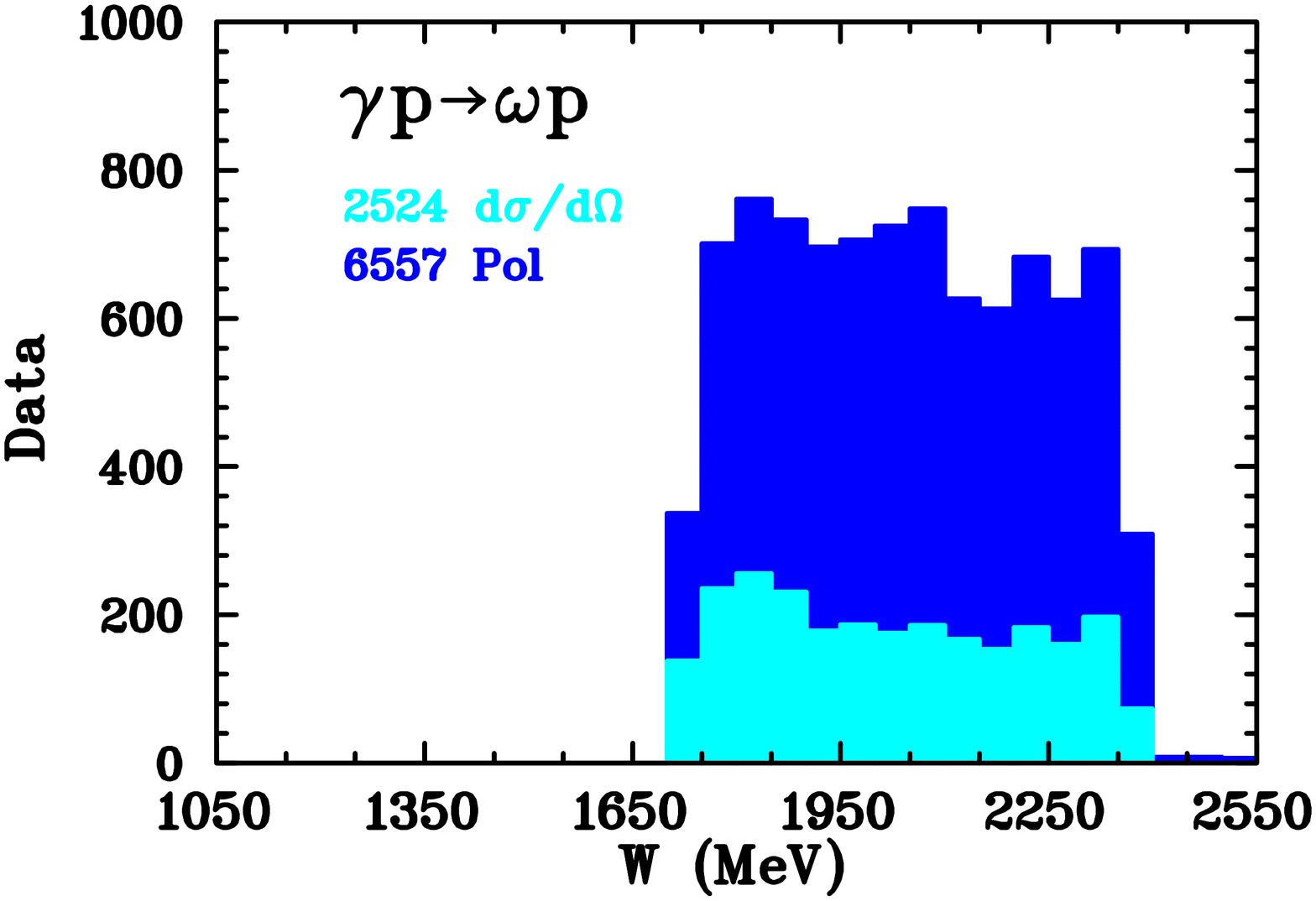}
\includegraphics[width=0.4\columnwidth,trim=4cm 2.5cm 0 0]{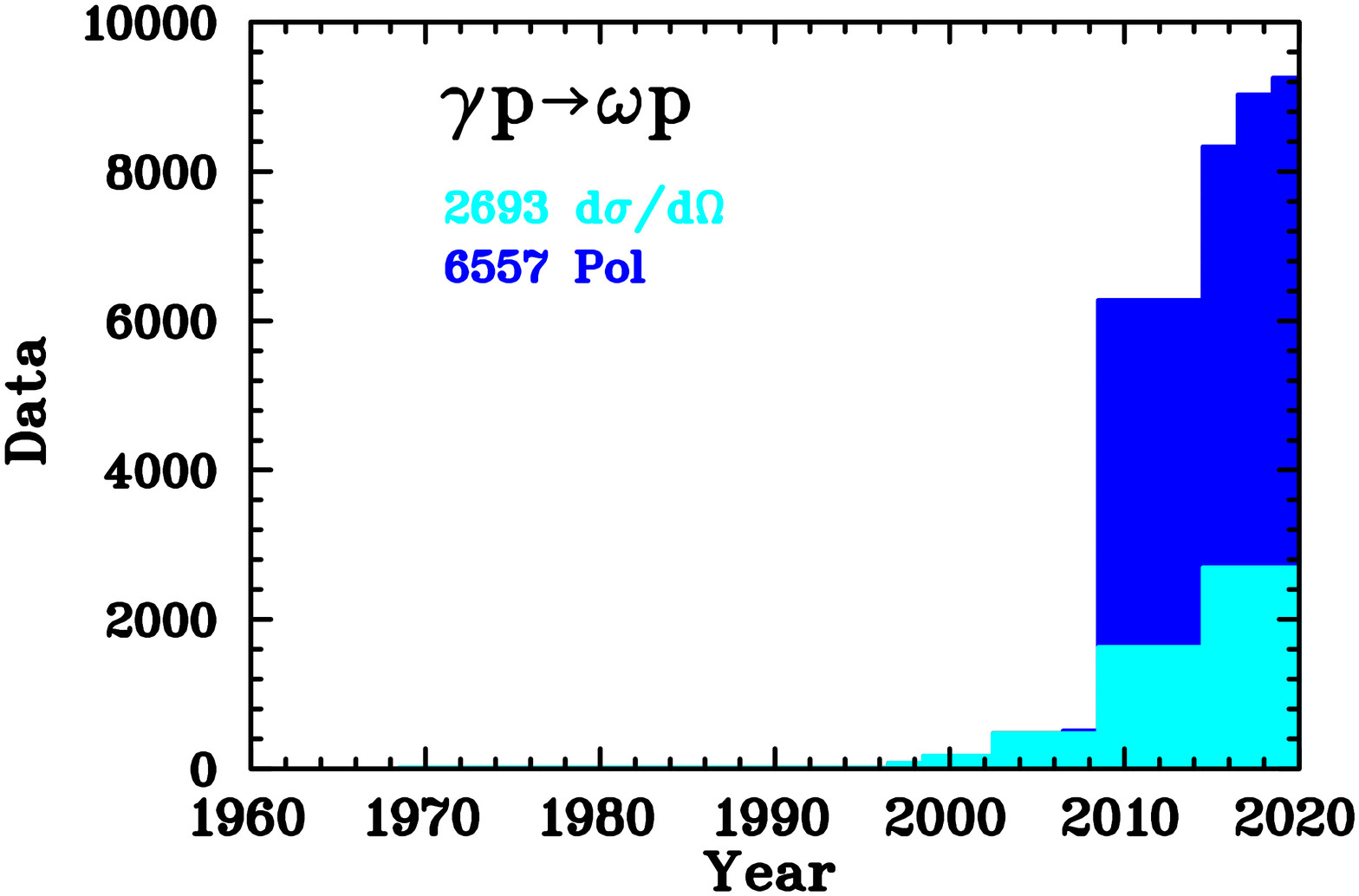}
	\protect\caption{\label{fig:omegap} Database for $\gamma p\to\omega 
    p$. The notation is the same as in Figure~\protect\ref{fig:pi0p}.}
\end{center}
\end{figure}

For convenience we list the thresholds for the relevant photoproduction reactions. They can be found in Table~\ref{tab:thr}. Since most of the photoproduction data were obtained within last two decades, we concentrate on this period. We also limit discussion to the center of mass energies $W \leq 2.55$~GeV (E$_\gamma\leq 3$~GeV). Figures~\ref{fig:pi0p} through~\ref{fig:omegap} show energy distribution for 1996 through 2018 (left) and time distributions (right). Tables~\ref{tab:sumpi0p} through \ref{tab:pomega} provide references to all relevant experiments from 1996 through 2018. They are organized by reaction and include observable, energy and angular range, number of the experimental data and a reference to original publication. We have not included total cross sections because they were not directly measured but obtained by integration of differential ones and depend on the angular range of differential quantities measurements and extrapolation procedure. For the reaction channels with limited amount of measurements we show only tables. For double meson production we don't provide tables but rather just list experiments,  their energy range and extracted observables. The reason for this is following. Since these are not binary reaction there are many possibles choices of the kinematic variables The same data can be binned differently depending on what is the goal of the analysis. In many cases the event by event likelihood analysis was used without any binning.


\vspace{-\parskip}
{
\setlength\extrarowheight{-2pt}
\begin{longtable}{L{3cm} C{2.5cm} C{3cm}}
\multicolumn{3}{r}{\emph{Continued on next page}}
\endfoot
\bottomrule
\endlastfoot
\captionsetup{width=.75\textwidth}
\caption{Threshold energies.}\\
\toprule
Reaction                  & W (MeV) & E$_\gamma$ (MeV)\\ \nopagebreak
\midrule
\endhead
$\gamma p\to\pi^0p$       &  1073.2  & 144.7 \\ \nopagebreak
$\gamma n\to\pi^0n$       &  1074.5  & 144.7 \\ \nopagebreak
$\gamma n\to\pi^-p$       &  1077.8  & 148.4 \\ \nopagebreak
$\gamma p\to\pi^+n$       &  1079.1  & 151.4 \\ \nopagebreak
\midrule 
$\gamma p\to\eta p$       &  1487.4  & 707.6 \\ \nopagebreak
$\gamma n\to\eta n$       &  1486.1  & 707.8 \\ \nopagebreak
\midrule 
$\gamma p\to K^+\Lambda$  &  1609.4  & 911.1 \\ \nopagebreak
$\gamma n\to K^0\Lambda$  &  1613.3  & 915.3 \\ \nopagebreak
\midrule 
$\gamma p\to K^+\Sigma^0$ &  1686.3  & 1046.2 \\ \nopagebreak
$\gamma p\to K^0\Sigma^+$ &  1687.0  & 1047.4 \\ \nopagebreak
$\gamma n\to K^0\Sigma^0$ &  1690.2  & 1050.6 \\ \nopagebreak
$\gamma n\to K^+\Sigma^-$ &  1691.1  & 1052.1 \\ \nopagebreak
\midrule 
$\gamma n\to\omega n$     &  1722.2  & 1108.6 \\ \nopagebreak
$\gamma p\to\omega p$     &  1720.9  & 1109.1 \\ \nopagebreak
\midrule 
$\gamma p\to\eta^\prime p$ &  1896.0  & 1446.6 \\ \nopagebreak
\midrule 
$\gamma p\to\pi^0\pi^0p$ &  1208.2  & 308.8 \\ \nopagebreak
$\gamma p\to\pi^+\pi^-p$ &  1217.4  & 320.7 \\ \nopagebreak
$\gamma p\to\pi^0\eta p$ &  1621.1  & 931.3 \\ \nopagebreak
\end{longtable}\label{tab:thr}}

\subsection{Single Pion Photoproduction \label{subsec:nninter}}

The first experimental study of single pion photoproduction have started just two years after discovery of pion. It has the lowest threshold and at low energies it is dominated by $\Delta$. The amount of data vs. energy essentially follows the cross section
For pion photoproduction, there is a dis-balance between $\pi^0$p and $\pi^+$n measurements, $\pi^+$n/$\pi^0$p = 20\%. While pion photoproduction on the neutron much less known vs on the proton, n/p = 31\%~\cite{SAID}.

{%
\setlength\extrarowheight{-2pt}
\centering
\begin{longtable}{C{3cm} L{2.5cm} C{2cm} r L{2.5cm} C{1cm}}
\multicolumn{6}{r}{\emph{Continued on next page}}
\endfoot
\bottomrule
\endlastfoot
\captionsetup{width=.75\textwidth}
\caption{Data for $\gamma p\to\pi^0p$ below W = 2.55~GeV 
	(E$_\gamma$ = 3~GeV).\\ $\Delta_{13} = (d\sigma/d\Omega)_{1/2}-	  		
    (d\sigma/d\Omega)_{3/2}$. Experimental data are from the SAID 
    database~\protect\cite{SAID} selected for 1996 through
    2018. Polarized data contribution is 40\%.}\\ 
\toprule
Observable & W (MeV)& $\theta$ (deg) & Data & Lab & Ref \\
\midrule
\endhead
\multirow{14}{*}{$d\sigma/d\Omega$}
&1074-1091&10-170&171&MAMI& \cite{SC01} \\ \nopagebreak
&1075-1136&18-162&600&MAMI& \cite{HO13} \\ \nopagebreak
&1122-1537&3-178&1129&MAMI& \cite{BE06} \\ \nopagebreak
&1131-1227&70-130&73&BNL& \cite{BL01} \\ \nopagebreak
&1136-1957&15-165&7978&MAMI& \cite{AD15} \\ \nopagebreak
&1209-1376&55-120&67&MAMI& \cite{AH04} \\ \nopagebreak
&1217-2439&32-148&1089&ELSA& \cite{BA05} \\ \nopagebreak
&1277-1277&70-178&24&MAMI& \cite{SC10} \\ \nopagebreak
&1386-1942&45-168&861&GRAAL& \cite{BA051} \\ \nopagebreak
&1390-1531&45-119&97&MAMI& \cite{AH02} \\ \nopagebreak
&1455-1538&26-154&799&MAMI& \cite{KR99} \\ \nopagebreak
&1465-2505&41-148&620&CEBAF& \cite{DU07} \\ \nopagebreak
&1810-2542&34-80&580&CEBAF& \cite{KU18} \\ \nopagebreak
&1934-2300&129-167&112&Spring-8& \cite{SU07} \\ \nopagebreak
\midrule
\multirow{11}{*}{$\Sigma$}
&1075-1126&25-155&220&MAMI& \cite{HO13} \\ \nopagebreak
&1086-1086&30-150&7&MAMI& \cite{SC01} \\ \nopagebreak
&1131-1306&60-150&84&BNL& \cite{BL01} \\ \nopagebreak
&1154-1306&11-170&353&MAMI& \cite{BE06} \\ \nopagebreak
&1216-1448&31-158&1403&MAMI& \cite{GA16} \\ \nopagebreak
&1349-1702&85-125&158&Yerevan& \cite{AD01} \\ \nopagebreak
&1384-1910&45-171&441&GRAAL& \cite{BA051} \\ \nopagebreak
&1523-1869&37-156&135&ELSA& \cite{EL09} \\ \nopagebreak
&1621-1998& 5-165&249&ELSA& \cite{SP10} \\ \nopagebreak
&1717-2091&32-148&700&CEBAF& \cite{DU13} \\ \nopagebreak
&1946-2280&129-167&48&Spring-8& \cite{SU07} \\ \nopagebreak
\midrule 
\multirow{3}{*}{P}
&1471-1613&51-163&152&ELSA& \cite{HA15} \\ \nopagebreak
&1527-2349&59-135&29&CEBAF& \cite{WI02} \\ \nopagebreak
&2084-2468&96-143&3&CEBAF& \cite{LU12} \\ \nopagebreak
\midrule
\multirow{4}{*}{T}
&1073-1291&5-175&4343&MAMI& \cite{SC15} \\ \nopagebreak
&1179-1398&53-127&52&ELSA& \cite{BO98} \\ \nopagebreak
&1306-1888&30-162&397&MAMI& \cite{CBTAPS} \\ \nopagebreak
&1471-2479&29-163&601&ELSA& \cite{HA15} \\ \nopagebreak
\midrule
\multirow{2}{*}{G}     
&1232-1232&70-110&3&MAMI& \cite{AH05} \\ \nopagebreak
&1438-1822&19-161&318&ELSA& \cite{TH17} \\ \nopagebreak
\midrule
{H}     
&1472-1613&51-163&154&ELSA& \cite{HA15} \\ \nopagebreak
\midrule
{F} 
&1306-1888&30-162&397&MAMI& \cite{AN16} \\ \nopagebreak
\midrule
{E}      
&1426-2259&22-158&456&ELSA& \cite{GO14} \\ \nopagebreak
\midrule
\multirow{2}{*}{$\Delta_{13}$}
&1209-1376&59-122&62&MAMI& \cite{AH04} \\ \nopagebreak
&1390-1531&44-123&78&MAMI& \cite{AH02} \\ \nopagebreak
\midrule
\multirow{3}{*}{$C_{x'}$}         
&1322-1841&75-140&45&MAMI& \cite{SI14} \\ \nopagebreak
&1527-2349&59-135&28&CEBAF& \cite{WI02} \\ \nopagebreak
&2084-2468&96-143&3&CEBAF& \cite{LU12} \\ \nopagebreak
\midrule
\multirow{2}{*}{$C_{z'}$}
&1527-2349&59-135&25&CEBAF& \cite{WI02} \\ \nopagebreak
&2084-2468&96-143&3&CEBAF& \cite{LU12} \\ \nopagebreak
\end{longtable}}\label{tab:sumpi0p}

{%
\setlength\extrarowheight{-2pt}
\begin{longtable}{C{3cm} L{2.5cm} C{2cm} r L{2.5cm} C{1cm}}
\multicolumn{6}{r}{\emph{Continued on next page}}
\endfoot
\bottomrule
\endlastfoot
\captionsetup{width=.75\textwidth}
\caption{Data for $\gamma p\to\pi^+n$ below W = 2.55~GeV (E$_\gamma$ = 3~GeV). 
	$\Delta_{13} = (d\sigma/d\Omega)_{1/2}-(d\sigma/d\Omega)_{3/2}$. 
    Experimental data are from the SAID database~\protect\cite{SAID} 
    selected for 1996 through 2018. Polarized data contribution is 
    51\%.}\\
\toprule
Observable & W (MeV)& $\theta$ (deg) & Data & Lab & Ref \\
\midrule
\endhead
\multirow{9}{*}{$d\sigma/d\Omega$}
&1080-1081&46-134&45&TRIUMF/SAL& \cite{KO99} \\ \nopagebreak
&1104-1313&31-157&205&MAMI& \cite{AH04} \\ \nopagebreak
&1162-1277&72-143&39&MAMI& \cite{BR00} \\ \nopagebreak
&1178-1292&45-135&160&MAMI& \cite{BE00} \\ \nopagebreak
&1193-2201&112-179&1267&ELSA& \cite{DA01} \\ \nopagebreak
&1323-1533&45-155&203&MAMI& \cite{AH06} \\ \nopagebreak
&1497-2505&32-148&618&CEBAF& \cite{DU09} \\ \nopagebreak
&1714-2354&50-90&10&CEBAF& \cite{ZH05} \\ \nopagebreak
&1934-2524&11-49&174&Spring-8& \cite{KO17} \\ \nopagebreak
\midrule
\multirow{6}{*}{$\Sigma$}
&1178-1292&20-170&85&BNL& \cite{BL01} \\ \nopagebreak
&1178-1292&45-135&160&MAMI& \cite{BE00} \\ \nopagebreak
&1416-1688&48-154&92&GRAAL& \cite{AJ00} \\ \nopagebreak
&1543-1901&47-160&237&GRAAL& \cite{BA02} \\ \nopagebreak
&1722-2091&32-148&386&CEBAF& \cite{DU13} \\ \nopagebreak
&1946-2496&11-49&84&Spring-8& \cite{KO17} \\ \nopagebreak
\midrule
{G}          
&1232-1232&30-130&6&MAMI& \cite{AH05} \\ \nopagebreak
\midrule
{E}
&1250-2230&20-148&900&CEBAF& \cite{ST15a} \\ \nopagebreak
\midrule
\multirow{2}{*}{$\Delta_{13}$}    
&1104-1313&35-153&129&MAMI& \cite{AH04} \\ \nopagebreak
&1323-1524&50-150&102&MAMI& \cite{AH06} \\ \nopagebreak
\end{longtable}}\label{tab:sumpi+n}

\clearpage

{%
\setlength\extrarowheight{-2pt}
\begin{longtable}{C{3cm} L{2.5cm} C{2cm} r L{2.5cm} C{1cm}}
\multicolumn{6}{r}{\emph{Continued on next page}}
\endfoot
\bottomrule
\endlastfoot
\captionsetup{width=.75\textwidth}
\caption{Data for $\gamma n\to\pi^-p$ below W = 2.55~GeV (E$_\gamma$ = 3.1~GeV). 
	Experimental data from the SAID database~\protect\cite{SAID} selected for 1996 through 2018. Polarized data contribution is 4\%.}
    \\
\toprule
Observable & W (MeV)& $\theta$ (deg) & Data & Lab & Ref \\
\midrule
\endhead
\midrule
\multirow{5}{*}{$d\sigma/d\Omega$}
&1191-1526&41-148&300&BNL& \cite{SH04} \\ \nopagebreak
&1203-1318&58-133&104&MAMI& \cite{BR12} \\ \nopagebreak
&1311-2366&26-135&8428&CEBAF& \cite{MA17} \\ \nopagebreak
&1690-2551&33-157&699&CEBAF& \cite{Chen2009,Chen2012} \\ \nopagebreak
&1720-2356&50-90&1&CEBAF& \cite{ZH05} \\ \nopagebreak
\midrule
{$\Sigma$}  
&1516-1894&33-163&99&GRAAL & \cite{MA10} \\ \nopagebreak
\midrule
{E}
&1500-2300&26-154&266&CEBAF& \cite{HO17} \\ \nopagebreak
\end{longtable}}\label{tab:sumpi-p}

\pagebreak

{%
\setlength\extrarowheight{-2pt}
\begin{longtable}{C{3cm} L{2.5cm} C{2cm} r L{2.5cm} C{1cm}}
\multicolumn{6}{r}{\emph{Continued on next page}}
\endfoot
\bottomrule
\endlastfoot
\captionsetup{width=.75\textwidth}
\caption{Data for $\gamma n\to\pi^0n$ below W = 	  
	2.55~GeV (E$_\gamma$ = 3~GeV). Experimental data are from the SAID 
    database~\protect\cite{SAID} selected for 1996 through 2018. 
    Polarized data contribution is 24\%.}\\
\toprule
Observable & W (MeV)& $\theta$ (deg) & Data & Lab & Ref \\
\midrule
\endhead
{$d\sigma/d\Omega$}
&1300-1900&32-162&969&MAMI& \cite{DI17} \\ \nopagebreak
\midrule
{$\Sigma$} 
&1484-1912&53-164&216&GRAAL& \cite{DI09} \\ \nopagebreak
\midrule
{E}
&1312-1888&46-154&151&MAMI& \cite{DI17} \\ 
\end{longtable}}\label{tab:sumpi0n}

\subsection{$\eta$ and $\eta^\prime$ photoproduction}

Since $\eta$ and $\eta^\prime$ are iso-singlets their photoproduction may not be directly coupled to $\Delta$ resonances but only to the excitation of $N^\ast$

{%
\setlength\extrarowheight{-2pt}
\begin{longtable}{C{3cm} L{2.5cm} C{2cm} r L{2.5cm} C{1cm}}
\multicolumn{6}{r}{\emph{Continued on next page}}
\endfoot
\bottomrule
\endlastfoot
\captionsetup{width=.75\textwidth}
\caption{Data for $\gamma p\to\eta p$ below W = 
	2.55~GeV (E$_\gamma$ = 3.1~GeV).  $\Delta_{13} = (d\sigma/d\Omega)_{1/2}-
	(d\sigma/d\Omega)_{3/2}$. Experimental data are from the SAID
  	database~\protect\cite{SAID} selected for 1996 through 2018. 
    Polarized data contribution is 6\%.}\\
\toprule
Observable & W (MeV)& $\theta$ (deg) & Data & Lab & Ref \\
\midrule
\endhead
\multirow{10}{*}{$d\sigma/d\Omega$} 
&1488-1870&18-162&2400&MAMI& \cite{MC10} \\ \nopagebreak
&1488-1957&17-163&5880&MAMI& \cite{KA17} \\ \nopagebreak
&1490-1911&32-162&487&GRAAL& \cite{BA07} \\ \nopagebreak
&1492-1739&26-154&180&LNS& \cite{NA06} \\ \nopagebreak
&1528-2120&46-134&190&CEBAF& \cite{DU02} \\ \nopagebreak
&1528-2120&33-148&1012&CEBAF& \cite{WI09} \\ \nopagebreak
&1533-2510&18-139&631&ELSA& \cite{CR05} \\ \nopagebreak
&1533-1537&70-70&2&MAMI& \cite{AR03} \\ \nopagebreak
&1685-2370&18-162&680&ELSA& \cite{CR09} \\ \nopagebreak
&1994-2300&130-162&32&Spring-8& \cite{SU09} \\ \nopagebreak
\midrule
\multirow{3}{*}{$\Sigma$}
&1496-1909&33-161&150&GRAAL& \cite{BA07} \\ \nopagebreak
&1569-1845&51-148&34&ELSA& \cite{ER07} \\ \nopagebreak
&1700-2080&46-134&201&CEBAF& \cite{CO17a} \\ \nopagebreak
\midrule
\multirow{2}{*}{T}  
&1492-1719&33-145&50&ELSA& \cite{BO98} \\ \nopagebreak
&1497-1848&24-156&144&MAMI& \cite{AK14} \\ \nopagebreak
\midrule
{F}     
&1497-1848&24-156&144&MAMI& \cite{AK14} \\ \nopagebreak
\midrule
{E}     
&1525-2125&46-154&69&CEBAF& \cite{SC16} \\ \nopagebreak
\midrule
{$\Delta_{13}$}   
&1533-1537&70-70&129&MAMI& \cite{AR03} \\ \nopagebreak
\end{longtable}}\label{tab:sumetap}


{%
\setlength\extrarowheight{-2pt}
\begin{longtable}{C{3cm} L{2.5cm} C{2cm} r L{2.5cm} C{1cm}}
\multicolumn{6}{r}{\emph{Continued on next page}}
\endfoot
\bottomrule
\endlastfoot
\captionsetup{width=.75\textwidth}
\caption{Data for $\gamma n\to\eta n$ below W = 
	2.55~GeV (E$_\gamma$ = 3~GeV). Experimental data are from the SAID 	
	database~\protect\cite{SAID} selected for 1996 through 2018. 
    Polarized data contribution is 15\%.}\\
\toprule
Observable & W (MeV)& $\theta$ (deg) & Data & Lab & Ref \\
\midrule
\endhead
\multirow{3}{*}{$d\sigma/d\Omega$}
&1483-2322&26-154&200&ELSA& \cite{JA08}\\ \nopagebreak
&1487-2070&51-151&279&ELSA& \cite{WI17}\\ \nopagebreak
&1492-1875&18-162&880&MAMI& \cite{WI13}\\ \nopagebreak
\midrule
{$\Sigma$}
&1506-1894&32-165&99&GRAAL& \cite{FA08}\\ \nopagebreak
\midrule
{E}
&1505-1882&37-143&135&MAMI& \cite{WI16}\\ \nopagebreak
\end{longtable}}\label{tab:sumetan}

{%
\setlength\extrarowheight{-2pt}
\begin{longtable}{C{3cm} L{2.5cm} C{2cm} r L{2.5cm} C{1cm}}
\multicolumn{6}{r}{\emph{Continued on next page}}
\endfoot
\bottomrule
\endlastfoot
\captionsetup{width=.75\textwidth}
\caption{ Data for $\gamma p\to\eta^\prime p$ below W 
	= 2.55~GeV (E$_\gamma$ = 3~GeV). Experimental data are from the SAID 
    database~\protect\cite{SAID} selected for 1996 through 2018. 
    Polarized data contribution is 7\%.}\\
\toprule
Observable & W (MeV)& $\theta$ (deg) & Data & Lab & Ref \\
\midrule
\endhead
\multirow{5}{*}{$d\sigma/d\Omega$} 
&1898-1956&26-154&120&MAMI& \cite{KA17} \\ \nopagebreak
&1917-2336&37-143&34&ELSA& \cite{PL98} \\ \nopagebreak
&1925-2380&32-146&524&CEBAF& \cite{WI09} \\ \nopagebreak
&1934-2350&26-154&200&ELSA& \cite{CR09} \\ \nopagebreak
&1935-2249&46-134&111&CEBAF& \cite{DU06} \\ \nopagebreak
\midrule
\multirow{2}{*}{$\Sigma$}
&1903-1912&20-159&14&GRAAL& \cite{LE15} \\
&1904-2080&46-134&60&CEBAF& \cite{CO17a} \\\nopagebreak
\end{longtable}}\label{tab:sumetapn}


\subsection{Kaon photoproduction}

Whilst the cross section for kaon photoproduction is a couple of orders of magnitude smaller than pion photoproduction, these channels have been seen as the ``golden'' channels in recent times for a number of reasons. A different coupling of kaons to light baryon resonances had been hypothesized as a means of discovering more resonances~\cite{capstick_strange_1998}. More importantly, especially with the $K\Lambda$ final state, the self-analyzing property of the $\Lambda$ through its weak decay means that information on the recoil polarization is readily obtainable in the final state. Together with the advances in photon beam and target polarization, this has meant that a large number of polarization observables have been extracted across the resonance region. Such data have been shown to be extremely useful in fitting model parameters and establishing the existence of resonances.

\begin{figure}[H] 
\begin{center}
\includegraphics[width=0.4\columnwidth,trim=4cm 2.5cm 0 0]{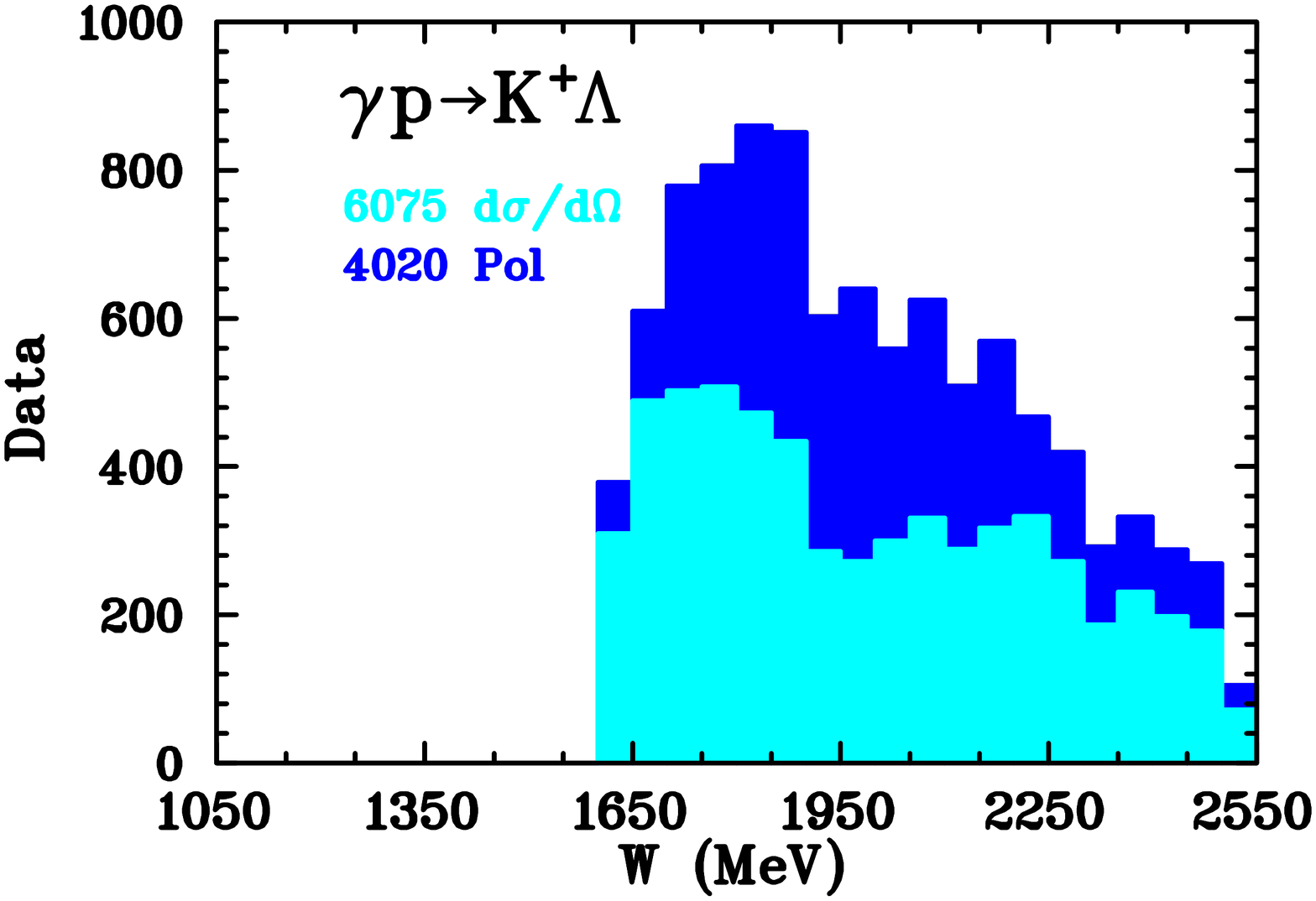}
\includegraphics[width=0.4\columnwidth,trim=4cm 2.5cm 0 0]{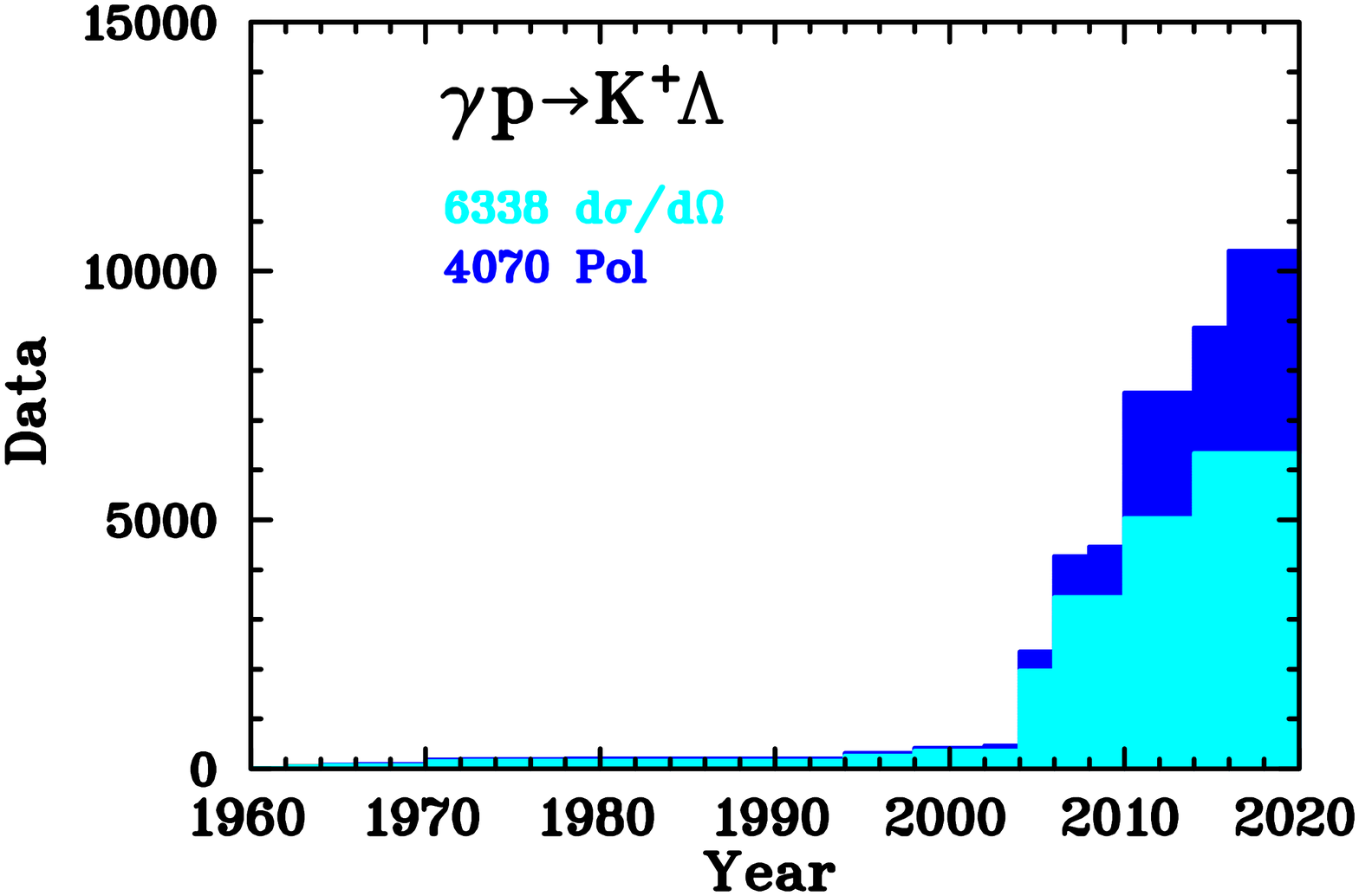}
	\protect\caption{\label{fig:K+L}Database for $\gamma p\to K^+
    \Lambda$. The notation is the same as in Figure~\protect\ref{fig:pi0p}.}
\end{center}
\end{figure}

The plot in Figure~\ref{fig:K+L} indicates that very few kaon photoproduction data were available before the start of the century. Initial measurements by SAPHIR~\cite{TR98,GL04}, 
SPring-8~\cite{SU06,ZE03,HI07} and GRAAL~\cite{LL07,LL09} have been added to by a comprehensive campaign of measurements by CLAS~\cite{MC04,BR07,MR10,DE10,PA16}.

It should be noted that, at the time of writing, a recently published paper by the BES Collaboration~\cite{ablikim_polarization_2019}, and a study of kaon photoproduction at CLAS~\cite{ireland_kaon_2019} have cast doubt on the previously quoted value of the weak decay parameter $\alpha_-$ of the $\Lambda$. The value obtained by both analyses is significantly higher than the number quoted in the current PDG~\cite{PDG_2018}. As such, this means that the polarization observables that depend on $\alpha_-$ (beam asymmetry, beam-recoil observables) could be systematically too high, and analyses that depend on a fit to them should be examined to establish whether this change would make a difference to the final results.

{%
\setlength\extrarowheight{-2pt}
\begin{longtable}{C{3cm} L{2.5cm} C{2cm} r L{2.5cm} C{1cm}}
\multicolumn{6}{r}{\emph{Continued on next page}}
\endfoot
\bottomrule
\endlastfoot
\captionsetup{width=.75\textwidth}
\caption{ Data for $\gamma p\to K^+\Lambda$ below W 	
	= 2.55~GeV (E$_\gamma$ = 3~GeV). Experimental data are from the SAID 
    database~\protect\cite{SAID} selected for 1996 through 2018. 
    Polarized data contribution is 40\%.}\\
\toprule
Observable & W (MeV)& $\theta$ (deg) & Data & Lab & Ref \\
\midrule
\endhead
\multirow{7}{*}{$d\sigma/d\Omega$} 
&1610-2390&18-162&701&ELSA& \cite{GL04} \\ \nopagebreak
&1612-1896&66-143&1306&MAMI& \cite{JU14} \\ \nopagebreak
&1617-2290&32-148&920&CEBAF& \cite{MC04} \\ \nopagebreak
&1617-2108&26-154&90&ELSA& \cite{TR98} \\ \nopagebreak
&1625-2395&27-154&1674&CEBAF& \cite{MR10} \\ \nopagebreak
&1628-2533&26-143&1377&CEBAF& \cite{BR06} \\ \nopagebreak
&1934-2310&13-41&78&Spring-8& \cite{SU06} \\ \nopagebreak
\midrule
\multirow{5}{*}{$\Sigma$}      
&1649-1906&31-144&66&GRAAL& \cite{LL07} \\ \nopagebreak
&1721-2180&37-134&314&CEBAF& \cite{PA16} \\ \nopagebreak
&1946-2300&13-49&45&Spring-8& \cite{ZE03} \\ \nopagebreak
&1946-2280&13-49&30&Spring-8& \cite{SU06} \\ \nopagebreak
&2041-2238&18-32&4&Spring-8& \cite{HI07} \\ \nopagebreak
\midrule
\multirow{6}{*}{P}              
&1617-2290&26-154&233&CEBAF& \cite{MC04} \\ \nopagebreak
&1625-2545&26-143&1497&CEBAF& \cite{MR10} \\ \nopagebreak
&1649-1906&31-144&66&GRAAL& \cite{LL07} \\ \nopagebreak
&1660-2017&41-139&12&ELSA& \cite{TR98} \\ \nopagebreak
&1660-2280&34-146&30&ELSA& \cite{GL04} \\ \nopagebreak
&1721-2180&37-134&314&CEBAF& \cite{PA16} \\ \nopagebreak
\midrule
\multirow{2}{*}{T}             
&1649-1906&31-144&66&GRAAL& \cite{LL09} \\ \nopagebreak
&1721-2180&37-134&314&CEBAF& \cite{PA16} \\ \nopagebreak
\midrule
{$C_{x'}$}            
&1678-2454&32-139&144&CEBAF& \cite{BR07} \\ \nopagebreak
\midrule
{$C_{z'}$}            
&1678-2454&32-139&146&CEBAF& \cite{BR07} \\ \nopagebreak
\midrule
\multirow{2}{*}{$O_{x}$}            
&1649-1906&31-144&66&GRAAL& \cite{LL09} \\ \nopagebreak
&1721-2180&37-134&314&CEBAF& \cite{PA16} \\ \nopagebreak
\midrule
\multirow{2}{*}{$O_{z}$}          
&1649-1906&31-144&66&GRAAL& \cite{LL09} \\ \nopagebreak
&1721-2180&37-134&314&CEBAF& \cite{PA16} \\ \nopagebreak
\end{longtable}}\label{tab:kplambda}

{%
\setlength\extrarowheight{-2pt}
\begin{longtable}{C{3cm} L{2.5cm} C{2cm} r L{2.5cm} C{1cm}}
\multicolumn{6}{r}{\emph{Continued on next page}}
\endfoot
\bottomrule
\endlastfoot
\captionsetup{width=.75\textwidth}
\caption{ Data for $\gamma p\to K^+\Sigma^0$ below 
	W = 2.55~GeV (E$_\gamma$ = 3~GeV). Experimental data are from the SAID 
    database~\protect\cite{SAID} selected for 1996 through 2018. 
    Polarized data contribution is 42\%.}\\
\toprule
Observable & W (MeV)& $\theta$ (deg) & Data & Lab & Ref \\
\midrule
\endhead
\multirow{9}{*}{$d\sigma/d\Omega$} 
&1695-2545&26-180&1576&CEBAF& \cite{DE10} \\ \nopagebreak
&1695-2390&18-162&656&ELSA& \cite{GL04} \\ \nopagebreak
&1702-2290&32-139&778&CEBAF& \cite{MC04} \\ \nopagebreak
&1703-1896&66-143&1130&MAMI& \cite{JU14} \\ \nopagebreak
&1713-2533&26-143&1279&CEBAF& \cite{BR06} \\ \nopagebreak
&1716-2370&26-154&120&ELSA& \cite{LA05} \\ \nopagebreak
&1716-2097&26-154&920&ELSA& \cite{TR98} \\ \nopagebreak
&1934-2310&13-41&78&Spring-8& \cite{SU06} \\ \nopagebreak
&1934-2310&18-49&144&Spring-8& \cite{KO06} \\ \nopagebreak
\midrule
\multirow{6}{*}{$\Sigma$}         
&1737-2170&37-124&127&CEBAF& \cite{PA16} \\ \nopagebreak
&1755-1906&18-138&42&GRAAL& \cite{LL07} \\ \nopagebreak
&1822-2185&37-143&10&ELSA& \cite{LA05} \\ \nopagebreak
&1946-2300&13-49&45&Spring-8& \cite{ZE03} \\ \nopagebreak
&1946-2280&13-49&30&Spring-8& \cite{SU06} \\ \nopagebreak
&1946-2300&13-49&72&Spring-8& \cite{KO06} \\ \nopagebreak
\midrule
\multirow{6}{*}{P}                
&1728-2550&27-163&355&CEBAF& \cite{DE10} \\ \nopagebreak
&1737-2170&37-124&127&CEBAF& \cite{PA16} \\ \nopagebreak
&1743-2029&41-139&12&ELSA& \cite{TR98} \\ \nopagebreak
&1743-2280&41-139&16&ELSA& \cite{GL04} \\ \nopagebreak
&1756-2290&26-134&97&CEBAF& \cite{MC04} \\ \nopagebreak
&1762-1851&39-130& 8&GRAAL& \cite{LL07} \\ \nopagebreak
\midrule
{T}          
&1737-2170&37-124&127&CEBAF&\cite{PA16} \\ \nopagebreak
\midrule
{$C_{x}$}    
&1787-2454&37-134&71&CEBAF& \cite{BR07} \\ \nopagebreak
\midrule
{$C_{z}$}    
&1787-2454&37-134&72&CEBAF& \cite{BR07} \\ \nopagebreak
\midrule
{$O_{x}$}    
&1737-2170&37-124&127&CEBAF& \cite{PA16} \\ \nopagebreak
\midrule
{$O_{z}$}    
&1737-2170&37-124&127&CEBAF& \cite{PA16} \\ \nopagebreak
\end{longtable}}\label{tab:k+sigma0}

{%
\setlength\extrarowheight{-2pt}
\begin{longtable}{C{3cm} L{2.5cm} C{2cm} r L{2.5cm} C{1cm}}
\multicolumn{6}{r}{\emph{Continued on next page}}
\endfoot
\bottomrule
\endlastfoot
\captionsetup{width=.75\textwidth}
\caption{ Data for $\gamma p\to K^0\Sigma^+$ below 		
	W = 2.55~GeV (E$_\gamma$ = 3~GeV). Experimental data are from the SAID 
    database~\protect\cite{SAID} selected for 1996 through 2018. 
    Polarized data contribution is 21\%.}\\
\toprule
Observable & W (MeV)& $\theta$ (deg) & Data & Lab & Ref \\
\midrule
\endhead
\multirow{3}{*}{$d\sigma/d\Omega$} 
&1730-1885&29-151&50&MAMI& \cite{AG13} \\ \nopagebreak
&1743-1898&20-156&18&ELSA& \cite{GO99} \\ \nopagebreak
&2062-2263&46-134&48&ELSA& \cite{LA05} \\ \nopagebreak
\midrule
\multirow{3}{*}{P}               
&1730-1885&29-151&49&MAMI& \cite{AG13} \\ \nopagebreak
&1822-1822&30-150&4&ELSA& \cite{GO99} \\ \nopagebreak
&2073-2073&30-150&4&ELSA& \cite{LA05} \\ \nopagebreak
\end{longtable}}\label{tab:k0sigma+}


{%
\setlength\extrarowheight{-2pt}
\begin{longtable}{C{3cm} L{2.5cm} C{2cm} r L{2.5cm} C{1cm}}
\multicolumn{6}{r}{\emph{Continued on next page}}
\endfoot
\bottomrule
\endlastfoot
\captionsetup{width=.75\textwidth}
\caption{Data for $\gamma n\to K^+\Sigma^-$ below 
    W = 2.55~GeV (E$_\gamma$ = 3~GeV). Experimental data are from the SAID 
    database~\protect\cite{SAID} selected for 1996 through 2018. 
    Polarized data contribution is 9\%.}\\
\toprule
Observable & W (MeV)& $\theta$ (deg) & Data & Lab & Ref \\
\midrule
\endhead
\multirow{2}{*}{$d\sigma/d\Omega$} 
&1745-2535&34-151&285&CEBAF& \cite{PE10} \\ \nopagebreak
&1934-2310&18-49& 144&Spring-8& \cite{KO06} \\ \nopagebreak
\midrule
{$\Sigma$}          
&1946-2300&13-49& 36&Spring-8& \cite{KO06} \\ \nopagebreak
\end{longtable}}\label{tab:k+sigma-}

{%
\setlength\extrarowheight{-2pt}
\begin{longtable}{C{3cm} L{2.5cm} C{2cm} r L{2.5cm} C{1cm}}
\multicolumn{6}{r}{\emph{Continued on next page}}
\endfoot
\bottomrule
\endlastfoot
\captionsetup{width=.75\textwidth}
\caption{Data for $\gamma n\to K^0 \Lambda$ below W 		
	= 2.55~GeV (E$_\gamma$ = 3~GeV). Experimental data are from the SAID 
    database~\protect\cite{SAID} selected for 1996 through 2018. There 
    are no unpolarized measurements.}\\
\toprule
Observable & W (MeV)& $\theta$ (deg) & Data & Lab & Ref \\
\midrule
\endhead
{$d\sigma/d\Omega$} 
&1645-2516&41-130&360&CEBAF& \cite{CO17} \\ \nopagebreak
\midrule
{E}
&1700-2020&53-127&6&CEBAF& \cite{HO18} \\ \nopagebreak
\midrule
\end{longtable}}\label{tab:sumK0L}

{%
\setlength\extrarowheight{-2pt}
\begin{longtable}{C{3cm} L{2.5cm} C{2cm} r L{2.5cm} C{1cm}}

\multicolumn{6}{r}{\emph{Continued on next page}}
\endfoot
\bottomrule
\endlastfoot
\captionsetup{width=.75\textwidth}
\caption{Data for $\gamma n\to K^0\Sigma^0$ below W 		
	= 2.55~GeV (E$_\gamma$ = 3~GeV). Experimental data are from the SAID 
    database~\protect\cite{SAID} selected for 1996 through 2018. 
    There are no unpolarized measurements.}\\
\toprule
Observable & W (MeV)&$\theta$ (deg) & Data & Lab & Ref \\
\midrule
\endhead
{E}        
&1700-2020&53-127&6&CEBAF& \cite{HO18}\\ \nopagebreak
\midrule
\end{longtable}}\label{tab:sumK0S0}

\subsection{$\omega$ photoproduction}

There was no $\omega$ photoproduction data before 2003. A substantial amount of data was accumulated since the. All major facilities (CLAS, CBELSA, Crysta Ball at MAMI, GRAAL) made their contributions. 
Based on these data it was found that excitation of nucleon resonance plays important role in $\omega$ photoproduction. \pagebreak[4] The quality of the data near threshold gives access to a variety of interesting physics aspects. As an example, an estimation of the $\omega$N scattering length $\alpha_{\omega p}$ is provided~\cite{ST15}.

{%
\setlength\extrarowheight{-2pt}
\begin{longtable}{C{3cm} L{2.5cm} C{2cm} r L{2.5cm} C{1cm}}
\multicolumn{6}{r}{\emph{Continued on next page}}
\endfoot
\bottomrule
\endlastfoot
\captionsetup{width=.75\textwidth}
\caption{Data for $\gamma p\to\omega p$ below W = 
	2.55~GeV (E$_\gamma$ = 3~GeV). SDME is spin-density matrix element. 
	Experimental data are from the SAID database~\protect\cite{SAID} 
	selected for 1996 through 2018. Polarized data contribution is 
	72\%.}\\
\toprule
Observable & W (MeV)& $\theta$ (deg) & Data & Lab & Ref \\
\midrule
\endhead
\multirow{5}{*}{$d\sigma/d\Omega$} 
&1723-2380&13-159&307&ELSA& \cite{BA03} \\ \nopagebreak
&1725-2545&24-147&1148&CEBAF& \cite{WI09a} \\ \nopagebreak
&1725-1872&21-159&300&MAMI& \cite{ST15} \\ \nopagebreak
&1736-2131&18-139&121&ELSA& \cite{DI15} \\ \nopagebreak
&1756-2350&11-162&648&ELSA& \cite{WI15} \\ \nopagebreak
\midrule
\multirow{4}{*}{$\Sigma$}  
&1720-2017&19-151&31&ELSA& \cite{KL08} \\ \nopagebreak
&1743-2174&15-145&81&CEBAF& \cite{RO18} \\ \nopagebreak
&1744-2098&32-148&492&CEBAF& \cite{CO17b} \\ \nopagebreak
&1750-1903&13-167&28&GRAAL& \cite{AJ06} \\ \nopagebreak
\midrule
{P}
&1150-2050&53-180&50&CEBAF& \cite{RO19} \\ \nopagebreak
\midrule
{T}
&1796-2458&37-180&143&CEBAF& \cite{RO18} \\ \nopagebreak
\midrule
{G}
&1778-1778&37-141&5&ELSA& \cite{EB15} \\ \nopagebreak
\midrule
\multirow{2}{*}{E} 
&1743-2300&29-151&104&CEBAF& \cite{AK17} \\ \nopagebreak
&1749-2256&28-151&95&ELSA& \cite{EB15} \\ \nopagebreak
\midrule
{F}
&1250-2750&37-180&160&CEBAF& \cite{RO19} \\ \nopagebreak
\midrule
{H}
&1150-2050&53-180&50&CEBAF& \cite{RO19} \\ \nopagebreak
\midrule
\multirow{2}{*}{SDME}
&1725-2545&23-147&4592&CEBAF& \cite{WI09} \\ \nopagebreak
&1756-2350&18-151&891&ELSA& \cite{WI15} \\ \nopagebreak
\midrule
\end{longtable}}\label{tab:pomega}
{%
\setlength\extrarowheight{-2pt}
\begin{longtable}{C{3cm} L{2.5cm} C{2cm} r L{2.5cm} C{1cm}}
\multicolumn{6}{r}{\emph{Continued on next page}}
\endfoot
\bottomrule
\endlastfoot
\captionsetup{width=.75\textwidth}
\caption{\label{tab:p:omega}Data for $\gamma n\to\omega n$ below 
    W = 2.55~GeV (E$_\gamma$ = 3~GeV). Experimental data are from the SAID 
    database~\protect\cite{SAID} selected for 1996 through 2018. 
    There are no polarized measurements.}\\
\toprule
Observable & W (MeV)& $\theta$ (deg) & Data & Lab & Ref \\
\midrule
\endhead
{$d\sigma/d\Omega$} 
&1762-2136&18-139&91&ELSA& \cite{DI15} \\ \nopagebreak
\end{longtable}}\label{tab:nomega}

\subsection{Photoproduction of two pseudoscalar mesons}

As photon energy increases all the single meson production cross sections decline, but the two pion cross section increases followed by $\eta\pi$ etc. Once we get above 1.6~GeV two pion production becomes dominant. The two meson final state provides a link to the final states $N\rho$, $N\sigma$, and more complex states such as $N^\ast\pi$, $\Delta\pi$, $\Delta\eta$ etc. The latter final states may result from the excitation of a higher mass resonance, with a sequential decay chain to an intermediate lighter resonance and one meson, followed by the decay to the ground state nucleon and a second meson. 

The first total cross section  measurements of $\pi^+\pi^-$ photoproduction were carried out in the late 1960s with untagged photon beams of energies up to 1~GeV incident on bubble chambers~\cite{ABBHHM69,Gialanella69}. The critical requirement for double meson production experiments is large solid angle coverage, the capability of detecting multiparticle events and high energy beams of tagged photons. This only became available in mid 90s. The first "new era" electronic experiment measuring two pion photoproduction was performed with DAPHNE at MAMI~\cite{Braghieria}. This experiment extracted total cross sections for three double pion channels: $\sigma_{tot}(p\pi^+\pi^-)$, $\sigma_{tot}(n\pi^+\pi^0)$, and $\sigma_{tot}(p\pi^0\pi^0)$. The measurements were done for photon energies from 400 to 800~MeV. SAPHIR extended the photon energy range for $\pi^+\pi^-$ up to 2.6~GeV. In this experiment they were able to extract differential cross sections and use Dalitz-plot analysis to isolate different contributions~\cite{SAPHIR_Wu05}. The first polarization measurements for this reaction were done by CLAS~\cite{Strauch05}. That experiment used circularly polarized photon beam and extracted the helicity asymmetry $I^c$ for photon energies from 1.35 to 2.30~GeV. The latest measurements of this channel were done by CLAS~\cite{Golovach19}. This experiment covered the range of the center of mass energies from 1.6 to 2.0~GeV. High statistics allowed for the first time the extraction of nine 1-fold differential cross section and the determination of photocouplings of some known resonances.

For the $\pi^0\pi^0$ channel a series of  experiments were performed at MAMI-B with TAPS on a proton target~\cite{Harter97,Wolf2000} and a deuteron target~\cite{Kleber2000} from threshold to 820~MeV photon energies. Then measurements were continued with the  Crystal Ball/TAPS~\cite{Zehr12} combination. The addition of the Crystal Ball allowed the access of $\pi^0\pi^+$ channel as well. With an extended energy reach of MAMI-C, the measurement with Crystal Ball/TAPS was performed up to 1.4~GeV~\cite{Kashevarov12,Dieterle15}. GRAAL extended measurements up to 1.5~GeV photon energies and in addition to the cross section they also took advantage of the linearly polarized photon beam and extracted $\Sigma$ beam asymmetry for this reaction~\cite{Assafei03}. Meanwhile the CBELSA collaboration did not stand aside and joined the effort~\cite{Thoma08,Sarantsev08,Thiel2015,Sokhoyan15a}, further extending the energy reach up to 2.5~GeV. They also contributed polarization measurements of $I^s$ and $I^c$~\cite{Sokhoyan15a}.

The natural next step after $\pi^0\pi^0$ from the experimental point of view was to study $\pi^0\eta$, which has similar topology. The first measurement of this reaction channel was reported by GRAAL. As usual for the GRAAL photon energy range up to 1.5~GeV, they presented total and differential cross section together with beam asymmetry  $\Sigma$~\cite{Ajaka08}. This was followed up by CBELSA in a series of measurements covering photon energies up to 2.5~GeV~\cite{Horn08a,Horn08b,Gutz10,Gutz14}. This experiment produced total and differential cross sections together with polarization observables $\Sigma$, $I^s$ and $I^c$. Crystal Ball/TAPS at MAMI-C measured total and differential cross sections~\cite{Kashevarov09}, which was followed by beam-target polarization measurements~\cite{Annand15}.

\section{\label{sec:conclusion} What Have We Learned from these Data so far}

\begin{figure}[h]
\begin{center}
\includegraphics[width=0.8\columnwidth]	
{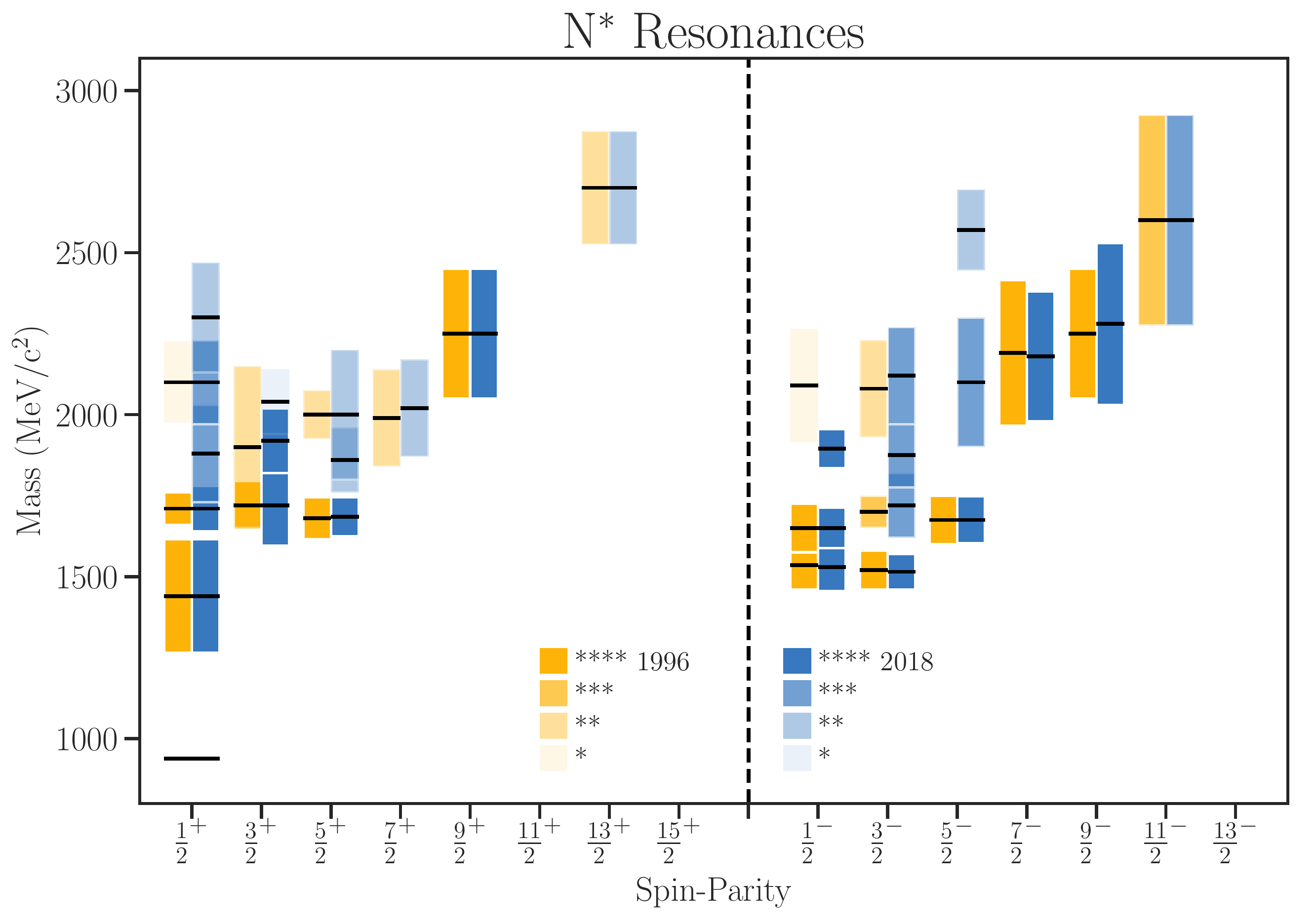}\protect\caption{\label{fig:nstar-spectrum} 
	Comparison of the $N^\ast$ spectrum from the PDG of 1996 with 2018 editions.}
\end{center}
\end{figure}

\begin{figure}
\begin{center}
\includegraphics[width=0.8\columnwidth]{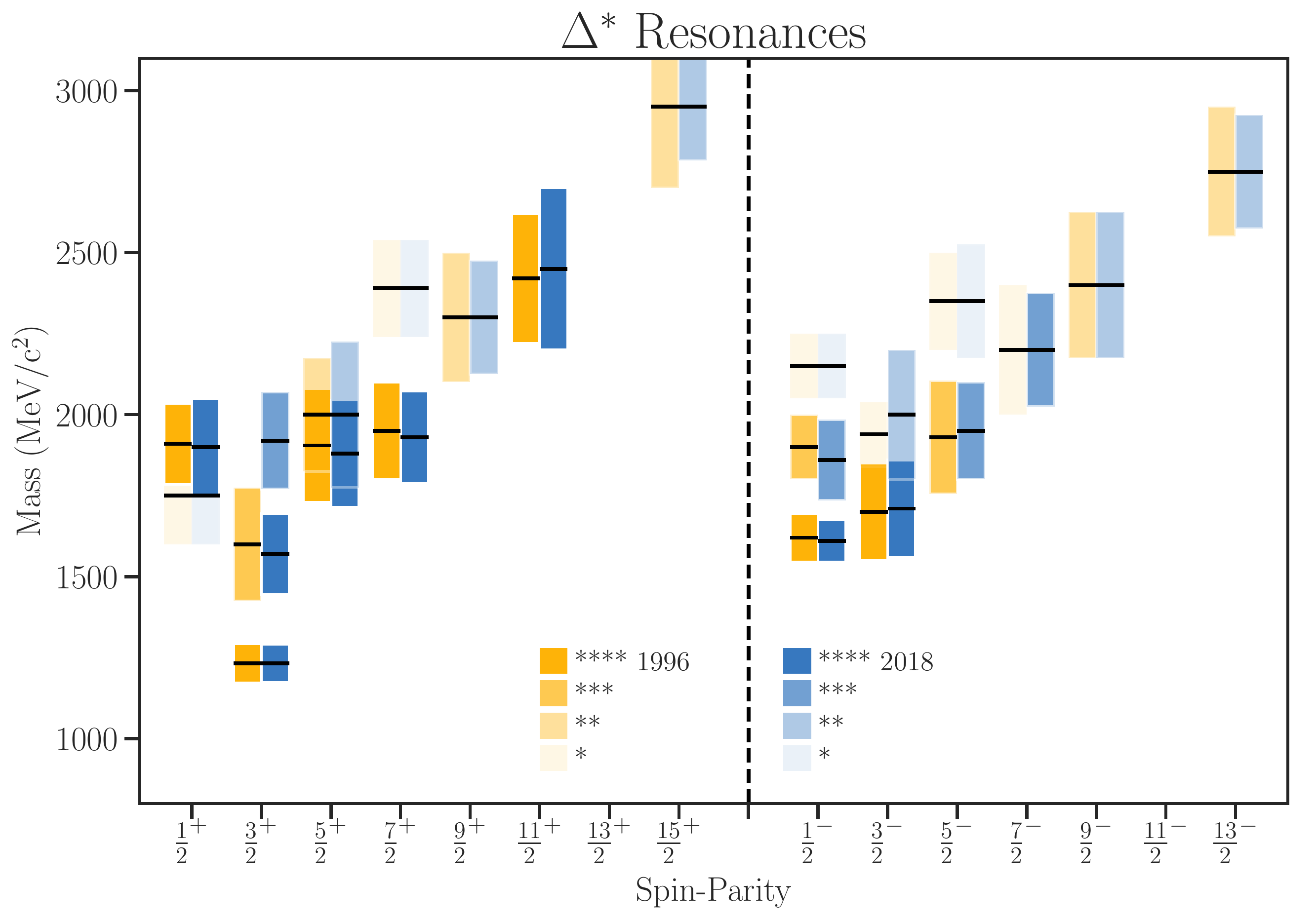}\protect\caption{\label{fig:delta-spectrum} 
	 Comparison of the $\Delta^\ast$ spectrum from the PDG of 1996 
	 with 2018 editions.}
\end{center}
\end{figure}

In the previous sections of this review, we have presented all the experimental photoproduction data obtained in the last two decades. We conclude by summarizing how this plethora of data has expanded our knowledge of nucleon excited states. Tables~\ref{tab:nstar} and \ref{tab:delta} compare the non-strange baryon summary tables from the PDG for the 1996~\cite{PDG_1996} and 2018~\cite{PDG_2018} editions. 
Figures~\ref{fig:nstar-spectrum} and \ref{fig:delta-spectrum} complement the tables by showing the spectra of states graphically, where masses and widths are represented by solid lines and boxes, respectively, and the star rating is represented by the shading.  The first thing one notices while looking at these tables and figures is that \emph{none} of the listed states has been left untouched, with one exception alone: the nucleon ground state. The tables show only the "star status" of the resonances. Quite often the knowledge of the resonance parameters improves while "star status" remains unchanged. The latest edition of PDG lists nine new states. Three states which have not received confirmation have been removed. The most of the changes are in $N^\ast$ table, and not so much in the table of $\Delta^\ast$'s. Most new information on nucleon resonances over the last two decades has come from photoproduction experiments,  while in the past it was mostly from $\pi N$ scattering. 

Nature gives us an additional powerful tool: an isospin filter. Photoproduction of the final states with isospin $I = 0$ mesons ($\eta$, $\eta^\prime$, $\omega $), or $I = 0$ baryons, $\Lambda$'s, cannot be directly coupled to  $\Delta$'s. As can be seen from the Table~\ref{tab:nstar},  most of the changes come exactly for these final states. New columns for $N\omega$ and $N\eta^\prime$ have been added. Couplings to these states were not known previously. Double meson production established couplings of several resonances to the $\sigma N$ decay channel, which again was not known previously. Double meson production data also allowed the identification of sequential decays and established couplings of some of the higher mass $\Delta^\ast$-resonances to $\Delta\eta$, which were not known before. 

These advances did not occur easily. It took time and effort for the information in the newly accumulated data sets to be translated into new knowledge of the baryon spectrum. As we described earlier, the renaissance of photoproduction started around mid 1990's. The first major overhaul of the non-strange baryon table happened in 2012~\cite{PDG_2012}. This represented the point at which the amount of new data needed to make an impact reached a critical mass. One remarkable example is new evidence for the $\Delta(2200){7/2}^-$. This was a poorly known ``1-star" state with only visible couplings to $N\pi$. New high accuracy polarization data from pion photoproduction were then added to the database. A coupled channel analysis  revealed this resonance coupling to many channels: $\pi^+n$, $\pi^0p$, $K\Sigma$, $\pi^0\pi^0p$, $\pi^0\eta p$~\cite{Anisovich2017}. In the latest edition of PDG, its status was upgraded to ``3-star". This example also demonstrates the strength of of the coupled channel approach to the data.

To conclude, it would no be exaggeration to say that non-strange baryon spectroscopy is quite healthy today. Several ``missing" resonances have been found. New photoproduction data keep coming and there are no signs of a decline any time soon. 


{
\setlength{\tabcolsep}{5pt}
\renewcommand*{\arraystretch}{0.8}
\setlength\extrarowheight{5pt}
\centering\small
\begin{longtable}{l l| l| l| l l l l l l l l l l }
\FourStar  & \multicolumn{13}{l}{Existence is certain}\\
\ThreeStar & \multicolumn{13}{l}{Existence is very likely}\\
\TwoStar   & \multicolumn{13}{l}{Evidence of existence is fair}\\
\OneStar   & \multicolumn{13}{l}{Evidence of existence is poor}\\
\midrule
\multicolumn{13}{r}{\emph{Continued on next page}}
\endfoot
\FourStar  & \multicolumn{13}{l}{Existence is certain}\\
\ThreeStar & \multicolumn{13}{l}{Existence is very likely}\\
\TwoStar   & \multicolumn{13}{l}{Evidence of existence is fair}\\
\OneStar   & \multicolumn{13}{l}{Evidence of existence is poor}\\
\bottomrule
\endlastfoot
\caption{Comparison of $N^\ast$  summary tables from PDG for the years 1996 and 2018. `` --- '' means the cell is not
    present for that year.}\\
\hline
\hline
\multirow{2}{*}{Particle} & \multirow{2}{*}{$J^P$ }& \multirow{2}{*}{Year}&Overall&\multicolumn{9}{c}{Status as seen in}\\ 
& &  & status & $N\gamma$ & $N\pi$ & $\Delta\pi$ & $N\sigma$ & $N\eta$ & $\Lambda K$ & $\Sigma K$ & $N\rho$ & $N\omega$ & $N\eta^\prime$ \\
\midrule
\endhead \label{tab:nstar}
\multirow{2}{*}{$N$} & \multirow{2}{*}{$1/2^+$} & 1996 & \FourStar  & &&&&&&&&&\\
                                                     & & 2018 & \FourStar & &&&&&&&&&\\
\midrule
\multirow{2}{*}{$N(1440)$} & \multirow{2}{*}{$1/2^+$}
          & 1996 & \FourStar  & \ThreeStar  & \ThreeStar  & \ThreeStar &     ---          & \OneStar   &                &               & \OneStar    &    ---       &   ---    \\
\nopagebreak       & & 2018 & \FourStar  & \FourStar    & \FourStar    & \FourStar   &                  &                 &                 &               &                  &               &                 \\  
\midrule                                                                                
\multirow{2}{*}{$N(1520)$} & \multirow{2}{*}{$3/2^-$} 
           & 1996 & \FourStar  & \FourStar   & \FourStar   & \FourStar   &    ---           & \OneStar   &                &             & \FourStar  &    ---        &   ---             \\
\nopagebreak        & & 2018 & \FourStar  & \FourStar   & \FourStar   & \FourStar   & \TwoStar    & \FourStar   &                &             &                 &                &                    \\                                                                        
\midrule                                                                                
\multirow{2}{*}{$N(1535)$} & \multirow{2}{*}{$1/2^-$} 
            & 1996 & \FourStar  & \ThreeStar & \FourStar     & \OneStar    &    ---           & \FourStar  &                &             & \TwoStar    &    ---        &   ---             \\
\nopagebreak         & & 2018 & \FourStar  & \FourStar   & \FourStar    & \ThreeStar  & \OneStar     & \FourStar  &                &             &                 &                &                   \\                                                                        
\midrule
\multirow{2}{*}{$N(1650)$} & \multirow{2}{*}{$1/2^-$} 
             & 1996 & \FourStar  & \ThreeStar & \FourStar   & \ThreeStar &    ---           & \OneStar  & \ThreeStar & \TwoStar  & \TwoStar    &    ---        &   ---        \\
\nopagebreak          & & 2018 & \FourStar  & \FourStar   & \FourStar    & \ThreeStar  & \OneStar  & \FourStar &  \OneStar  &             &                 &                &                \\                                                                        
\midrule
\multirow{2}{*}{$N(1675)$} & \multirow{2}{*}{$5/2^-$} 
             & 1996 & \FourStar  & \ThreeStar & \FourStar    & \FourStar &    ---         & \OneStar  & \OneStar  &                   & \OneStar    &    ---        &   ---        \\
\nopagebreak          & & 2018 & \FourStar  & \FourStar   & \FourStar    & \FourStar& \ThreeStar & \FourStar &  \OneStar & \OneStar  &                 &                &                \\                                                                        
\midrule
\multirow{2}{*}{$N(1680)$} & \multirow{2}{*}{$5/2^+$} 
             & 1996 & \FourStar  & \FourStar   & \FourStar    & \FourStar &    ---         &                &                 &                   & \FourStar    &    ---        &   ---        \\
\nopagebreak          & & 2018 & \FourStar  & \FourStar   & \FourStar    & \FourStar& \ThreeStar & \FourStar &  \OneStar & \OneStar  &                 &                &                \\                                                                        
\midrule
\multirow{2}{*}{$N(1700)$} & \multirow{2}{*}{$3/2^-$} 
             & 1996 & \ThreeStar  & \TwoStar   & \ThreeStar & \TwoStar &    ---         & \OneStar & \TwoStar   & \OneStar    & \OneStar    &    ---        &   ---        \\
\nopagebreak          & & 2018 & \ThreeStar  & \TwoStar & \ThreeStar   & \ThreeStar& \OneStar & \OneStar &                &                    & \OneStar    &                &                \\                                                                        
\midrule
\multirow{2}{*}{$N(1710)$} & \multirow{2}{*}{$1/2^+$} 
             & 1996 & \ThreeStar & \ThreeStar & \ThreeStar & \TwoStar &    ---         & \TwoStar  & \TwoStar   & \OneStar    & \OneStar    &    ---        &   ---        \\
\nopagebreak          & & 2018 & \FourStar   & \FourStar   & \FourStar   & \OneStar&                & \ThreeStar & \TwoStar & \OneStar    & \OneStar    & \OneStar &                \\                                                                        
\midrule
\multirow{2}{*}{$N(1720)$} & \multirow{2}{*}{$3/2^+$} 
             & 1996 & \FourStar  & \TwoStar    & \FourStar  & \OneStar &    ---         & \OneStar  & \TwoStar   & \OneStar    & \TwoStar    &    ---        &   ---        \\
\nopagebreak          & & 2018 & \FourStar  & \FourStar   & \FourStar   & \ThreeStar& \OneStar & \OneStar & \FourStar & \OneStar    & \OneStar    & \OneStar &                \\                                                                        
\midrule
\multirow{2}{*}{$N(1860)$} & \multirow{2}{*}{$5/2^+$} 
             & 1996 & ---             & ---               & ---               & ---            &    ---         & ---           & ---              & ---             & ---             &    ---        &   ---        \\
\nopagebreak          & & 2018 & \TwoStar  & \OneStar    & \TwoStar   &                & \OneStar & \OneStar &                   &                 &                 &               &                \\                                                                        
\midrule
\multirow{2}{*}{$N(1875)$} & \multirow{2}{*}{$3/2^-$} 
             & 1996 & ---             & ---               & ---               & ---            &    ---         & ---           & ---              & ---             & ---             &    ---        &   ---        \\
\nopagebreak          & & 2018 & \ThreeStar  & \TwoStar    & \TwoStar  & \OneStar  & \TwoStar & \OneStar & \OneStar   & \OneStar  & \OneStar  & \OneStar &                \\                                                                        
\midrule
\multirow{2}{*}{$N(1880)$} & \multirow{2}{*}{$1/2^+$} 
             & 1996 & ---             & ---               & ---               & ---            &    ---         & ---           & ---              & ---             & ---             &    ---        &   ---        \\
\nopagebreak          & & 2018 & \ThreeStar  & \TwoStar    & \OneStar  & \TwoStar  & \TwoStar & \OneStar & \TwoStar   & \TwoStar  &                & \TwoStar &                \\                                                                        
\midrule
\multirow{2}{*}{$N(1895)$} & \multirow{2}{*}{$1/2^-$} 
             & 1996 & ---             & ---               & ---               & ---            &    ---         & ---           & ---              & ---             & ---             &    ---        &   ---        \\
\nopagebreak          & & 2018 & \FourStar  & \FourStar    & \OneStar  & \OneStar   & \OneStar & \FourStar & \TwoStar   & \TwoStar  & \OneStar  & \TwoStar & \FourStar \\                                                                        
\midrule
\multirow{2}{*}{$N(1900)$} & \multirow{2}{*}{$3/2^+$} 
             & 1996 & \TwoStar  &                   & \TwoStar     &               &    ---         &                &                 &                & \TwoStar    &    ---        &   ---        \\
\nopagebreak          & & 2018 & \FourStar  & \FourStar   & \TwoStar     & \TwoStar & \OneStar & \OneStar & \TwoStar   & \TwoStar &                  & \OneStar & \TwoStar \\                                                                        
\midrule
\multirow{2}{*}{$N(1900)$} & \multirow{2}{*}{$7/2^+$} 
             & 1996 & \TwoStar  & \OneStar     & \TwoStar     &               &    ---         & \OneStar & \OneStar   & \OneStar  &                  &    ---        &   ---        \\
\nopagebreak          & & 2018 & \TwoStar  & \TwoStar   & \TwoStar     &                 &                & \OneStar & \OneStar   & \OneStar &                  &               &                \\                                                                        
\midrule
\multirow{2}{*}{$N(2000)$} & \multirow{2}{*}{$5/2^+$} 
             & 1996 & \TwoStar  &                   & \TwoStar     & \OneStar &    ---         & \OneStar & \OneStar   & \OneStar  & \TwoStar   &    ---        &   ---        \\
\nopagebreak          & & 2018 & \TwoStar  & \TwoStar   & \OneStar     & \TwoStar  & \OneStar  & \OneStar &                &                  &                  & \OneStar &                \\                                                                        
\midrule
\multirow{2}{*}{$N(2040)$} & \multirow{2}{*}{$3/2^+$} 
             & 1996 & ---             & ---               & ---               & ---            &    ---         & ---           & ---              & ---             & ---             &    ---        &   ---        \\
\nopagebreak          & & 2018 & \OneStar  & \OneStar   &                     &                &                &                &                &                  &                  &               &                \\                                                                        
\midrule
\multirow{2}{*}{$N(2060)$} & \multirow{2}{*}{$5/2^+$} 
             & 1996 & ---             & ---               & ---               & ---            &    ---         & ---           & ---              & ---             & ---             &    ---        &   ---        \\
\nopagebreak          & & 2018 & \ThreeStar & \ThreeStar & \TwoStar   & \OneStar & \OneStar  & \OneStar   & \OneStar  &\OneStar  & \OneStar & \OneStar &                \\                                                                        
\midrule
\multirow{2}{*}{$N(2080)$} & \multirow{2}{*}{$3/2^-$} 
             & 1996 & \TwoStar  & \OneStar     & \TwoStar     &              &    ---         & \OneStar & \OneStar   &                  &                 &    ---        &   ---        \\
\nopagebreak           && 2018 & ---             & ---               & ---               & ---            &    ---         & ---           & ---              & ---             & ---             &    ---        &   ---        \\
\midrule
\multirow{2}{*}{$N(2090)$} & \multirow{2}{*}{$1/2^-$} 
             & 1996 & \OneStar  &                   & \OneStar     &              &    ---         &                  &                  &                  &                 &    ---        &   ---        \\
\nopagebreak           && 2018 & ---             & ---               & ---               & ---            &    ---         & ---           & ---              & ---             & ---             &    ---        &   ---        \\
\midrule
\multirow{2}{*}{$N(2100)$} & \multirow{2}{*}{$1/2^+$} 
             & 1996 & \TwoStar  &                   & \TwoStar     & \OneStar &    ---         & \OneStar & \OneStar   & \OneStar  & \TwoStar   &    ---        &   ---        \\
\nopagebreak          & & 2018 & \TwoStar  & \TwoStar   & \OneStar     & \TwoStar  & \OneStar  & \OneStar &                &                  &                  & \OneStar &                \\                                                                        
\midrule
\multirow{2}{*}{$N(2120)$} & \multirow{2}{*}{$3/2^-$} 
             & 1996 & ---             & ---               & ---               & ---            &    ---         & ---           & ---              & ---             & ---             &    ---        &   ---        \\
          & & 2018 & \ThreeStar & \ThreeStar & \TwoStar   & \TwoStar & \TwoStar  &               & \TwoStar  &\OneStar  &                   & \OneStar & \OneStar \\                                                                        
\midrule
\nopagebreak \multirow{2}{*}{$N(2190)$} & \multirow{2}{*}{$7/2^-$} 
             & 1996 & \FourStar  &  \OneStar  & \FourStar     &                &    ---         & \OneStar & \OneStar   & \OneStar  & \TwoStar   &    ---        &   ---        \\
\nopagebreak 
          & & 2018 & \FourStar  & \FourStar   & \FourStar     & \FourStar& \TwoStar  & \OneStar & \TwoStar & \OneStar  & \OneStar    & \OneStar &                \\                                                                        
\midrule
\nopagebreak 
\multirow{2}{*}{$N(2200)$} & \multirow{2}{*}{$5/2^-$} 
             & 1996 & \TwoStar  &          & \TwoStar     &              &    ---         & \OneStar &  \OneStar&                  &                 &    ---        &   ---        \\
\nopagebreak 
           && 2018 & ---        & ---      & ---               & ---            &    ---         & ---           & ---              & ---             & ---             &    ---        &   ---        \\
\midrule
\multirow{2}{*}{$N(2220)$} & \multirow{2}{*}{$9/2^+$} 
             & 1996 & \FourStar  &                   & \FourStar     &                &    ---         & \OneStar &                &                  &                 &    ---        &   ---        \\
\nopagebreak          & & 2018 & \FourStar  & \TwoStar    & \FourStar     &                &                & \OneStar & \OneStar & \OneStar  &                   &               &                \\                                                                        
\midrule
\multirow{2}{*}{$N(2250)$} & \multirow{2}{*}{$9/2^-$} 
             & 1996 & \FourStar  &                   & \FourStar     &                &    ---         & \OneStar &                &                  &                 &    ---        &   ---        \\
\nopagebreak          & & 2018 & \FourStar  & \TwoStar    & \FourStar     &                &                & \OneStar & \OneStar & \OneStar  &                   &               &                \\                                                                        
\midrule
\multirow{2}{*}{$N(2300)$} & \multirow{2}{*}{$1/2^+$} 
             & 1996 & ---             & ---               & ---               & ---            &    ---         & ---           & ---              & ---             & ---             &    ---        &   ---        \\
\nopagebreak          & & 2018 & \TwoStar    &     &    \TwoStar               &                &                &               &                   &                 &                   &             &              \\                                                                        
\midrule
\multirow{2}{*}{$N(2570)$} & \multirow{2}{*}{$5/2^-$} 
             & 1996 & ---             & ---               & ---               & ---            &    ---         & ---           & ---              & ---             & ---             &    ---        &   ---        \\
\nopagebreak          & & 2018 & \TwoStar    &    &   \TwoStar                 &                &                &               &                   &                 &                   &             &              \\                                                                        
\midrule
\multirow{2}{*}{$N(2600)$} & \multirow{2}{*}{$11/2^-$} 
             & 1996 & \ThreeStar  &                   & \ThreeStar     &                &    ---         &          &                &                  &                 &    ---        &   ---        \\
\nopagebreak           && 2018 & \ThreeStar  &                   & \ThreeStar     &                &    ---         &          &                &                  &                 &    ---        &   ---        \\
\midrule
\multirow{2}{*}{$N(2700)$} & \multirow{2}{*}{$13/2^+$} 
             & 1996 & \TwoStar  &                   & \TwoStar     &                &    ---         &          &                &                  &                 &    ---        &   ---        \\
\nopagebreak           && 2018 & \TwoStar  &                   & \TwoStar     &                &    ---         &          &                &                  &                 &    ---        &   ---        \\
\midrule
\end{longtable}}

{
\setlength{\tabcolsep}{5pt}
\renewcommand*{\arraystretch}{0.8}
\setlength\extrarowheight{5pt}
\begin{longtable}{l l| l| l| l l l l l l l l l l }
\centering\small
\FourStar   & \multicolumn{10}{l}{Existence is certain}\\
\ThreeStar  & \multicolumn{10}{l}{Existence is very likely}\\
\TwoStar    & \multicolumn{10}{l}{Evidence of existence is fair}\\
\OneStar    & \multicolumn{10}{l}{Evidence of existence is poor}\\
\midrule
\multicolumn{10}{r}{\emph{Continued on next page}}
\endfoot
\FourStar   & \multicolumn{10}{l}{Existence is certain}\\
\ThreeStar  & \multicolumn{10}{l}{Existence is very likely}\\
\TwoStar    & \multicolumn{10}{l}{Evidence of existence is fair}\\
\OneStar    & \multicolumn{10}{l}{Evidence of existence is poor}\\
\bottomrule
\endlastfoot

\caption{Comparison of $\Delta^\ast$ summary tables from PDG for the years 1996 and 2018. `` --- '' means the cell is not present for that year.}\\
\hline
\hline
\multirow{2}{*}{Particle} & \multirow{2}{*}{$J^P$ }& \multirow{2}{*}{Year}&Overall&\multicolumn{6}{c}{Status as seen in}\\ 
 & &  & status & $N\gamma$ & $N\pi$ & $\Delta\pi$ & $\Sigma K$ &  $N\rho$ & $\Delta\eta$ \\
\midrule
\endhead \label{tab:delta}
\multirow{2}{*}{$\Delta(1232)$} & \multirow{2}{*}{$3/2^+$}
          & 1996 & \FourStar  & \FourStar  & \FourStar  &                &                 &                 &      ---            \\ \nopagebreak
       & & 2018 & \FourStar  & \FourStar    & \FourStar  &                &                  &                 &                   \\  \nopagebreak
\midrule
\multirow{2}{*}{$\Delta(1600)$} & \multirow{2}{*}{$3/2^+$}
          & 1996 & \ThreeStar  & \TwoStar  & \ThreeStar  & \ThreeStar &                & \OneStar &    ---              \\ \nopagebreak
       & & 2018 & \FourStar  & \FourStar    & \ThreeStar  & \FourStar  &                  &                 &                   \\ \nopagebreak  
\midrule
\multirow{2}{*}{$\Delta(1620)$} & \multirow{2}{*}{$1/2^-$}
          & 1996 & \FourStar  & \ThreeStar    & \FourStar     & \FourStar &                 & \FourStar &    ---          \\ \nopagebreak
       & & 2018 & \FourStar  & \FourStar    & \FourStar  & \FourStar  &                  &                 &                   \\ \nopagebreak 
\midrule
\multirow{2}{*}{$\Delta(1700)$} & \multirow{2}{*}{$3/2^-$}
          & 1996 & \FourStar  & \ThreeStar    & \FourStar   & \ThreeStar & \OneStar & \TwoStar   &    ---              \\ \nopagebreak
       & & 2018 & \FourStar  & \FourStar    & \FourStar  & \FourStar  & \OneStar      &  \OneStar &                   \\  \nopagebreak
\midrule
\multirow{2}{*}{$\Delta(1750)$} & \multirow{2}{*}{$1/2^+$}
          & 1996 & \OneStar  &                   & \OneStar   &                    &               &                  &    ---              \\ \nopagebreak
       & & 2018 & \OneStar    & \OneStar    & \OneStar   &                    & \OneStar &                  &                   \\ \nopagebreak 
\midrule
\multirow{2}{*}{$\Delta(1900)$} & \multirow{2}{*}{$1/2^-$}
          & 1996 & \ThreeStar  & \OneStar   & \ThreeStar  & \OneStar    & \OneStar & \TwoStar  &    ---              \\ \nopagebreak
       & & 2018 & \ThreeStar  & \ThreeStar& \ThreeStar   & \OneStar   & \TwoStar &  \OneStar &                   \\  \nopagebreak
\midrule
\multirow{2}{*}{$\Delta(1905)$} & \multirow{2}{*}{$5/2^+$}
          & 1996 & \FourStar  & \ThreeStar   & \FourStar   & \TwoStar    & \OneStar & \TwoStar  &    ---              \\ \nopagebreak
       & & 2018 & \FourStar  & \FourStar     & \FourStar   & \TwoStar   & \OneStar &  \OneStar & \TwoStar        \\ \nopagebreak  
\midrule
\multirow{2}{*}{$\Delta(1910)$} & \multirow{2}{*}{$1/2^+$}
          & 1996 & \FourStar  & \OneStar     & \FourStar   & \OneStar    & \OneStar & \OneStar  &    ---              \\ \nopagebreak
       & & 2018 & \FourStar  & \ThreeStar  & \FourStar   & \TwoStar    & \TwoStar &                & \OneStar        \\ \nopagebreak 
\midrule
\multirow{2}{*}{$\Delta(1920)$} & \multirow{2}{*}{$3/2^+$}
          & 1996 & \ThreeStar  & \OneStar     & \ThreeStar & \TwoStar    & \OneStar &               &    ---              \\ \nopagebreak
       & & 2018 & \ThreeStar  & \ThreeStar  & \ThreeStar & \ThreeStar& \TwoStar &                & \TwoStar        \\ \nopagebreak 
\midrule
\multirow{2}{*}{$\Delta(1930)$} & \multirow{2}{*}{$5/2^-$}
          & 1996 & \ThreeStar  & \TwoStar   & \ThreeStar &                   & \OneStar &               &    ---              \\ \nopagebreak
       & & 2018 & \ThreeStar  & \ThreeStar & \ThreeStar & \OneStar    & \OneStar &                &                    \\ \nopagebreak
\midrule
\multirow{2}{*}{$\Delta(1940)$} & \multirow{2}{*}{$3/2^-$}
          & 1996 & \OneStar  &                 & \OneStar     &                   &                &               &    ---              \\ \nopagebreak
       & & 2018 & \TwoStar     & \OneStar   & \TwoStar    & \OneStar    &                &              &  \OneStar     \\ \nopagebreak
\midrule
\multirow{2}{*}{$\Delta(1950)$} & \multirow{2}{*}{$7/2^+$}
          & 1996 & \FourStar    & \FourStar   & \FourStar    & \FourStar  &  \OneStar  & \OneStar   &    ---              \\ \nopagebreak
       & & 2018 & \TwoStar     & \FourStar   & \FourStar    & \TwoStar  & \ThreeStar &                  &           \\ \nopagebreak
\midrule
\multirow{2}{*}{$\Delta(2000)$} & \multirow{2}{*}{$5/2^+$}
          & 1996 & \TwoStar    &                  &                    &                   &              & \TwoStar   &    ---              \\ \nopagebreak
       & & 2018 & \TwoStar     & \OneStar   & \TwoStar    & \OneStar    &              &  \OneStar   &                     \\ \nopagebreak
\midrule
\multirow{2}{*}{$\Delta(2150)$} & \multirow{2}{*}{$1/2^-$}
          & 1996 & \OneStar    &                  &   \OneStar  &                   &              &                    &    ---              \\ \nopagebreak
       & & 2018 & \OneStar     &                 & \OneStar    &                   &              &                    &           \\ \nopagebreak
\midrule
\multirow{2}{*}{$\Delta(2200)$} & \multirow{2}{*}{$7/2^-$}
          & 1996 & \OneStar    &                  &   \OneStar  &                   &              &                    &    ---              \\ \nopagebreak
       & & 2018 & \ThreeStar &  \ThreeStar& \TwoStar    & \ThreeStar & \TwoStar&                    &           \\ \nopagebreak
\midrule
\multirow{2}{*}{$\Delta(2300)$} & \multirow{2}{*}{$9/2^+$}
          & 1996 & \TwoStar    &                  &   \TwoStar  &                   &              &                    &    ---              \\ \nopagebreak
       & & 2018 & \TwoStar     &                 & \TwoStar    &                   &              &                    &           \\ \nopagebreak
\midrule
\multirow{2}{*}{$\Delta(2350)$} & \multirow{2}{*}{$5/2^-$}
          & 1996 & \OneStar    &                  &   \OneStar  &                   &              &                    &    ---              \\ \nopagebreak
       & & 2018 & \OneStar     &                 & \OneStar    &                   &              &                    &           \\ \nopagebreak
\midrule
\multirow{2}{*}{$\Delta(2390)$} & \multirow{2}{*}{$7/2^+$}
          & 1996 & \OneStar    &                  &   \OneStar  &                   &              &                    &    ---              \\ \nopagebreak
       & & 2018 & \OneStar     &                 & \OneStar    &                   &              &                    &           \\ \nopagebreak
\midrule
\multirow{2}{*}{$\Delta(2400)$} & \multirow{2}{*}{$9/2^-$}
          & 1996 & \TwoStar    &                  &   \TwoStar  &                   &              &                    &    ---              \\ \nopagebreak
       & & 2018 & \TwoStar     & \TwoStar & \TwoStar    &                   &              &                    &           \\ \nopagebreak
\midrule
\multirow{2}{*}{$\Delta(2420)$} & \multirow{2}{*}{$11/2^+$}
          & 1996 & \FourStar    &                  &   \FourStar  &                   &              &                    &    ---              \\ \nopagebreak
       & & 2018 & \FourStar    & \OneStar   & \FourStar    &                   &              &                    &           \\ \nopagebreak
\midrule
\multirow{2}{*}{$\Delta(2750)$} & \multirow{2}{*}{$13/2^-$}
          & 1996 & \TwoStar    &                  &   \TwoStar  &                   &              &                    &    ---              \\ \nopagebreak
       & & 2018 & \TwoStar     &                 & \TwoStar    &                   &              &                    &           \\ \nopagebreak
\midrule
\multirow{2}{*}{$\Delta(2950)$} & \multirow{2}{*}{$15/2^+$}
          & 1996 & \TwoStar    &                  &   \TwoStar  &                   &              &                    &    ---              \\ \nopagebreak
       & & 2018 & \TwoStar     &                 & \TwoStar    &                   &              &                    &           \\ \nopagebreak
\midrule
\end{longtable}}

\pagebreak
\section*{Acknowledgements} 

The work of D.I. was supported by the United Kingdom's Science and Technology Facilities Council (STFC) from grant number ST/P004458/1.
The work of I.S. was supported in part by the US Department of Energy Grant DE--SC0016583. This material in part is based upon work supported by the U.S. Department of Energy, Office of Science, Office of Nuclear Physics under contract No. DE--AC05--06OR23177.
Notice: Authored by Jefferson Science Associates, LLC under U.S. DOE Contract No. DE--AC05--06OR23177. The U.S. Government retains a non-exclusive, paid-up, irrevocable, world-wide license to publish or reproduce this manuscript for U.S. Government purposes.


\section*{References}

\input refs.bbl
\end{document}

%% file: header.tex
\usepackage{amsmath}
\usepackage{amssymb}
\usepackage{array}
\usepackage{booktabs}
\usepackage{colortbl}
\usepackage{fancyvrb}
\usepackage[T1]{fontenc}
\usepackage{graphicx}
\usepackage[latin1]{inputenc}
\usepackage{longtable}
\usepackage{lscape}
\usepackage[fleqn]{mathtools}
\usepackage{lipsum}
\usepackage{multicol}
\usepackage{multirow}
\usepackage{pifont}
\usepackage{siunitx}
\usepackage{tabularx}
\usepackage{units}
\usepackage{caption}
\usepackage{float}
\usepackage{afterpage}

\newcommand{\be}{\begin{equation}}
\newcommand{\ee}{\end{equation}}
\newcommand{\bea}{\begin{eqnarray}}
\newcommand{\eea}{\end{eqnarray}}

\newcommand{\OneStar}{$*$}
\newcommand{\TwoStar}{$*$$*$}
\newcommand{\ThreeStar}{$*$$*$$*$}
\newcommand{\FourStar}{$*$$*$$*$$*$}
\newcommand{\obs}[1]{\textcolor{red}{\mathbf{#1}}}
\newcommand{\linpol}{P_L^{\gamma}}
\newcommand{\circpol}{P_{\odot}^{\gamma}}
\newcommand{\tarpol}{P^T}
\newcommand{\recpol}{P^R}

\newcommand{\lum}{{\cal L}}
\newcommand{\sett}{{\cal S}}
\newcommand{\eff}{\varepsilon}

\newcolumntype{R}{@{\extracolsep{1cm}}X@{\extracolsep{0pt}}}%
\newcolumntype{C}[1]{>{\centering\let\newline\\\arraybackslash\hspace{0pt}}m{#1}}
\newcolumntype{L}[1]{>{\let\newline\\\arraybackslash\hspace{0pt}}m{#1}}
\newcolumntype{Q}[1]{>{\raggedleft\arraybackslash}p{#1}} 
\newcounter{exno}
\renewcommand{\arraystretch}{1.5}  



%% file: ps-cross-section.tex
\begin{longtable}{L{5cm} C{1cm} p{9.5cm} >{(\refstepcounter{subexno}\theexno)} C{1cm}}\label{eq:xsec}

$\obs{d\sigma^{B,T,R}}(\vec{P^\gamma},\vec{P^T},\vec{P^R},\phi)$  
& $=$ 
&
$\frac{1}{2} \left\{ \obs{d\sigma_0} \left[ 1 - \linpol \tarpol_y \recpol_{y^\prime} \cos2(\alpha - \phi) \right] \right.$
& \label{eqn:cs1} \\ 
Single spin observables 
& &
$ \quad + \obs{\Sigma} \left [- \linpol \cos2(\alpha - \phi)  + \tarpol_y \recpol_{y^\prime}   \right]$ & \label{eqn:cs2}  \\
& &
$ \quad + \obs{T} \left[ \tarpol_y - \linpol \recpol_{y^\prime} \cos 2(\alpha - \phi) \right]$ & \label{eqn:cs3}  \\
& &
$ \quad + \obs{P} \left[ \recpol_{y^\prime} - \linpol \tarpol_y \cos 2(\alpha - \phi) \right]$  & \label{eqn:cs4} 
\\
Beam-Target observables
& &
$ \quad + \obs{E} \left[ -\circpol \tarpol_z + \linpol \tarpol_x \recpol_{y^\prime}\sin2(\alpha - \phi) \right]$ & \label{eqn:cs5} \\ 
& &
$ \quad + \obs{G} \left[ \linpol \tarpol_z \sin2(\alpha - \phi) + \circpol \tarpol_x \recpol_{y^\prime} \right]$ & \label{eqn:cs6} \\  
& & 
$ \quad + \obs{F} \left[ \circpol \tarpol_x + \linpol \tarpol_z \recpol_{y^\prime} \sin2(\alpha - \phi) \right]$ & \label{eqn:cs7} \\
& & 
$ \quad + \obs{H} \left[ \linpol \tarpol_x \sin2(\alpha - \phi) - \circpol \tarpol_x \recpol_{y^\prime} \right]$ & \label{eqn:cs8} 
\\
Beam-Recoil observables
& & 
$ \quad + \obs{C_{x^\prime}} \left[ \circpol \recpol_{x^\prime} - \linpol \tarpol_y \recpol_{z^\prime} \sin2(\alpha - \phi) \right]$ & \label{eqn:cs9} \\
& & 
$ \quad + \obs{C_{z^\prime}} \left[ \circpol \recpol_{z^\prime} - \linpol \tarpol_y \recpol_{x^\prime} \sin2(\alpha - \phi) \right]$ & \label{eqn:cs10} \\
& &
$ \quad + \obs{O_{x^\prime}} \left[ \linpol \recpol_{x^\prime} \sin2(\alpha - \phi) + \circpol \tarpol_y \recpol_{z^\prime} \right]$ & \label{eqn:cs11} \\
& & 
$ \quad + \obs{O_{z^\prime}} \left[ \linpol \recpol_{z^\prime} \sin2(\alpha - \phi) - \circpol \tarpol_y \recpol_{x^\prime} \right]$ & \label{eqn:cs12} 
\\
Target-Recoil observables
& &
$ \quad + \obs{L_{x^\prime}} \left[ \tarpol_z \recpol_{x^\prime} + \linpol \tarpol_x \recpol_{z^\prime}
\cos2(\alpha - \phi) \right]$ & \label{eqn:cs13} \\
& & 
$ \quad + \obs{L_{z^\prime}} \left[ \tarpol_z \recpol_{z^\prime} - \linpol \tarpol_x \recpol_{x^\prime}
\cos2(\alpha - \phi) \right]$ & \label{eqn:cs14} \\
& &
$ \quad + \obs{T_{x^\prime}} \left[ \tarpol_x \recpol_{x^\prime} + \linpol \tarpol_z \recpol_{z^\prime}
\cos2(\alpha - \phi) \right]$ & \label{eqn:cs15} \\
& &
$ \left. \quad + \obs{T_{z^\prime}} \left[ \tarpol_x \recpol_{z^\prime} - \linpol \tarpol_z \recpol_{x^\prime}
\cos2(\alpha - \phi) \right] \right\}$ & \label{eqn:cs16} 
\\
\end{longtable}

%% file: exp-config-table.tex


\begin{longtable}{C{2cm} C{2cm} C{1.5cm} p{9.5cm} >{(\refstepcounter{subexno}\theexno)} C{1cm}}
\captionsetup{width=.75\textwidth}
\caption{\label{tab:cross-sections} Expressions for cross sections for different experiments.}\\

\toprule
\multicolumn{3}{c}{\emph{Configuration}} &  
\hfil\multirow{2}[8]{*}{\emph{Cross section formula, $\sigma/\sigma_0$}}\hfill \\
\cmidrule(lr){1-3}
Beam & Target & Recoil &   \\ 

\endfirsthead
\midrule

Beam & Target & Recoil &  \hfil\emph{Cross section formula, $\sigma/\sigma_0$}\hfill \\
\midrule
\endhead

\multicolumn{5}{r}{\emph{Continued on next page}}
\endfoot
\endlastfoot

\midrule 

\multirow{6}*{Unpolarized}    
    & \multirow{2}*{Unpolarized}    
            & N      & 1 & \label{row1}\\  
\cmidrule(r){4-5}
    &       & Y      & $1 + \obs{P} \recpol_{y^\prime}$ & \label{row2} \\
\cmidrule[0.7\cmidrulewidth](lr){3-5}
    & \multirow{2}*{Longitudinal}
            & N     & 1 & \label{row3} \\ 
\cmidrule(r){4-5}
    &       & Y      & $
    					1 + \obs{P} \recpol_{y^\prime} + 
                        \left(
                        \obs{L_{x^\prime}} \recpol_{x^\prime} + \obs{L_{z^\prime}} \recpol_{z^\prime} 
                        \right) \tarpol_z
                       $
    					 & \label{row4}\\
\cmidrule[0.7\cmidrulewidth](lr){3-5}
    & \multirow{2}*{Transverse}
             & N      & $1 + \obs{T} \tarpol_T \sin (\beta -\phi) $ & \label{row5}\\ 
\cmidrule(r){4-5} \addlinespace[0.5em]
    &        & Y      &  
    					
    					 $\begin{aligned}
                         & 1 + \obs{P} \recpol_{y^\prime} 
                         + 
						 \left( 
                         \obs{\Sigma} \recpol_{y^\prime} + \obs{T} 
                         \right) \tarpol_T \sin (\beta -\phi) \\
                         & \enspace +
                         \left( 
                         \obs{T_{x^\prime}} \recpol_{x^\prime} + \obs{T_{z^\prime}} \recpol_{z^\prime} 
                         \right) \tarpol_T \cos (\beta - \phi)
                        \end{aligned}$
                        
                         & \label{row6} \\
\midrule 

\multirow{6}*[-4.0em]{Circular}    
    & \multirow{2}*{Unpolarized}    
            & N      & 1 & \label{row7} \\ 
\cmidrule(r){4-5}
    &       & Y      & $
                        1 + \obs{P} \recpol_{y^\prime} +
                        \left(
                        \obs{C_{x^\prime}} \recpol_{x^\prime} + \obs{C_{z^\prime}} \recpol_{z^\prime}
                        \right) \circpol 
                       $
                        & \label{row8} \\
\cmidrule[0.7\cmidrulewidth](lr){3-5}
    & \multirow{2}*{Longitudinal}
            & N      & $ 
                        1 - \obs{E} \circpol \tarpol_y 
                       $ & \label{row9} \\ 
\cmidrule(r){4-5} \addlinespace[0.5em]
    &       & Y      & $ 
    					\begin{aligned}
    					& 1 + \obs{P} \recpol_{y^\prime} + \left(
                        \obs{L_{x^\prime}} \recpol_{x^\prime} + \obs{L_{z^\prime}} \recpol_{z^\prime} 
                        \right) \tarpol_z \\
                        & \enspace + \left\{ 
                        \obs{C_{x^\prime}}  \recpol_{x^\prime} + \obs{C_{z^\prime}} \recpol_{z^\prime} 
                        - \left(
                        \obs{E} + \obs{H} \recpol_{y^\prime} 
                        \right) \tarpol_z 
                        \right\} \circpol
    					\end{aligned}
                       $ 
                        & \label{row10} \\ \addlinespace[0.5em]
\cmidrule[0.7\cmidrulewidth](lr){3-5}
    & \multirow{2}*{Transverse}
             & N      & $
                         1 + \obs{T} \tarpol_T \sin (\beta -\phi) +
                         \obs{F} \circpol \tarpol_T \cos (\beta -\phi)
                        $
                         & \label{row11} \\ 
\cmidrule(r){4-5} \addlinespace[0.5em]
    &        & Y      & $
                         \begin{aligned}
                         & 1 + \obs{P} \recpol_{y^\prime} 
						 + \left( 
                         \obs{\Sigma} \recpol_{y^\prime} + \obs{T} 
                         \right) \tarpol_T \sin (\beta -\phi) \\ 
                         & \enspace + \left( 
                         \obs{T_{x^\prime}} \recpol_{x^\prime} + \obs{T_{z^\prime}} \recpol_{z^\prime} 
                         \right) \tarpol_T \cos (\beta-\phi) \\
                         & \enspace +
                         \left\{
                         \obs{C_{x^\prime}} \recpol_{x^\prime} 
                         + \obs{C_{z^\prime}} \recpol_{z^\prime} 
                         + \left(
                         \obs{F}  + \obs{G} 
                         \recpol_{y^\prime}  \right) \tarpol_T \cos (\beta-\phi) \right.\\
                         & \left. \qquad + \left( 
                         \obs{O_{x^\prime}} \recpol_{z^\prime} - \obs{O_{z^\prime}} \recpol_{x^\prime} 
                         \right) \tarpol_T \sin (\beta-\phi)
                         \right\}\circpol
                         \end{aligned}
                        $
                         & \label{row12}\\ \addlinespace[0.5em]
\midrule 
\multirow{6}*[-5.0em]{Linear}    
    & \multirow{2}*{Unpolarized}    
            & N      & $1 - \obs{\Sigma} \linpol \cos 2(\alpha-\phi)$ & \label{row13}\\ 
\cmidrule(r){4-5} \addlinespace[0.5em]
    &       & Y      & 
                       $\begin{aligned}
                       & 1 + \obs{P} \recpol_{y^\prime}  -
                        \left\{
                        \obs{\Sigma} + \obs{T} \recpol_{y^\prime}
                        \right\} \linpol \cos 2(\alpha-\phi) \\
                        & \enspace +
                        \left\{
                        \obs{O_{x^\prime}} \recpol_{x^\prime} + \obs{O_{z^\prime}} \recpol_{z^\prime}
                        \right\} \linpol \sin 2(\alpha-\phi) \end{aligned} 
                       $ 
                        & \label{row14}\\ \addlinespace[0.5em]
\cmidrule[0.7\cmidrulewidth](lr){3-5}
    & \multirow{2}*{Longitudinal}
            & N      &  $1 - \obs{\Sigma} \linpol \cos 2(\alpha-\phi) +
                         \obs{G} \tarpol_y \linpol \sin 2(\alpha-\phi)$& \label{row15}\\ 
\cmidrule(r){4-5} \addlinespace[0.5em]
    &       & Y      & $ \begin{aligned}
                        &1 + \obs{P} \recpol_{y^\prime} + 
                        \left(
                        \obs{L_{x^\prime}} \recpol_{x^\prime} + 
                        \obs{L_{z^\prime}} \recpol_{z^\prime} 
                        \right) \tarpol_z \\
                        & \; - 
                        \left\{
                        \obs{\Sigma} + \obs{T} \recpol_{y^\prime} +   
                        \left(
                        \obs{T_{x^\prime}} \recpol_{z^\prime} - \obs{T_{z^\prime}} \recpol_{x^\prime} 
                        \right) \tarpol_z
                        \right\} \linpol \cos 2(\alpha-\phi) \\
                        & \; +
                        \left\{
                        \left(
                        \obs{F} \recpol_{y^\prime} + \obs{G} 
                        \right) \tarpol_z + 
                        \obs{O_{x^\prime}} \recpol_{x^\prime} + \obs{O_{z^\prime}} \recpol_{z^\prime}
                        \right\} \linpol \sin 2(\alpha-\phi)
                        \end{aligned}
                        $
                        & \label{row16} \\ \nopagebreak \addlinespace[0.5em]
\cmidrule[0.7\cmidrulewidth](lr){3-5} \addlinespace[0.5em]
    & \multirow{2}*[-2.0em]{Transverse}
             & N      & $\begin{aligned}
                         & 1 + \obs{T} \tarpol_T \sin (\beta-\phi) \\
                         & \enspace -
                         \left\{
                         \obs{\Sigma}+ 
                         \obs{P} \tarpol_T \sin(\beta-\phi) 
                         \right\} \linpol \cos 2(\alpha-\phi) \\
                         & \enspace +
                         \obs{H} \tarpol_T \cos(\beta-\phi) \linpol \sin 2(\alpha-\phi)
                         \end{aligned}
                        $
                         & \label{row17} \\ \addlinespace[0.5em]
\cmidrule(r){4-5} \addlinespace[0.5em]
    &        & Y      & $
                         \begin{aligned}
                         & 1 - \linpol \recpol_{y^\prime} \tarpol_T \sin(\beta-\phi) 
                         \cos 2(\alpha-\phi) + \obs{P} \recpol_{y^\prime} \\
                         & \enspace +
                         \left(
                         \obs{T_{x^\prime}} \recpol_{x^\prime} + \obs{T_{z^\prime}} \recpol_{z^\prime}
                         \right) \tarpol_T \cos(\beta-\phi) \\
                         & \enspace + 
                         \left(
                         \obs{\Sigma} \recpol_{y^\prime} + \obs{T}
                         \right) \tarpol_T \sin(\beta-\phi) \\
                         & \enspace -
                         \left\{
                          \obs{\Sigma} 
                         + \obs{T} \recpol_{y^\prime}
                         + \obs{P} \tarpol_T \sin(\beta-\phi) \right. \\
                         & \left. \qquad - \left(
                         \obs{L_{x^\prime}} \recpol_{z^\prime} - \obs{L_{z^\prime}} \recpol_{x^\prime}
                         \right) \tarpol_T \cos(\beta-\phi)
                         \right\} \linpol \cos 2(\alpha-\phi) \\
                         & \enspace +
                         \left\{
                         \obs{O_{x^\prime}} \recpol_{x^\prime} + \obs{O_{z^\prime}} \recpol_{z^\prime} 
                         + \left(
                         \obs{E} \recpol_{y^\prime}  + \obs{H} 
                         \right) \tarpol_T \cos(\beta-\phi) \right. \\
          	             & \left. \qquad - \left(
                         \obs{C_{x^\prime}} \recpol_{z^\prime} - \obs{C_{z^\prime}} \recpol_{x^\prime} 
                         \right) \tarpol_T \sin(\beta-\phi) 
                         \right\} \linpol \sin 2(\alpha-\phi)
                         \end{aligned}
						$
                         & \label{row18} \\ \addlinespace[0.5em]
\bottomrule 

\end{longtable}

%% file: asymm-config-table.tex


\begin{longtable}{C{1.5cm} C{1.5cm} C{3.5cm} p{8.5cm} >{(\refstepcounter{subexno}\theexno)} C{1cm}}
\captionsetup{width=.75\textwidth}
\caption{\label{tab:asymmetries} Expressions for asymmetries for different experiments. The definitions of angles are shown in figure \ref{fig:axes}. Configurations are labeled U, C and L for unpolarized, circular and linear polarized photon beams; U, L and T for unpolarized, longitudinal and transverse target polarization. Where degrees of polarization are labelled +ve or -ve, this refers to their direction with respect to an axis: lab $x$, $y$ or $z$ for target polarization; photon beam direction for circular photon polarization. } \\

\toprule
\multicolumn{3}{c}{\emph{Configuration}} &  
\hfil\multirow{2}[8]{*}{\emph{Asymmetry formula, $A_\sigma$}}\hfill \\
\cmidrule(lr){1-3}
Beam & Target & Settings &   \\ 

\endfirsthead
\midrule

Beam & Target & Settings &  \hfil\emph{Asymmetry formula, $A_\sigma$}\hfill \\
\midrule
\endhead

\multicolumn{5}{r}{\emph{Continued on next page}}
\endfoot
\endlastfoot

\midrule 
\addlinespace[0.5em]
\multirow{2}*[-1.5em]{U}  
    & L
    & $\tarpol_z\;+$ve; $\tarpol_z\;-$ve      
    & $\dfrac{\tarpol_z \left(\obs{L_{x^\prime}} \recpol_{x^\prime} + \obs{L_{z^\prime}} \recpol_{z^\prime} \right)}{1 + \obs{P} \recpol_{y^\prime}}
                       $
    					 & \label{a:row1} \\
\addlinespace[1.0em]                         

    & T       
    & $\beta = 0;\;\beta=\pi$      
    & 
    $\begin{aligned}
    \dfrac{ \tarpol_T}
    {1 + \obs{P} \recpol_{y^\prime}}
    & \left\{ 
    \left( \obs{T_{x^\prime}} \recpol_{x^\prime} +   \obs{T_{z^\prime}} \recpol_{z^\prime} \right) \cos \phi
        \right. \\
    & \left. \enspace -
    \left( \obs{\Sigma} \recpol_{y^\prime} + \obs{T} 
       \right) \sin \phi 
    \right\}
    \end{aligned}$
                         & \label{a:row2} \\
\addlinespace[0.5em]
\midrule 
\addlinespace[0.5em]
\multirow{3}*[-3.45em]{C}    
    & U    
    & $\circpol\;+$ve; $\circpol\;-$ve      
    & $\dfrac{\circpol \left(\obs{C_{x^\prime}} \recpol_{x^\prime} + \obs{C_{z^\prime}} \recpol_{z^\prime} \right)}{1 + \obs{P} \recpol_{y^\prime}}$
    & \label{a:row3} \\

\addlinespace[1.0em]

    & L
    & ($\circpol\;+$ve, $\tarpol_z\;+$ve |
      $\circpol\;-$ve, $\tarpol_z\;-$ve); 
      ($\circpol\;+$ve, $\tarpol_z\;-$ve |
      $\circpol\;-$ve, $\tarpol_z\;+$ve) 
    & $\dfrac{\circpol \tarpol_z \left(\obs{E} + \obs{H} \recpol_{y^\prime} \right)}{1 + \obs{P} \recpol_{y^\prime}}$
    & \label{a:row4} \\ 
    
\addlinespace[1.0em]
    
    & T
    & ($\circpol\;+$ve, $\beta=0$ |
      $\circpol\;-$ve, $\beta=\pi$); 
      ($\circpol\;+$ve, $\beta=0$ |
      $\circpol\;-$ve, $\beta=\pi$)   
    & 
    $\begin{aligned}
    \dfrac{ \circpol \tarpol_T}
    {1 + \obs{P} \recpol_{y^\prime}}
    & \left\{ 
    \left(  \obs{F}  + \obs{G} \recpol_{y^\prime} \right) 
    \cos \phi
        \right. \\
    & \left. \enspace -
    \left( \obs{O_{x^\prime}} \recpol_{z^\prime} - \obs{O_{z^\prime}} \recpol_{x^\prime}
       \right)
    \sin \phi 
    \right\}
    \end{aligned}$
    & \label{a:row5}\\ 
    \addlinespace[0.5em]
\midrule 
\addlinespace[0.5em]

\pagebreak
\addlinespace[0.5em]
\multirow{3}*[-4.2em]{L}    
    & U    
    & $\alpha=0$; $\alpha=\frac{\pi}{2}$      
    & 
    $\begin{aligned}
    \dfrac{ - \linpol }
    {1 + \obs{P} \recpol_{y^\prime}}
    & \left\{ 
    \left(  \obs{\Sigma}  + \obs{T} \recpol_{y^\prime} \right) 
    \cos 2 \phi
        \right. \\
    & \left. \enspace +
    \left( \obs{O_{x^\prime}} \recpol_{x^\prime} -      
    \obs{O_{z^\prime}} \recpol_{z^\prime}
       \right) 
    \sin 2 \phi
    \right\}
    \end{aligned}$
    & \label{a:row6}\\ 
\addlinespace[1.0em]
    & L
    & ($\alpha=0$, $\tarpol_z\;+$ve |
      $\alpha=\frac{\pi}{2}$, $\tarpol_z\;-$ve); 
      ($\alpha=0$, $\tarpol_z\;-$ve |
      $\alpha=\frac{\pi}{2}$, $\tarpol_z\;+$ve)      
    & 
    $\begin{aligned}
    \dfrac{ - \linpol \tarpol_z}
    {1 + \obs{P} \recpol_{y^\prime}}
    & \left\{ 
    \left(  \obs{T_{x^\prime}} \recpol_{z^\prime} - \obs{T_{z^\prime}} \recpol_{x^\prime} \right) 
    \cos 2 \phi
        \right. \\
    & \left. \enspace +
    \left( \obs{F} \recpol_{y^\prime} + \obs{G}
       \right) 
    \sin 2 \phi
    \right\}
    \end{aligned}$
                        & \label{a:row7} \\ 
                        
    \addlinespace[1.0em]
    
    & T
    & ($\alpha=0$, $\beta=0$ |
      $\alpha=\frac{\pi}{2}$, $\beta=\pi$); 
      ($\alpha=0$, $\beta=0$ |
      $\alpha=\frac{\pi}{2}$, $\beta=\pi$)      
    & $\begin{aligned}
    \dfrac{ \linpol \tarpol_T}
    {1 + \obs{P} \recpol_{y^\prime}}
    & \left\{ 
    \left( \recpol_{y^\prime} + \obs{P}  \right) 
    \sin\phi \cos 2 \phi 
    \right. \\
    & +
     \left( \obs{L_{x^\prime}} \recpol_{z^\prime} - \obs{L_{z^\prime}} \recpol_{x^\prime}\right) 
    \cos\phi \cos 2 \phi 
    \\
    & +
     \left( \obs{C_{x^\prime}} \recpol_{z^\prime} - \obs{C_{z^\prime}} \recpol_{x^\prime} \right) 
    \sin\phi \sin 2 \phi 
    \\    
    & \left. +
     \left( \obs{E} \recpol_{y^\prime}  + \obs{H} \right) 
    \cos\phi \sin 2 \phi 
    \right\}
    \end{aligned}
    $
    & \label{a:row8} \\ 
    \addlinespace[0.5em]
\bottomrule 

\end{longtable}